\documentstyle[times,mathptm,epsfig,here,amsfonts,12pt,twoside]{mbook}

\begin{document}
\newcommand{\pc}{\mbox{\'c}}
\newcommand{\pC}{\mbox{\'C}}
\newcommand{\pn}{\mbox{\'n}}
\newcommand{\pN}{\mbox{\'N}}
\newcommand{\po}{\mbox{\'o}}
\newcommand{\pO}{\mbox{\'O}}
\newcommand{\ps}{\mbox{\'s}}
\newcommand{\pS}{\mbox{\'S}}
\newcommand{\px}{\mbox{\'z}}
\newcommand{\pX}{\mbox{\'Z}}
\newcommand{\pz}{\mbox{\.z}}
\newcommand{\pZ}{\mbox{\.Z}}
\newcommand{\pl}{\mbox{\l}}
\newcommand{\pL}{\mbox{\L}}
\newcommand{\pa}{\mbox{\c{a}}}
\newcommand{\pA}{\mbox{\c{A}}}
\newcommand{\pe}{\mbox{\c{e}}}
\newcommand{\pE}{\mbox{\c{E}}}
\newcommand{\sie}{\mbox{si{\pe} }}
\newcommand{\ze}{\mbox{{\pz}e }}
\newcommand{\iz}{\mbox{i{\pz} }}
\newcommand{\Kosciola}{\mbox{Ko\'scio{\l}a }}
\newcommand{\kosciola}{\mbox{ko\'scio{\l}a }}
\newcommand{\Kosciol}{\mbox{Ko\'sci{\po}{\l} }}
\newcommand{\kosciol}{\mbox{ko\'sci{\po}{\l} }}
\newcommand{\Kosciolowi}{\mbox{Ko\'scio{\l}wi }}
\newcommand{\kosciolowi}{\mbox{ko\'scio{\l}wi }}
\newcommand{\Panstwowych}{\mbox{Pa\'nstwowych }}
\newcommand{\panstwowych}{\mbox{pa\'nstwowych }}
\newcommand{\Panstwa}{\mbox{Pa\'nstwa }}
\newcommand{\panstwa}{\mbox{pa\'nstwa }}
\newcommand{\Panstwo}{\mbox{Pa\'nstwo }}
\newcommand{\panstwo}{\mbox{pa\'nstwo }}
\newcommand{\Panstwu}{\mbox{Pa\'nstwu }}
\newcommand{\panstwu}{\mbox{pa\'nstwu }}


\pagestyle{empty}

\begin{center}
{\Large  Jagiellonian University }  \\
{\Large  The Faculty of Physics, Astronomy} \\
{\Large  and Applied Computer Science }\\
{\Large  Marian Smoluchowski Institute of Physics}

\vspace{3.5cm}

{\bf {\Huge Isospin dependence of the $\eta^{\prime}$ meson production
 in nucleon--nucleon collisions }}

\vspace{0.5cm}

{\bf {\Large Joanna Klaja}}
\end{center}

\vspace{3.5cm}

\begin{flushright}
Ph.D. dissertation prepared at the Nuclear Physics Department \\
of the Marian Smoluchowski Institute of Physics
of the Jagiellonian University \\
and at the Institute for Nuclear Physics in the Research Centre J{\"u}lich\\
guided by: Prof. Pawe{\l} Moskal  \\
\end{flushright}
\vspace{1.5cm}
\begin{center}
    Cracow 2009
\end{center}

   \cleardoublepage

   \pagestyle{empty}

\begin{center}
{\bf Abstract}\\
\end{center}

\vspace{0.7cm}

The upper limit of the total cross section for quasi-free $pn \to pn\eta^{\prime}$
reaction has been determined in the excess energy range near the kinematical threshold.\\
The measurement has been
carried out at the COSY--11 detection setup using a proton beam and a deuteron cluster target.
The identification of the $\eta^{\prime}$ meson has been performed using the missing mass
technique. The energy dependence of the upper limit of the cross section is extracted with a fixed
proton beam momentum of $p_{beam}=3.35$~GeV/c and exploiting the Fermi
momenta of nucleons inside the deuteron. The data cover a range of centre-of-mass
excess energies from 0 to 24~MeV.\\
The experimentally determined upper limit of the ratio
$R_{\eta^{\prime}}~=~{{\sigma(pn \to pn\eta^{\prime})} \over {\sigma(pp \to pp\eta^{\prime})}} $,
which is smaller than the ratio for the $\eta$ meson, excludes the excitation of the
S$_{11}$(1535) resonance as a dominant production mechanism of the $\eta^{\prime}$ meson in
nucleon-nucleon collisions.
At the same time, the determined upper limits of R$_{\eta^{\prime}}$ go in the direction
of what one would expect in the glue production and production via mesonic currents.
For quantitative tests of these mechanisms an order of magnitude larger statistics and
a larger energy range would be required. This can be reached with the WASA-at-COSY facility.

\begin{center}
{\bf Streszczenie}\\
\end{center}

\vspace{0.7cm}

G{\'o}rna granica ca{\l}kowitego przekroju czynnego dla kwazi-swobodnej reakcji
$pn~\to~pn\eta^{\prime}$ zosta{\l}a wyznaczona w przyprogowym obszarze energii.\\
Pomiar zosta{\l} przeprowadzony wykorzystuj{\pa}c uk{\l}ad detekcyjny COSY--11,
wi{\pa}zk{\pe} protonow{\pa} oraz klastrow{\pa} tarcz{\pe} deuteronow{\pa}.
Mezon $\eta^{\prime}$ zosta{\l} zidentyfikowany przy u{\.z}yciu techniki masy brakuj{\pa}cej.
Zale{\.z}no{\'s}{\'c} g{\'o}rnej granicy ca{\l}kowitego przekroju czynnego
od energii (funkcj{\pe} wzbudzenia)
przy sta{\l}ej warto{\'s}ci p{\pe}du wi{\pa}zki $p_{beam}~=~3.35$~GeV/c, uzyskano
dzi{\pe}ki p{\pe}dowi Fermiego nukleon{\'o}w w deuteronie. Dane zosta{\l}y zmierzone
w przedziale energii nadprogowej Q od 0~MeV do 24~MeV.\\
Otrzymana g{\'o}rna granica stosunku R$_{\eta^{\prime}}$ ca{\l}kowitych przekroj{\'o}w
czynnych dla reakcji $pn~\to~pn\eta^{\prime}$
i $pp~\to~pp\eta^{\prime}$, kt{\'o}rej warto{\'s}{\'c} okaza{\l}a si{\pe}
by{\'c} znacznie mniejsza od analogicznego stosunku zmierzonego dla mezonu $\eta$, wyklucza
hipotez{\pe} i{\.z} dominuj{\pa}cym procesem w produkcji mezonu $\eta^{\prime}$ jest wzbudzenie
rezonansu S$_{11}$(1535). R{\'o}wnocze{\'s}nie, uzyskany wynik nie wyklucza innych
mechanizm{\'o}w produkcji mezonu $\eta^{\prime}$ takich jak pr{\pa}dy mezonowe czy wzbudzenie gluon{\'o}w.\\
Dla przeprowadzenia ilo{\'s}ciowych test{\'o}w mo{\.z}liwych mechanizm{\'o}w produkcji mezonu $\eta^{\prime}$
wymagane jest wykonanie eksperymentu z wi{\pe}ksz{\pa} statystyk{\pa} oraz w
wi{\pe}kszym zakresie energii. Tak{\pa} mo{\.z}liwo{\'s}{\'c} daje uk{\l}ad detekcyjny WASA-at-COSY.


   \tableofcontents
   
\chapter{Introduction}

\pagestyle{myheadings}
\markboth{Introduction}{Introduction}

The main aim of the studies presented in this thesis is the determination of
the excitation function of the total cross section for the quasi-free
$pn \to pn\eta^{\prime}$ reaction near the kinematical threshold.\\

Despite the fact that the $\eta$ and $\eta^{\prime}$ mesons - which are members
of the ground-state pseudoscalar nonet~\cite{amsler-pdg} - were discovered many decades ago, they are still
subject of many theoretical and experimantal investigations.\\
According to the quark model, $\eta$ and $\eta^{\prime}$ mesons can be described as a
mixture of the singlet and octet states of the SU(3) - flavour pseudoscalar meson nonet.
Within the one mixing angle scheme, a small mixing angle ($\theta~=~-15.5^{\circ}$) implies
that the masses of $\eta$ and $\eta^{\prime}$ mesons should be almost equal. However, masses
of these mesons differ by about a factor of two. Additionaly, the mass of the
$\eta^{\prime}$ meson does not fit into the SU(3) scheme and it is thought to be induced by the
gluonic component in its wave function. This hypothesis is strenghtened by the decay scheme of
mesons like $B^{+}$ or $D^{+}_{s}$ since the branching ratios of the decay of
these mesons into some channels involving $\eta^{\prime}$ are significantly higher than the analogous
for $\eta$ especially in processes requiring the involvement of gluons~\cite{du-prd59,ball-plb365,fritzsch-plb415}. \\
The properties of the $\eta^{\prime}$ meson should manifest itself in the production mechanism
in the collisions of nucleons.
At present there is not much known about the relative contribution of the possible reaction mechanisms
of the $\eta^{\prime}$ meson production in nucleon-nucleon collisions. It is expected that the
$\eta^{\prime}$ meson can be produced through heavy meson exchange, through the excitation of
an intermediate resonance or via the fusion of virtual mesons~\cite{nakayama-prc61,kampfer-ep,cao-prc78}.
However, it is not possible to judge about the mechanism responsible for the $\eta^{\prime}$
meson production only from the total cross section of the $pp \to pp\eta^{\prime}$ reaction~\cite{moskal-hab}.
Therefore, one has to investigate the $\eta^{\prime}$ production more detailed by e.g. selecting
separate channels in the relevant degrees of freedom like the isospin which means a comparison of both,
proton-proton and proton-neutron scattering. This conclusion motivated our investigations, which are
presented in this thesis.\\
A comparison of the close-to-threshold total cross section for the $\eta^{\prime}$ production
in both $pp \to pp\eta^{\prime}$ and $pn \to pn\eta^{\prime}$ reactions constitutes a tool not only
to investigate the production of the $\eta^{\prime}$ meson in channels of isospin $I~=~1$ and $I~=~0$
but also may provide insight into the flavour-singlet (perhaps also into gluonium) content of the
$\eta^{\prime}$ meson and the relevance of quark-gluon or hadronic degrees of freedom in the
creation process.
It is also possible that the $\eta^{\prime}$ meson is produced from excited glue
in the interaction region of the colliding nucleons, which couple to the $\eta^{\prime}$
meson directly via its gluonic component or through its SU(3)-flavour-singlet
admixture~\cite{bass-pst99,bass-app11}. As suggested in reference~\cite{bass-plb463},
$\eta^{\prime}$ production via the colour-singlet object does not depend on the
total isospin of the colliding nucleons and should lead to the same production amplitude
for the $\eta^{\prime}$ in the $pn \to pn\eta^{\prime}$ and $pp \to pp\eta^{\prime}$ reactions.
In case of the $\eta$ meson, the ratio of the total cross sections for the reactions
$pn \to pn\eta$ and $pp \to pp\eta$ was determined to be $R_{\eta} =$ 6.5
in the excess energy range from $\approx 15$~MeV to $\approx 160$~MeV~\cite{calen-prc58},
what suggests the dominance of isovector meson exchange in the $\eta$ production in nucleon-nucleon
collisions.
Since the fractional amounts of different quark flavours of $\eta$ and $\eta^{\prime}$
mesons are very similar,
in case of the dominant isovector meson exchange -- by the analogy to
the $\eta$ meson production -- we can expect that the ratio $R_{\eta^{\prime}}$
should also be about 6.5. If however the $\eta^{\prime}$ meson is produced via its
flavour-blind gluonium component from the colour-singlet glue excited
in the interaction region the ratio should approach unity after corrections
for the initial and final state interactions.
The close--to--threshold
excitation function for the $pp \to pp\eta^{\prime}$ reaction has been determined in
previous experiments~\cite{moskal-plb474,khoukaz-epj,balestra-plb491,moskal-prl80,hibou-plb438}
and the determination of the total cross section for the $\eta^{\prime}$ meson production in
the proton-neutron interaction constitutes the main motivation for the experiment
which is subject of this thesis.\\
It is worth to strees that the $pn \to pn\eta^{\prime}$ reaction was never investigated so far.
Such studies are challenging experimentally because the total cross section is expected to be about
a factor of fourty to hundred less than in case of the $\eta$ meson and additionally
(in comparison to the $\eta$ near threshold production) the cross section of the multi-pion
background grows significantly. For some channels (like eg. 3$\pi$) even by more than a few
orders of magnitude~\cite{zielinski-dt}. The measurement of the $pn \to pn\eta^{\prime}$
reaction is also much more difficult in comparison to the $pp \to pp\eta^{\prime}$ reaction
due to the neutron in the final state and the lack of a pure neutron target.

The experiment described in this thesis has been performed by the COSY--11 collaboration
by means of the COSY--11 facility at the Cooler Synchrotron COSY at the Research Centre J{\"u}lich
in Germany.
A quasi-free proton-neutron reaction was induced by a proton beam
impinging on a deuteron target. For the data analysis
the proton from the deuteron is considered as a spectator which does not interact
with the bombarding proton, but escapes untouched and hits the detector
carrying the Fermi momentum possessed at the time of the reaction.
The experiment is based on the registration of all outgoing nucleons
from the $pd\to ppnX$ reaction.
Protons moving forward are measured in two drift
chambers and scintillator detectors and the neutron is registered
in the neutral particle detector.
Protons considered as a spectator are measured by a dedicated
silicon-pad detector. The total energy available for the quasi-free proton-neutron
reaction can be calculated for each event from the vector of the momenta
of the spectator and beam protons.
The absolute momentum of the neutron is determined from the time-of-flight between
the target and the neutron detector.
Application of the missing mass technique allows to identify events
of the creation of the meson under investigation.\\

The thesis is divided into ten chapters. The second chapter - following the introduction -
describes briefly the motivation for investigating of the $NN \to NN\eta^{\prime}$ process, in particular
in view of the study of the $\eta^{\prime}$ production mechanism and its structure. In this chapter
the most interesting issues concerning the $\eta^{\prime}$ physics are presented,
and different possible production mechanisms and the predictions of the total cross section
for the $pn \to pn\eta^{\prime}$ reaction are given.\\
The spectator model is introduced in chapter three, with the description of the quasi-free meson production.\\
An introduction of the cooler synchrotron COSY and the COSY--11 detection setup is presented in the
fourth chapter.\\
The fifth chapter is devoted to the calibration of the detectors. A special emphasis is put on the neutral
particle  and spectator detectors, two new devices which enabled to measure the quasi-free
$pn \to pnX$ reactions.\\
A detailed description of the data analysis and the result of the identification of the $pn \to pn\eta^{\prime}$
reaction are presented in chapter six.\\
The seventh chapter is devoted to the determination of the luminosity based on the quasi-free
proton-proton elastic scattering.\\
Upper limits of the total cross section as well as of the ratio
$R_{\eta^{\prime}}$~=~$\sigma(pn \to pn\eta^{\prime})$ / $\sigma(pp \to pp\eta^{\prime})$
are given in chapter eight.\\
The results of the analysis are compared to theoretical predictions in chapter nine.\\
The tenth chapter comprises the summary and perspectives. In particular the
possibility  to study of the $\eta^{\prime}$ production in the pure isospin $I=$~0 channel
via the $pn \to d\eta^{\prime}$ is discussed.


\chapter{Motivation}
\pagestyle{myheadings}

\markboth{\bf Motivation}
         {\bf Motivation}

Understanding of the structure of hadrons is a long standing challenge.
Quantum chromodynamics (QCD) is the well established theory of
strong interactions that describes the underlying forces of coloured
quarks which bind them together to form the colour neutral hadrons
observed in nature. However, due to the increasing coupling constant with
decreasing energy this theory does not allow at present for the exact description
of the hadron structure in the low energy domain. One of the hadrons which is especially intriguing
is the $\eta^{\prime}$ meson. \\
The $\eta^{\prime}$ meson was first observed in 1964~\cite{kablfleisch-plr12,goldberg-prl12} in the reaction
$K^{-}p \to \Lambda \eta^{\prime}$.
Although the $\eta$ and $\eta^{\prime}$ mesons -- the members of the nonet
of the lightest pseudoscalar mesons -- have been discovered many decades ago,
they are still  subject of considerable interest of theoretical as well as experimental
studies. Particularly in case of the $\eta^{\prime}$ meson, despite more than fourty years
of investigations, its structure and
properties, as well as the production mechanism in collisions involving hadrons
are still not well determined.

According to the quark model, the two physical states of the $\eta$ and $\eta^{\prime}$
mesons are considered as a mixture of the SU(3) pseudoscalar octet ($\eta_8$) and
singlet ($\eta_1$) states with the pseudoscalar mixing angle $\theta_P$:
\begin{equation}
\label{eta_etaprime}
\begin{array}{l}
|\eta\rangle=
\cos\theta_{P}|\eta_{8}\rangle-\sin\theta_{P}|\eta_{1}\rangle\ ,\\[1ex]
|\eta\prime\rangle=
\sin\theta_{P}|\eta_{8}\rangle+\cos\theta_{P}|\eta_{1}\rangle\ ,
\end{array}
\end{equation}
where, following the notation introduced by Gilman and Kaufman~\cite{gilman-prd36} and
previous work by Rosner~\cite{rosner-prd27}, the SU(3) pseudoscalar octet state
$\eta_8$ and singlet state $\eta_1$ are:
\begin{equation}
\label{eta8_eta1}
|\eta_{8}\rangle=\frac{1}{\sqrt{6}}|u\bar u+d \bar d-2s \bar s\rangle\ ,\ \
|\eta_{1}\rangle=\frac{1}{\sqrt{3}}|u\bar u+d \bar d+s \bar s\rangle\ .
\end{equation}
The value of the $\eta-\eta^{\prime}$ mixing angle in the pseudoscalar meson nonet has
been discussed many times in the last years~\cite{bramon-epjc7}. The most up to date value
of the mixing angle $\theta_P$, averaged over all present experimental results,
amounts to $\theta_P = -15.5^{\circ} \pm 1.3^{\circ}$~\cite{bramon-epjc7}.
Such small mixing angle implies
similar amounts of strange and nonstrange quark content:
\begin{equation}
\label{eta_etaprime_usd}
\begin{array}{l}
|\eta\rangle=0.77 \cdot \frac{1}{\sqrt{2}}(u\bar u+d \bar d) - 0.63 \cdot s \bar s\ ,\\[1ex]
|\eta\prime\rangle=0.63 \cdot \frac{1}{\sqrt{2}}(u\bar u+d \bar d) - 0.77 \cdot s \bar s\ .
\end{array}
\end{equation}
This suggests that the masses of both $\eta$ and $\eta^{\prime}$ mesons should be almost
equal. However, the values of these masses differ by about a factor of
two. Concurrently, the mass of the $\eta^{\prime}$ meson does not fit utterly to the SU(3)
scheme. More surprisingly, the masses of all the pseudescalar mesons, vector mesons and
baryons are well described in terms of the naive quark model.
\par
There are more differences in the physical properties of both mesons. For example:
excited states of nucleons exists which decay via the emission of the $\eta$ meson,
yet none of the baryon resonances decay via the emission of the $\eta^{\prime}$
meson~\cite{amsler-pdg}. Very high $\eta^{\prime}$ apperance in
the decays of {\it B} and $D_S$ mesons~\cite{behrens-prl80,jessop-prd58,brandenburg-prl75}
are observed.
The branching ratio for $B^{+} \to K^{+}\eta^{\prime} = (6.5\pm1.7) \cdot 10^{-5}$ is much larger than the one
for the corresponding $\eta$ channel $B^{+} \to K^{+}\eta < 1.4 \cdot 10^{-5}$~\cite{behrens-prl80}.
Similar relations are found in the decay of the D$_S$ meson where
${\Gamma(D^{+}_{S} \to \eta^{\prime}\rho^{+}) \over \Gamma(D^{+}_{S} \to \eta^{\prime}e^{+}\nu)}~
=~12.0 \pm 4.3$ exceeds
${\Gamma(D^{+}_{S} \to \eta\rho^{+}) \over \Gamma(D^{+}_{S} \to \eta e^{+}\nu)}~=~
4.4 \pm 1.2$ by a factor of about three~\cite{jessop-prd58,brandenburg-prl75}.
It is worth to stress that the observed branching
ratios do not agree with the predictions which ignore the gluonic content of
the $\eta^{\prime}$~\cite{du-prd59}.
\par
There exist also essential differences between the production of $\eta$ and $\eta^{\prime}$ meson
in proton-proton collisions close-to-threshold.
The total cross section for the $pp \to pp\eta^{\prime}$ reaction is by a factor
of fourty smaller than the cross section for the $pp \to pp\eta$ reaction at corresponding
values of the excess energy. The shape of the excitation function is also different indicating
that the $\eta$ meson interaction with nucleons is much stronger than the $\eta^{\prime}$-nucleon
one~\cite{moskal-hab}. Thus, it is expected that not only physical properties but also
the production mechanisms of these mesons should differ from each other.
\par
\begin{figure}[H]
\centerline{\includegraphics[height=.15\textheight]{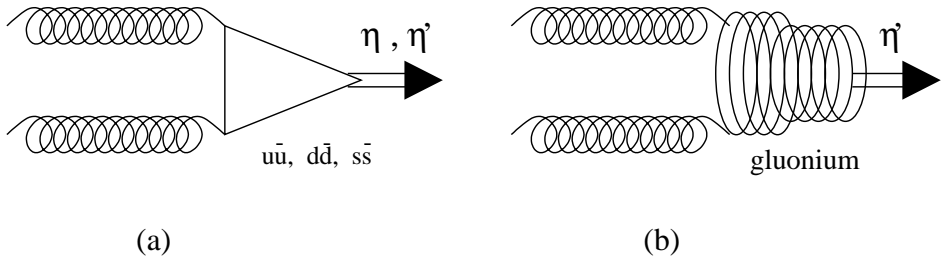}}
\caption{Coupling of $\eta$ and $\eta^{\prime}$ to two gluons through quark and
         anti-quark triangle loop (a) and through gluonic admixture (b).
         Figure and caption are adapted from reference~\cite{kou-prd63}.}
\label{kou_gluony}
\end{figure}

As was already discussed, the $\eta^{\prime}$ meson with its mass of $m_{\eta^{\prime}} \sim 958~MeV$
is far from being "light" and its mass  is almost three times larger than the value
expected if this meson would be a pure Goldstone boson associated with spontaneously broken
chiral symmetry~\cite{weinberg-prd11}. The much larger mass of the $\eta^{\prime}$ meson is
thought to be induced by the non-perturbative gluon
dynamics~\cite{hooft-prl37,witten-npb156,veneziano-npb159}
and the axial anomaly $U_{A}(1)$ ~\cite{adler-pr177}. \\
A gluonic component of the $\eta^{\prime}$ meson is introduced as a flavour singlet
state additionally to the $\eta_{1}$, which couples directly to the glue~\cite{kou-prd63}.
Thus, the $\eta^{\prime}$ meson can couple to gluons not only via the
quark and antiquark triangle loop but also directly through its gluonic admixture as shown
in figure~\ref{kou_gluony}.
Experimental indications exist that the gluonic content of the $\eta^{\prime}$ meson
constitutes a significant fraction of the $\eta^{\prime}$ meson wave function. For example,
the unexpected large
branching ratio measured for the decay of beauty particles,
$B \to \eta^{\prime}+X$~\cite{behrens-prl80}, has been interpreted as a possible experimental
evidence in this respect~\cite{atwood-plb405}.
Recenty, the KLOE collaboration
has estimated the gluonium fractional content of the $\eta^{\prime}$ meson
to be  $14\% \pm 4\%$~\cite{ambrosino-plb648}
by looking for the radiative decays $\phi \to \eta^{\prime}\gamma$ and $\phi \to \eta\gamma$.
\par
The most remarkable feature -- in the frame of the quark model -- distinguishing the
$\eta^{\prime}$ meson from all other pseudoscalar and vector ground state mesons, is the
fact, that the $\eta^{\prime}$ is predominantly a flavour-singlet combination of quark-antiquark
pairs and therefore can mix with purely gluonic states.
In any case, if there is a strong coupling of the $\eta^{\prime}$ meson to gluons, quark-gluon
degrees of freedom may play a significant role in the production dynamics of this meson, especially
close-to-thereshold where the $\eta^{\prime}$ production requires a large
momentum transfer between the nucleons and hence can occur only at short distances $\sim$~0.3~fm
in the nucleon-nucleon collision.
The role of gluonic degrees of freedom in the $\eta^{\prime}$--nucleon system can be investigated
for example through the flavour-singlet Goldberger-Treiman relation~\cite{shore-npb381}:
\begin{equation}
2 m_{N} G_{A}(0) = F g_{NN\eta^{\prime}}(0) + \frac{F^2}{2N_F} m_{\eta^{\prime}} g_{NNG}(0),
\end{equation}
which relates the nucleon-nucleon-$\eta^{\prime}$ coupling constant $g_{NN\eta^{\prime}}$ with the
flavour-singlet axial constant $G_{A}(0)$. The $g_{NNG}$ describes the coupling of
the nucleon to the gluons arising from contributions violating the Okubo-Zweig-Iizuka
rule~\cite{okubo-pl5,zweig-cernrep412,iizuka-ptps37}. The coupling constatnt $g_{NNG}$ is in part
related to the contribution of gluons to the proton spin~\cite{bass-ksiazka}.
As shown by the measurements of the EMC
collaboration omission of the spin carried by gluons in polarized protons leads to the so-called
"spin crisis"~\cite{ashman-plb206}. A small value of $G_{A}(0) \sim$~0.20-0.35 extracted from
measurements of the deep inelastic muon-proton scattering~\cite{ashman-plb206} and a large mass of
the $\eta^{\prime}$ meson would be explained by the positive $g_{NNG}~\sim$~2.45 value.
However, neither $g_{NN\eta^{\prime}}$ nor $g_{NNG}$ have been measured directly in experiments so far.
There is only an estimation of the upper limit of $g_{NN\eta^{\prime}} <$~2.5 derived from the
close-to-threshold total cross section for the $pp \to pp\eta^{\prime}$ reaction~\cite{moskal-prl80}.
From this point of view an investigation of processes where the $\eta^{\prime}$ meson is produced
directly off a nucleon, such as the $NN \to NN\eta^{\prime}$ reaction may be considered as a tool
for supplying the information about the above coupling constants and the role of the gluons in
dynamical chiral symmetry breaking in low-energy QCD.
\par
\begin{figure}[H]
\parbox[c]{0.5\textwidth}{\centering\includegraphics[width=.45\textwidth]{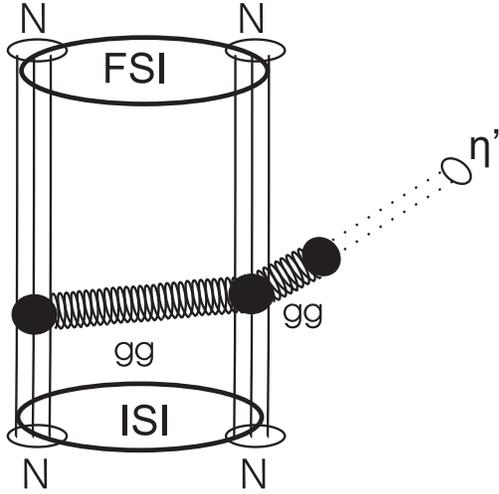}}\hfill
\parbox{.4\textwidth}{\caption{ Diagram depicting possible quark-gluon dynamics of the
                     $NN \to NN\eta^{\prime}$ reaction. Production via a rescattering
                     of a "low energy pseudoscalar pomeron like object" (gluonic current) is isospin independent.
                     \label{grafy_gluony} }}
\end{figure}

As suggested by Bass~\cite{bass-pst99,bass-app11} the $\eta^{\prime}$ meson can be created via
a {\it contact interaction} from the glue which is excited in the interaction region of the
colliding nucleons. A gluon-induced {\it contact interaction} contributing to the close-to-threshold
$NN \to NN\eta^{\prime}$ cross section is derived in the frame of the U(1)-anomaly
extended chiral Lagrangian. The physical interpretation of the contact interaction is a very
short distance (0.2~fm) interaction where created gluons could couple
to the $\eta^{\prime}$ directly via its gluonic component or through its
flavour-singlet admixture $\eta_{1}$. This gluonic contribution to the total cross section
of the $NN \to NN\eta^{\prime}$ reaction is additional to the production mechanism associated with
meson exchange~\cite{nakayama-prc61,nakayama-prc69,kampfer-ep,cao-prc78}.
As proposed by Bass in reference~\cite{bass-plb463},
the $\eta^{\prime}$ meson production via the colour-singlet object does not depend on the
total isospin of the colliding nucleons (see figure~\ref{grafy_gluony})
and should lead to the same amplitudes of the production
for the $pn \to pn\eta^{\prime}$ and $pp \to pp\eta^{\prime}$ reactions. This observation
motivated the studies of the isospin dependence of the $\eta^{\prime}$ meson production
in the nucleon--nucleon collisions and in particular a measurement of the $pn \to pn\eta^{\prime}$
reaction presented in this thesis.\\

Treating proton and neutron as different states of the nucleon
distinguished only by the isospin projection, $+{{1}\over{2}}$ for the proton
and $-{{1}\over{2}}$ for the neutron, we may classify the $NN \to NNX$ reactions
according to the total isospin of the nucleon pair in the initial and final state.
A total isospin of two nucleons equals 1 for proton-proton and
neutron-neutron pairs, and may acquire the value 1 or 0 for the
neutron-proton system. Since the $\eta^{\prime}$ meson is isoscalar, there are only two pertinent
transitions for the $NN \to NNX$ reaction, provided that it occurs via an isospin
conserving interaction. Thus, it is sufficient  to measure two reaction channels for
an unambiguous determination of the isospin 0 and 1 cross sections~\cite{moskal-hab}.\\

\begin{figure}[H]
\centerline{\includegraphics[height=.4\textheight]{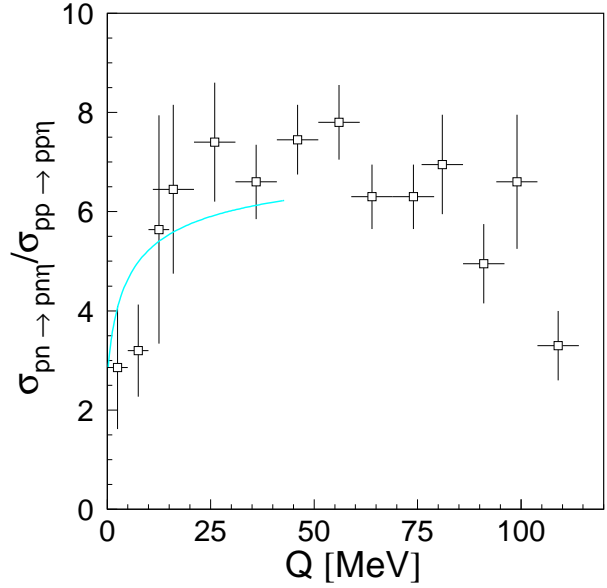}}
\caption{The ratio $(R_{\eta})$ of the total cross sections for the
         $pn \to pn\eta$ and $pp \to pp\eta$ reactions.
         The superimposed line indicates a result of the fit taking into account the final
         state interaction of nucleons~\cite{moskal-prc79}.}
\label{ratio_eta}
\end{figure}

The isospin dependence has been already established in case of the $\eta$ meson production,
and the total cross sections in both the proton-proton as well as the
proton-neutron reactions have been measured.
In case of the $\eta$ meson, the ratio of the total cross sections for the reactions
$pn \to pn\eta$ and $pp \to pp\eta$ was determined to be $R_{\eta}$~$\approx~$~6.5~\cite{calen-prc58}.
At the lower values of $Q$, close-to-threshold, the ratio falls down and amounts to
$R_{\eta}$~$\approx$~3~\cite{moskal-prc79,czyzykiewicz-app39}, (see figure~\ref{ratio_eta}).
As explained by Wilkin~\cite{wilkin-plb382,wilkin-prc56}, to a large extent this behaviour
may plausibly be assigned to the difference in the strength of the proton-proton and proton-neutron
final state interaction.

Since,
$$ \sigma(pp \to pp\eta) = \sigma_{I=1},$$
$$ \sigma(pn \to pn\eta) = {{ \sigma_{I=0} + \sigma_{I=1}} \over 2}$$
we have
$$ \sigma_{I=0} = (2R_{\eta} - 1)\sigma_{I=1}, $$
where {\it I} denotes the total isospin in the initial and final state
of the nucleon pair. The production of the $\eta$ meson in this reaction
with total isospin $I =$~0 exceeds the production with  isospin
$I =$~1 by a factor of 12.
This large difference of the total cross sections suggests the
dominance of isovector mesons exchange in the creation of the
$\eta$ meson in collisions of nucleons. A recent experimental study of the analysing
power of the $\vec p p \to pp\eta$ reaction~\cite{czyzykiewicz-prl98} indicates that the
$\pi$ meson exchange
between the colliding nucleons may be predominant.

Since the amount of different quark flavours in $\eta$ and $\eta^{\prime}$ mesons wave function
is similar,
in case of the dominant isovector meson exchange -- by the analogy to
the $\eta$ meson production -- we can expect that the ratio $R_{\eta^{\prime}}$
should be large. If however the $\eta^{\prime}$ meson is produced via its
flavour-blind gluonium component from the colour-singlet glue excited
in the interaction region the ratio should approach unity after corrections
for the initial and final state interactions.
The close--to--threshold
excitation function for the $pp \to pp\eta^{\prime}$ reaction has already been
established~\cite{moskal-plb474,khoukaz-epj,balestra-plb491,moskal-prl80,hibou-plb438}
and the determination of  the total cross section for the $\eta^{\prime}$
meson production in the proton-neutron
interaction motivated the work presented in this dissertation.
Recently Bass and Thomas~\cite{bass-thomas}
argued that the strength of the interaction  of $\eta$ and $\eta^{\prime}$ mesons
with nucleons is sensitive to the singlet-flavour component,
and hence to the glue content in these mesons. This makes a connection in our
endeavour to investigate the structure, the production dynamics, and
the interaction of the $\eta$ and $\eta^{\prime}$  mesons with nucleons~\cite{pklaja-phd}.

\begin{figure}[H]
\centerline{\includegraphics[height=.45\textheight]{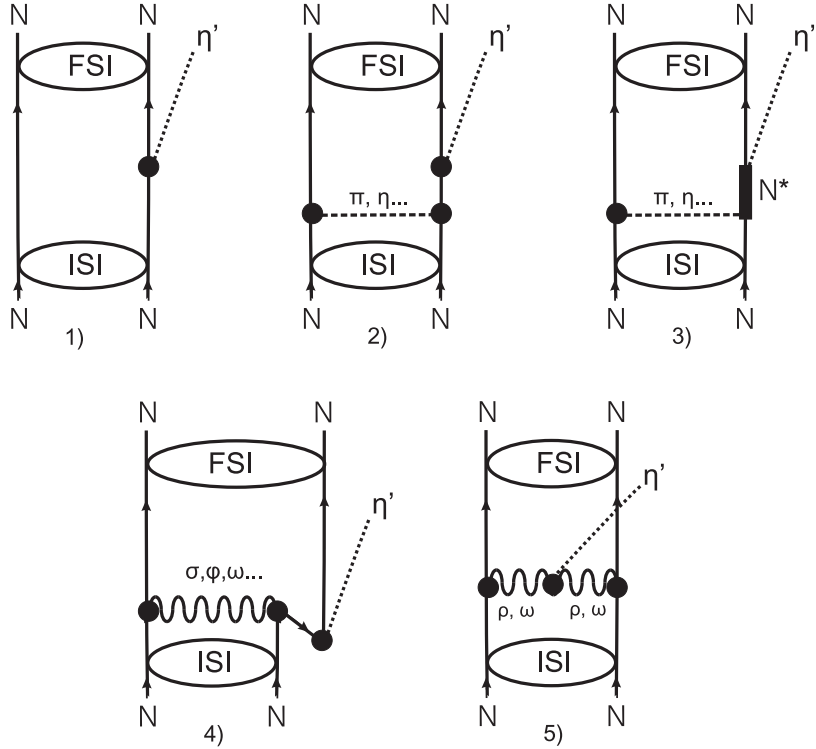}}
\caption{Diagrams for the $NN \to NN \eta^{\prime}$ reaction close-to-threshold:
         1$)$ - $\eta^{\prime}$-bremsstrahlung (nucleonic current), 2$)$ -
         "rescattering" term (nucleonic current), 3$)$ - excitation of an intermediate
          resonance (nucleon resonance current), 4$)$ - production through heavy meson
          exchange, 5$)$ - fusion of the virtual mesons (mesonic current).}
\label{grafy_meson}
\end{figure}
Certainly, other production mechanisms which are shown in figure~\ref{grafy_meson},
such as meson exchange and nucleon resonance
currents, must be taken into account before a role of gluons in the NN$\eta^{\prime}$
vertex is explored.
It is expected that the $\eta^{\prime}$ meson can be produced through
$\eta^{\prime}$-bremsstrahlung (nucleonic current), "rescattering" term (nucleonic
currents), excitation of an intermediate resonance (nucleon
resonance current),  heavy meson exchange or through fusion of the virtual mesons
(mesonic curent).
The two latter mechanisms, which are of short-range, are expected to contribute
even more significantly due to the large four-momentum transfer needed between nucleons
to create the $\eta^{\prime}$ meson.
Theoretical studies of the $\eta^{\prime}$ meson production mechanism have shown
that the existing data could be explained either by mesonic and nucleonic currents
or by a dominance of two new resonances S$_{11}$(1897) and P$_{11}$(1986)~\cite{nakayama-prc61}.
Moreover an extended study~\cite{nakayama-prc69} motivated by the updated data of the
$\gamma p \to \eta^{\prime} p$ and $pp \to pp \eta^{\prime}$ reactions indicated contributions from
resonances S$_{11}$(1650) and P$_{11}$(1870). However, it is premature to identify these states,
as these authors pointed out.\\
Recently, Cao and Lee~\cite{cao-prc78} have studied the near-threshold $\eta^{\prime}$
production mechanism in nucleon-nucleon and $\pi$-nucleon collisions under the assumption
that the resonance S$_{11}$(1535) is predominant. In an effective
Lagrangian approach which gives a resonable description to the $pN \to pN\eta$
and $\pi p \to p\eta$ reactions, it was found that the t-channel $\pi$ exchange
makes the dominant contribution to the $pN \to pN\eta^{\prime}$ process, and a value
of 6.5 for the ratio of $\sigma(pn \to pn \eta^{\prime})$ to $\sigma(pp \to pp \eta^{\prime})$
was predicted. \\
On the contrary, other authors~\cite{kampfer-ep} reproduced the magnitude of the total
cross section for the $pp \to pp \eta^{\prime}$ reaction including meson currents and
nucleon currents with the resonances S$_{11}$(1650), P$_{11}$(1710) and P$_{13}$(1720).
A resonable agreement with the $pp \to pp \eta^{\prime}$ data is achieved by the contribution of the
meson conversion currents indicating other terms to be less significant. In the frame of this model the
R$_{\eta^{\prime}}$ ratio is predicted to be 1.5.\\
The above considerations shows that our understanding of the $\eta^{\prime}$ meson
production mechanism is still unsatisfactory. Therefore it is important to test the discussed
mechanisms by confronting them with the experimental results on the isospin dependence of the
$\eta^{\prime}$ meson production.

 
\chapter{ Spectator model}
\markboth{\bf Spectator model}
         {\bf Spectator model}

The absence of pure neutron targets makes measurements of mesons
production in proton-neutron interactions very difficult, especially
very close-to-thereshold where the total cross section changes rapidly
with the excess energy~\cite{moskal-ppnp49}. Therefore, to measure the $pn \to pn \eta^{\prime}$
reaction we used a deuteron target as an alternative. \\
\begin{figure}[H]
\centerline{\includegraphics[height=.4\textheight]{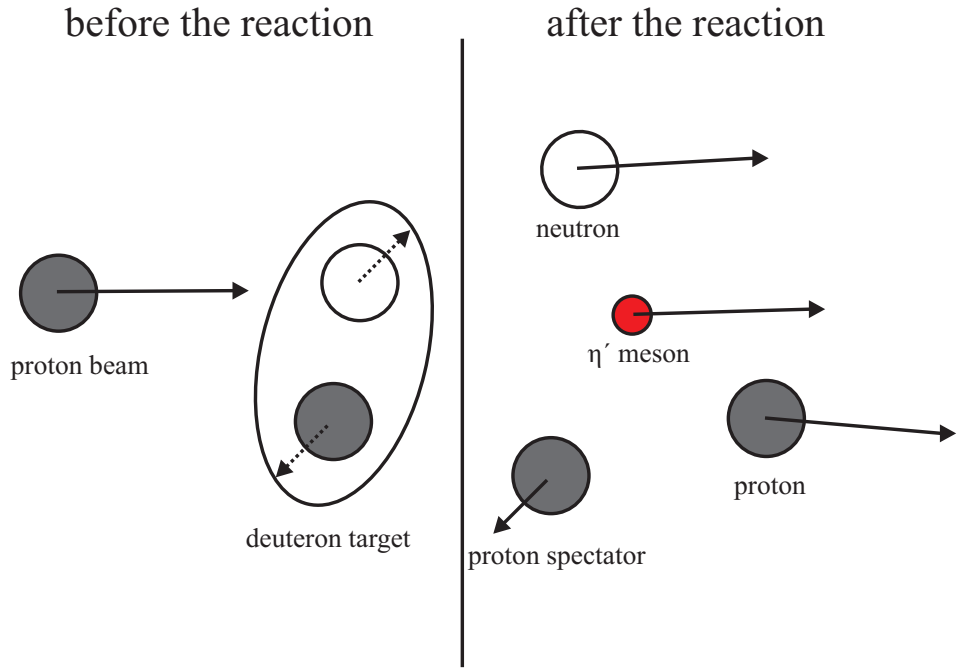}}
\caption{Schematic picture of the quasi-free $ pn \to pn\eta^{\prime}$ reaction.
         Grey circles represent
         protons, whereas neutrons are depicted as open circles. The Fermi
         momentum of the nucleons inside the deuteron is represented by the dotted arrows.}
\label{reaction_before_after}
\end{figure}

Figure~\ref{reaction_before_after} shows schematically the quasi-free
$pn \to pn\eta^{\prime}$ reaction.
In the experiment described in this thesis, a cooled proton beam with
a momentum of $p_{beam}~=~3.35$~GeV/c and
a deuteron cluster beam target as a source for neutrons have been used.
The applied technique is similar to investigations
of the $\pi^{0}$ and $\eta$ production in proton-neutron collisions carried out
by the WASA/PROMICE collaboration at the CELSIUS accelerator
in Uppsala~\cite{calen-prc58,calen-prl79,stepaniak-FZJ,haggstrom-PhD}.

\begin{figure}[H]
\centerline{\includegraphics[height=.4\textheight]{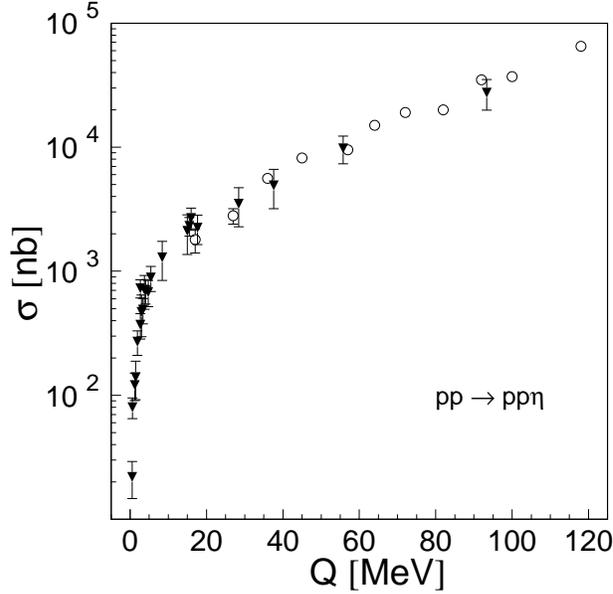}}
\caption{Total cross section for the $pp \to pp\eta$ reaction as a function
         of the excess energy measured using proton beam colliding with proton target
         (closed triangles)~\cite{smyrski-plb474,moskal-prc69,calen-plb366,chiavassa-plb322,
         bergdolt-prd48,hibou-plb438} and proton beam scattered on  deuteron target
         (open circles)~\cite{calen-prl79}.}
\label{uppsala_free_quasi}
\end{figure}

The main difficulty, when using deuterons as a neutron target is the fact
that nucleons inside the deuteron are not at rest but are moving with the
Fermi momentum. Therefore, in order to achieve the required resolution
for the determination of the excess energy in the proton-neutron reaction
it is mandatory to determine the neutron momentum inside the deuteron for
each event. The momentum of the neutron is derived indirectly based on the measurement
of the proton Fermi momentum. For the analysis a {\it spectator model} is applied.
Due to the  relatively small  binding energy of a deuteron $(E_B~=~2.2$~MeV$)$ which is more
than three orders of magnitude smaller
compared to the kinetic energy of the bombarding proton  $(T_{beam}~=~2540$~MeV$)$,
the neutron is considered as a free particle in the sense that the matrix element
for the quasi-free meson production on a bound neutron is equal to the one in
free production from an unbound neutron. The second assumption of the spectator model
used for the evaluation of the $pn \to pn\eta^{\prime}$ reaction is that
the proton from the deuteron target does not take part in the reaction and that
it is on its mass shell when the beam hits the target. The validity of these assumptions
was proven by measurements performed at CELSIUS~\cite{calen-prl79,stepaniak-FZJ,haggstrom-PhD},
TRIUMF~\cite{duncan-prl80} and COSY-TOF~\cite{abdelbary-epj}.
The comparison of the quasi-free and free production cross sections for the $pp \to pp\eta$
reaction done by the WASA/PROMICE collaboration has shown that there is no difference
between the total cross section for the free and quasi-free process
within the statistical errors.
The excitation function for the $pp \to pp\eta$
reaction measured with free and quasi-free proton-proton scattering is presented in
figure~\ref{uppsala_free_quasi}.
Similarly, investigations of pion production carried out at the TRIUMF facility have shown that
the experimental spectator momentum distribution is consistent with expectations
based upon the hypotheses of the spectator model. It was also shown, that the magnitude
of the differential cross sections for the quasi-free $pp \to d\pi^{+}$ reaction agrees on the
few percent level with the free differential cross sections.
Recently, the validity of the spectator model was proven also by the COSY-TOF collaboration.
The shape of the angular distribution for the quasi-free $np \to pp\pi^{-}$ and $pn \to pn$
reactions as well as the form of the momentum distribution of the spectator  have been measured.
The experimental data were consistent with calculations based upon the spectator model
with an accuracy better than 4\%~\cite{abdelbary-epj}.

\begin{figure}[H]
\includegraphics[height=.29\textheight]{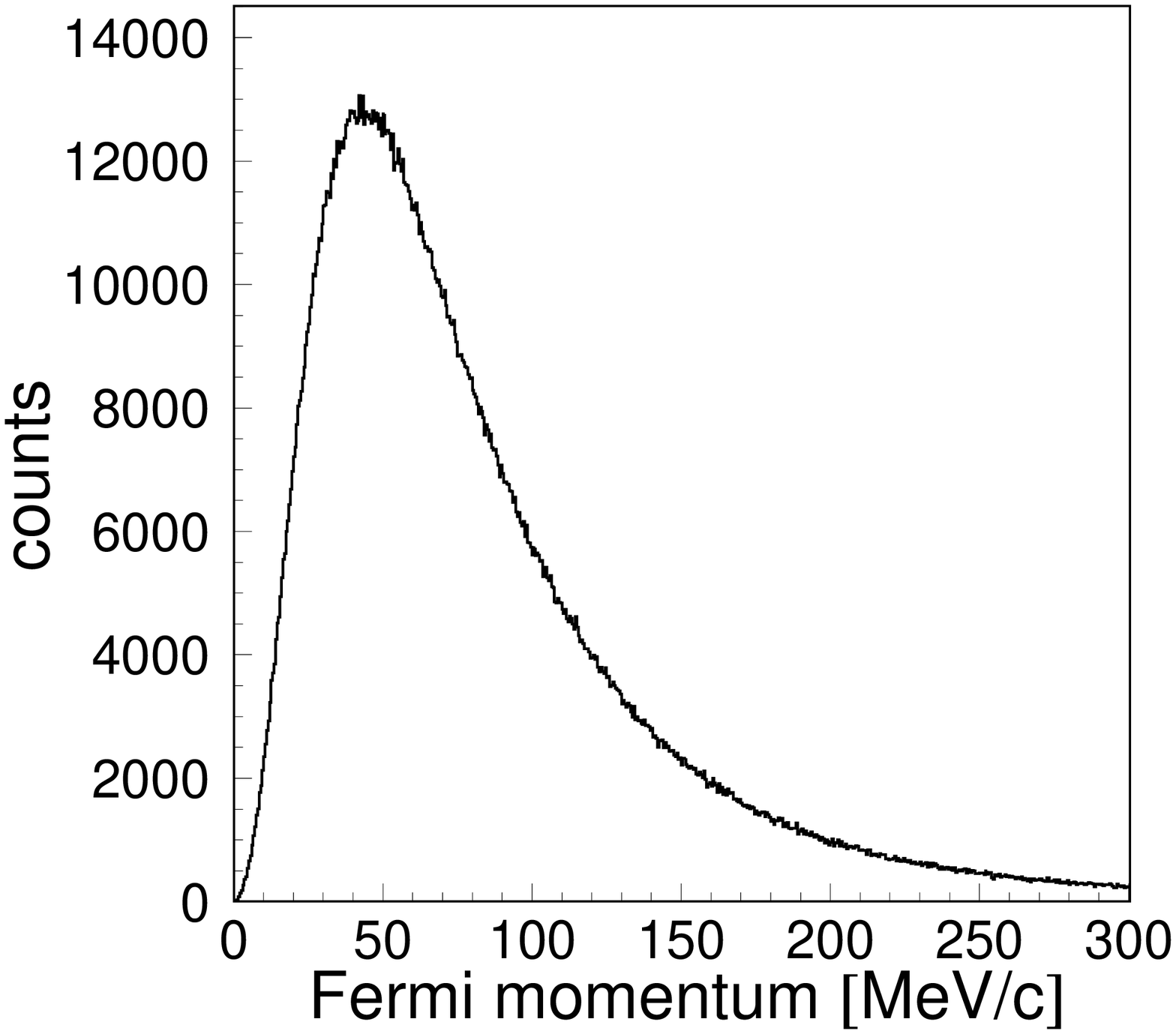}
\includegraphics[height=.29\textheight]{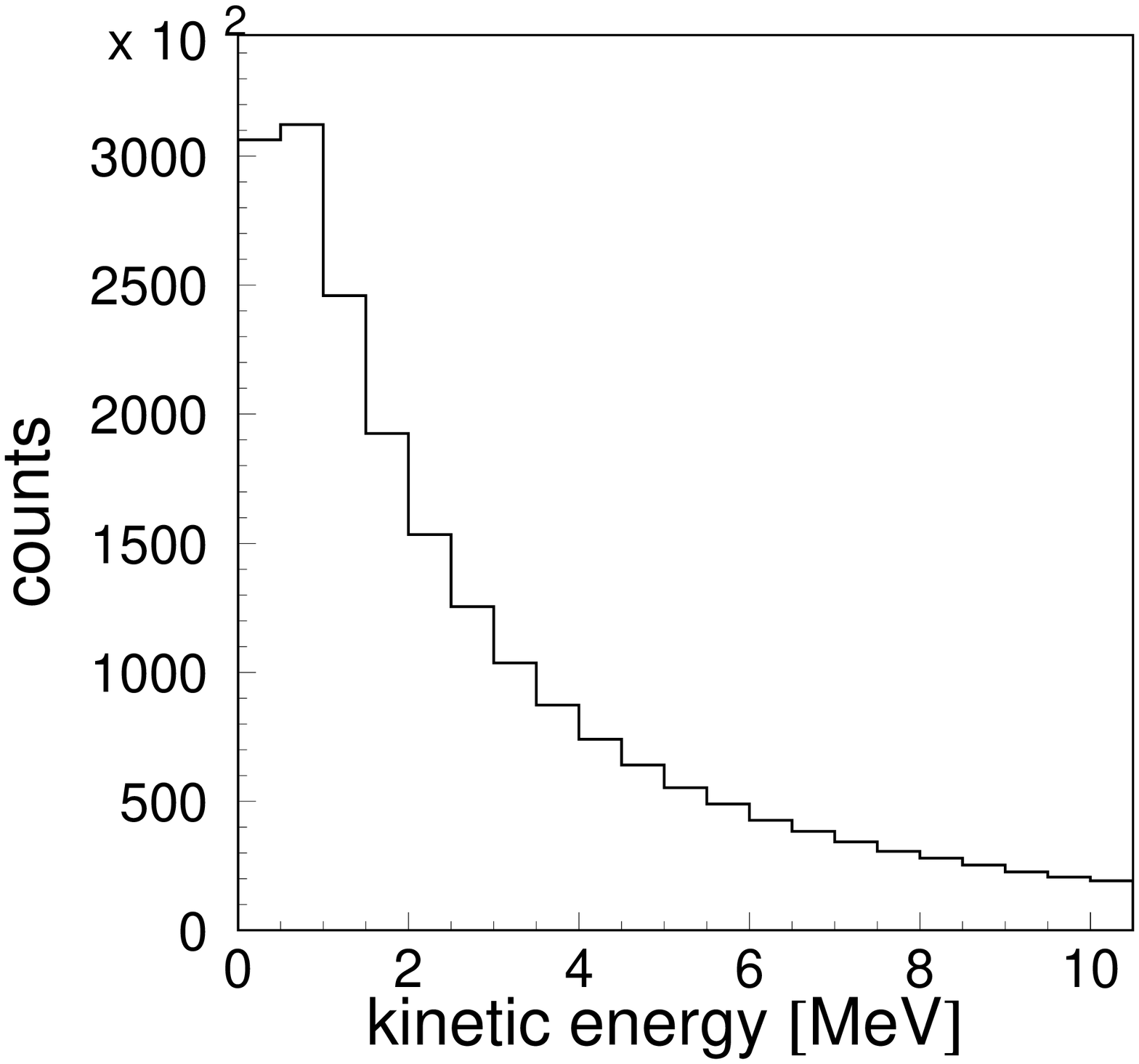}
\caption{ Distribution of the momentum {\bf (left)} and the kinetic energy
         {\bf (right)} of nucleons inside the deuteron generated
         according to the analytical parametrisation of the deuteron
         wave function derived from the Paris
         potential~\cite{lacombe-plb101,haggstrom-PhD}.}
\label{sym_spec_q}
\end{figure}

\begin{figure}[H]
\includegraphics[height=.29\textheight]{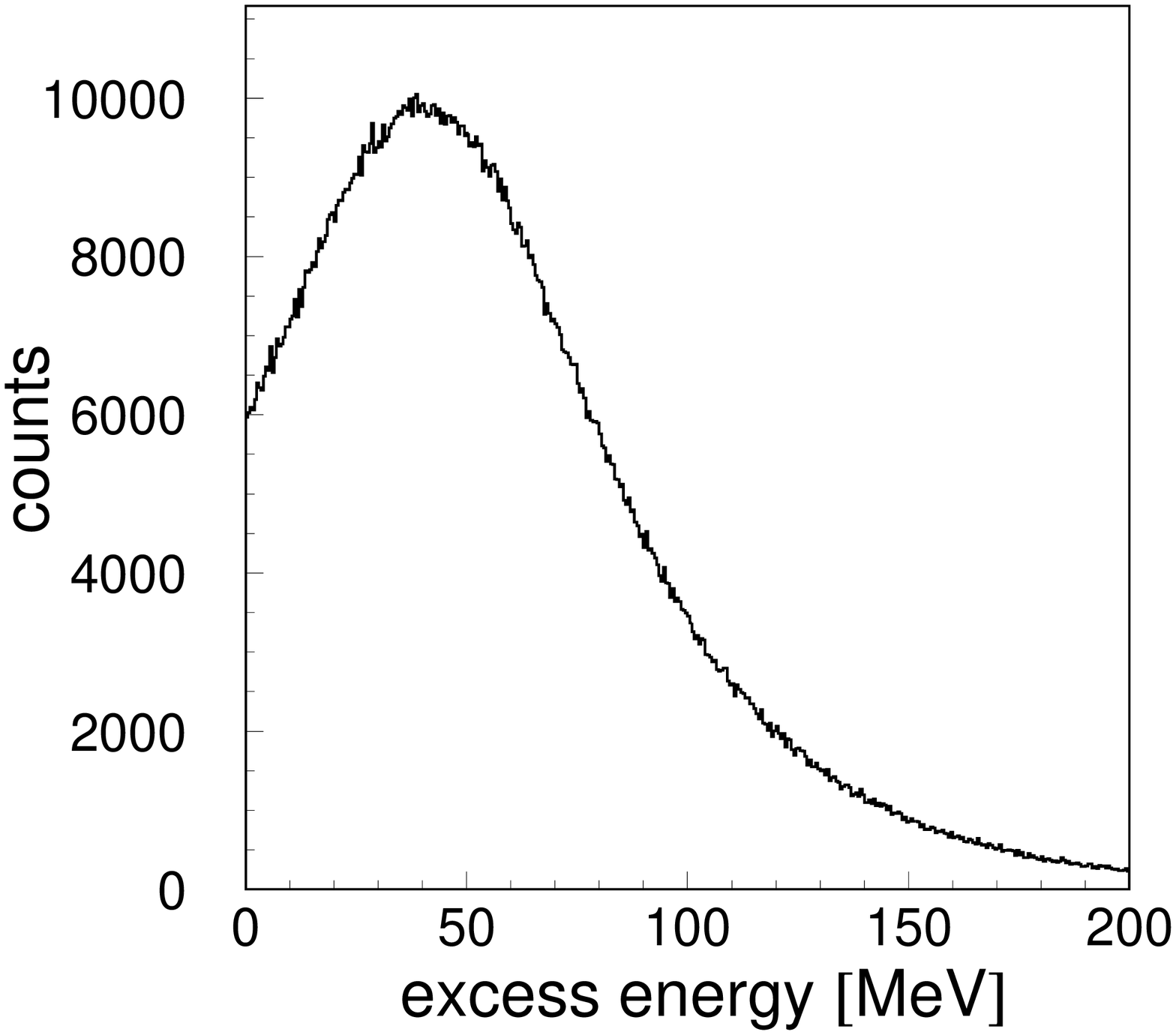}
\includegraphics[height=.29\textheight]{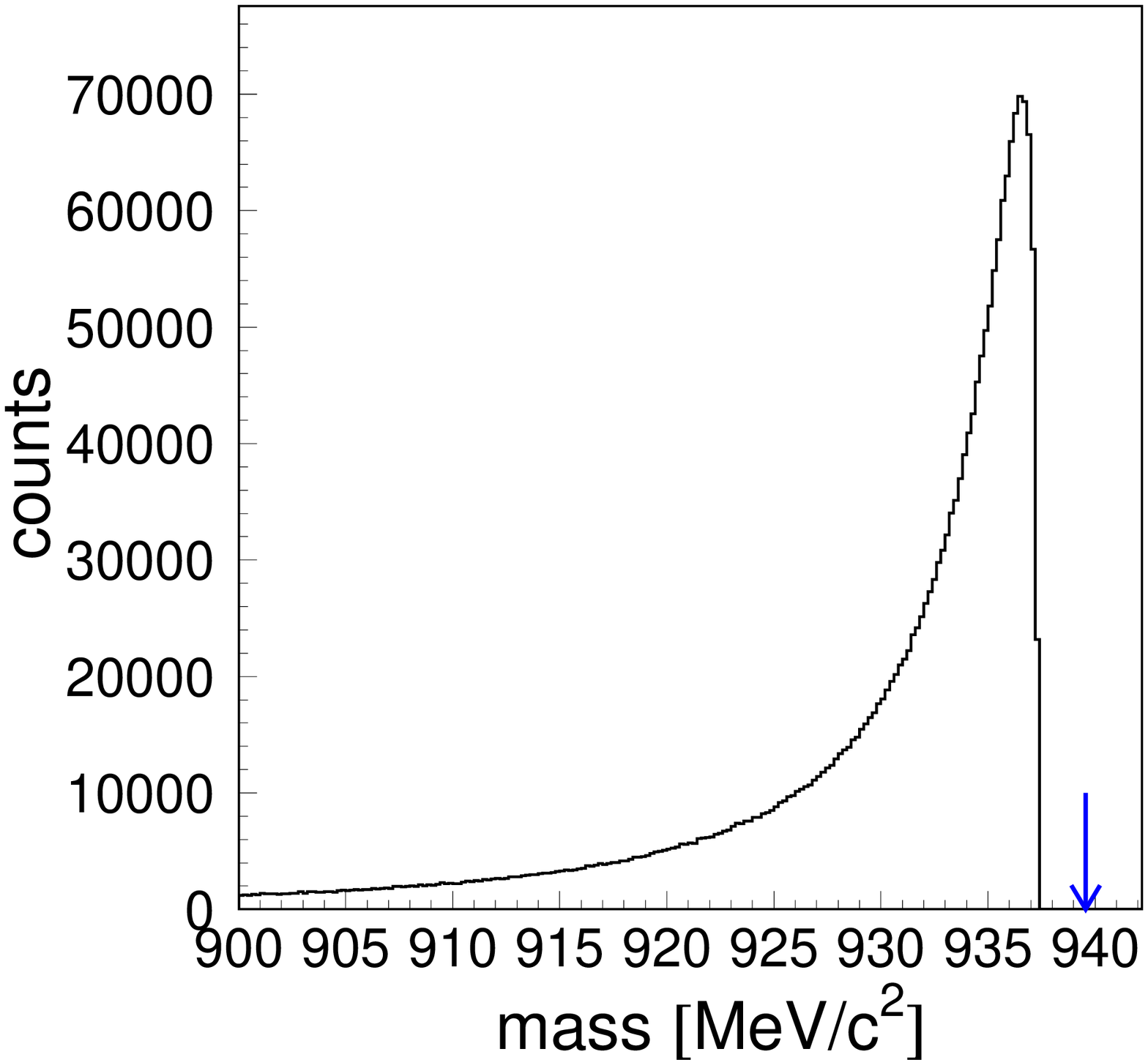}
\caption{{\bf Left:} Distribution of the excess energy $(Q_{cm})$ for the $pn\eta^{\prime}$
         system originating from the $pd \to pn\eta^{\prime} p_{spec}$
         reaction calculated with a proton beam momentum of $p_{beam}~=~3.35$~GeV/c
         and a neutron target momentum smeared out according to the Fermi distribution.
         {\bf Right:} Spectrum of the  mass of the interacting neutron. An arrow
         indicates a mass of the free neutron.}
\label{sym_excess}
\end{figure}

In case of the quasi-free
$pn~\to~pn\eta^{\prime}$ reaction the difference between the on-shell
and the off-shell total cross section should be even smaller than in case of
the quasi-free $\eta$ meson production since the total energy of colliding
nucleons is much larger for near threshold $\eta^{\prime}$ creation than for
the creation of the $\eta$ meson,
whereas the main difference between the on-shell and off-shell neutron mass
remains the same in both cases.\\
The momentum and kinetic energy distribution of a spectator proton is shown in
figure~\ref{sym_spec_q}. The momentum distribution is peaked at a value of 40~MeV/c
but due to the long tail the mean momentum is approximately equal to 100~MeV/c.
Since the neutron momentum changes from event to event, both the total energy in the
centre-of-mass system $(\sqrt{s})$ and the excess energy $(Q_{cm})$ vary also and
have to be determined for each event.\\
Figure~\ref{sym_excess}~(left) presents the distribution of the excess energy
$(Q_{cm})$ for the $pp\eta^{\prime}$ system originating from the $pd \to pn\eta^{\prime} p_{spec}$
reaction calculated with a proton beam momentum of $p_{beam}~=~3.35$~GeV/c
and a neutron target momentum smeared out according to the Fermi distribution.
Here $p_{spec}$ denotes the spectator proton.
The broad excess energy distribution enables
to scan a large range of the excess energy with a single beam momentum setting but
in parallel requires the reconstruction of the centre-of-mass energy for each event.
This can be done only if the four-momentum of target neutrons is known, what can be done
by measuring the four-momentum vector of the spectator proton, or by measuring all
four-momentum vectors of all outgoing nucleons and mesons (proton, neutron and $\eta^{\prime}$).
The first possibility requires the detection of the spectator proton in a suitable
detection unit as it was realized in the COSY--11 experimental facility.
For more details concerning the spectator detector
and calibration method see chapters 4. and 5.\\
As already mentioned, in the frame of the spectator model the proton in the deuteron is assumed to be an
untouched particle staying on-shell throughout the reaction. Thus, by measuring the spectator
proton momentum $(\vec{p}_{spec})$ one gets the target neutron momentum $(\vec{p}_{n})$ which is
equal to $- \vec{p}_{spec}$. The total energy of on-shell proton is equal to:
\begin{equation}
E_{spec} = \sqrt{m_{p}^2 + p_{spec}^2},
\end{equation}
and hence the total energy of the off-shell neutron can be calculated as:
\begin{equation}
E_{n}~=~m_d~-~E_{spec},
\end{equation}
where $m_{p}$ and $m_d$ denote the free proton and deuteron masses, respectively.
Even in case when the proton is at rest $(E_{spec}~=~m_{p})$ the neutron is off mass shell
due to the binding energy of the deuteron.\\

Figure~\ref{sym_excess}~(right) shows the distribution of the  mass of the
interacting neutron as calculated according to the relation:
\begin{equation}
m_{n}^{real} = \sqrt{E_{n}^2 - p_{spec}^2},
\end{equation}
where  $E_{n}$ is the total energy of the off-shell neutron.
As it is seen, the maximum of this distribution differs only by 2.2~MeV from the free
neutron mass $(m_{n}~=~939.565$~MeV/c$^2$).
\par
The quasi-free meson production is disturbed by some nuclear effects, namely
a shadow effect and a reabsorption of the produced meson by the spectator proton
but they are rather of minor importance. In the first case the reduction of the
beam flux originating in neutron shielding by a spectator proton
decreases the total cross section by about $4.5\%$ in case of $\eta$ meson
production~\cite{chiavassa-plb337}.
The same effect is expected for the $\eta^{\prime}$ meson production.
The absorption of the $\eta$ meson
reduces the total cross section by  $2\% - 4\%$~\cite{chiavassa-plb337} depending on the energy
of the $\eta$ meson which was produced. In case of a quasi-free $\eta^{\prime}$ meson
production this effect is much smaller since the proton-$\eta^{\prime}$
interaction is much weaker than the proton-$\eta$ one~\cite{moskal-hab}.


\chapter{ Experimental tools}
\markboth{\bf Experimental tools}
         {\bf Experimental tools}

The experiment described in this thesis has been performed using the COSY--11
facility~\cite{brauksiepe-nim,pklaja-aip796,smyrski-nim541}
an internal magnetic spectrometer installed at
the cooler synchrotron and storage ring COSY~\cite{maier-nim390} in J{\"u}lich, Germany.

\section{Cooler Synchrotron COSY}

The Cooler Synchrotron COSY stores and accelerates unpolarized as well as
polarized proton and deuteron beams in the momentum range between 0.6~GeV/c
and 3.7~GeV/c.
\begin{figure}[H]
\centerline{\includegraphics[height=.55\textheight]{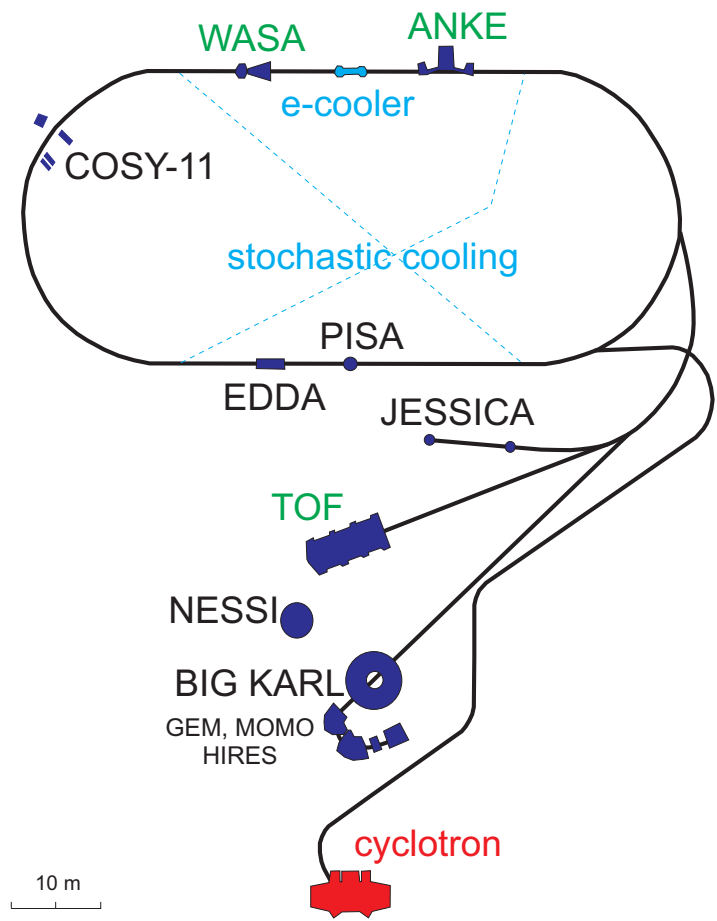}}
\caption{ Scheme of the COoler Synchrotron COSY with indicated internal experiments:
         (COSY--11~\cite{brauksiepe-nim}, WASA~\cite{wasa,adam-ep}, ANKE~\cite{anke},
         PISA~\cite{pisa}, EDDA~\cite{eddawww}) as well as external experiments
         (JESSICA~\cite{jessica}, TOF~\cite{tof}, NESSI~\cite{nessi}, GEM~\cite{gem},
         BIG KARL~\cite{gem}, MOMO~\cite{momo}, HIRES~\cite{hires}).}
\label{cosy}
\end{figure}
Cooling with electrons in the lower momentum range and stochastic
cooling in the upper momentum regime is used to decrease the spread of the
beam momentum and its emittance. Proton and deuteron beams
are supplied to experiments with the circulating beam -- "internal experiments"
or for experiments with the extracted beam -- "external experiments".
In case of internal experiments, at the highest beam momentum a few
times $10^{10}$ accelerated particles
pass through the target $\sim 10^6$ times per second.
The COSY synchrotron --  which is schematically depicted in figure~\ref{cosy} --
has a 180~m circumference including two half-circle sections connected by two 40~m
straight sections.\\
The experiment presented in this thesis is the first ever attempt to measure
the cross section of the $pn \to pn \eta^{\prime}$
reaction. It has been carried out at the COSY-11 detection setup using a deuteron
cluster target and a stochastically cooled proton beam with a momentum of 3.35~GeV/c.
The experiment was performed in August 2004 and the data have been collected
for 24 days.
It was based on the measurement of four-momenta of the outgoing
nucleons, whereas an unregistered short lived meson $\eta^{\prime}$ was identified via
the missing mass technique.

\section{COSY--11 facility}

\begin{figure}[H]
\centerline{\includegraphics[height=.81\textheight]{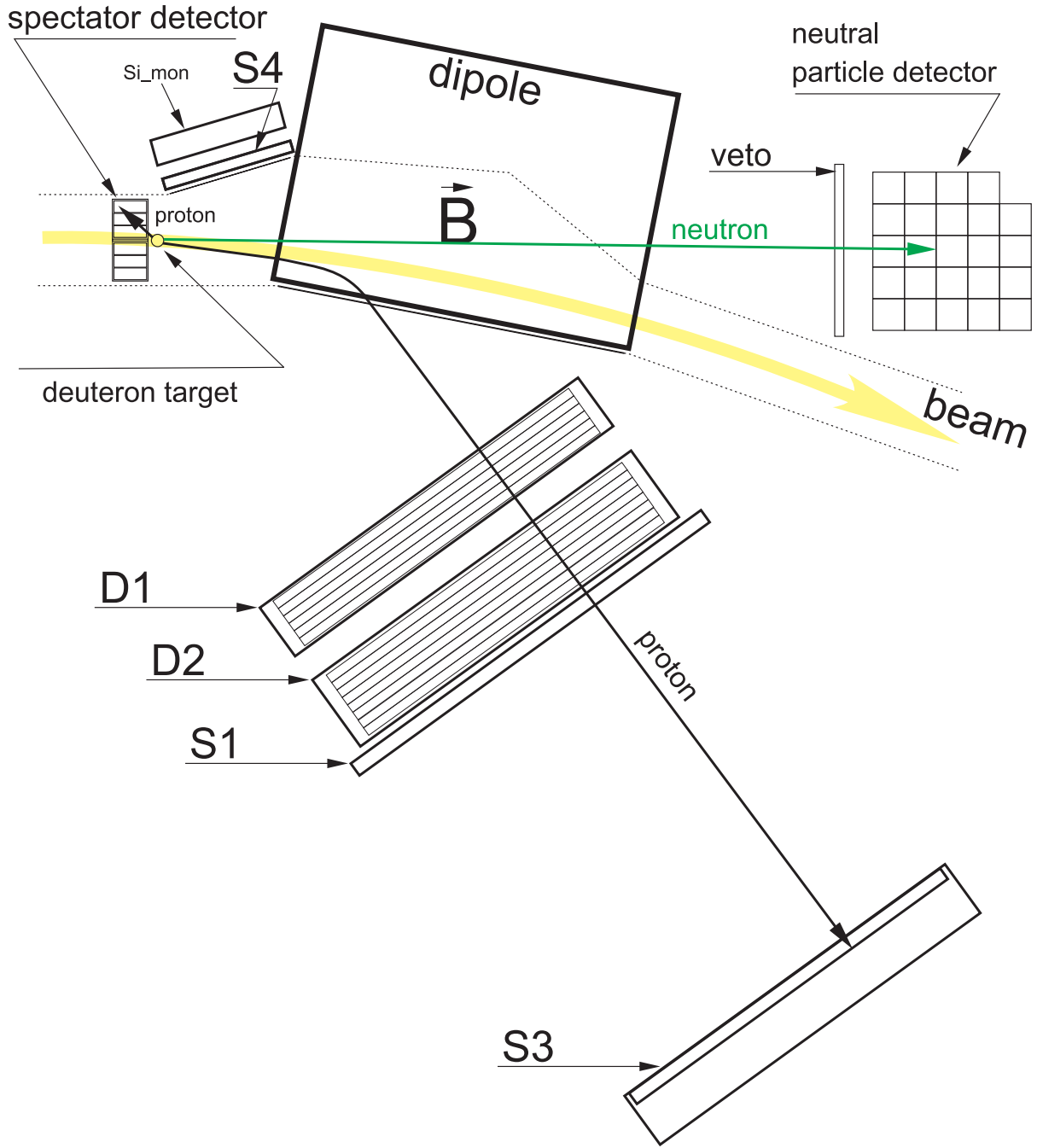}}
\caption{ Scheme of the COSY--11 detector system with superimposed
          tracks from the $pd \to p_{spec}(pn\eta^{\prime})$ reaction.
          Protons are registered in two drift chambers D1, D2
          and in the scintillator hodoscopes S1 and S3. The S1 detector is
          built out of 16 vertically arranged modules, whereas the S3 hodoscope
          is a non-segmented scintillator wall viewed by a matrix
          of 217 photomultipliers.
          An array of silicon pad detectors $(spectator~detector)$
          is used for the registration of the spectator protons.
          Neutrons are registered in the neutral particle detector
          consisting of 24 independent detection units.
          In order to distinguish neutrons from charged particles a veto
          detector is used. Elastically scattered protons are measured
          in the scintillator detector S4 and position sensitive silicon detector
          $Si_{mon}$. Detectors' size and their relative distances
          are not to scale.}
\label{cosy11}
\end{figure}

The COSY--11 detector system is schematically depicted in figure~\ref{cosy11}.
The detectors used for the registration of the spectator proton
and the neutron (silicon pad detector and neutral particle detector, recpectively)
will be described more detailed in the following sections. The details of the
other detector components used in this experiment and the method of
measurement can be found in
references~\cite{brauksiepe-nim,smyrski-nim541,moskal-phd,wolke-phd}.
Therefore, here the used experimental technique will be presented only briefly.
\par
\begin{figure}[H]
\centerline{\includegraphics[height=.4\textheight]{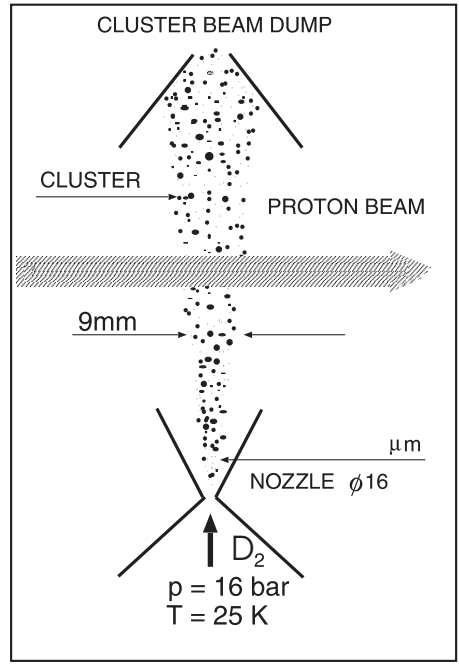}}
\caption{Schematic view of the cluster target. The figure is adapted from
         reference~\cite{moskal-phd}.}
\label{target}
\end{figure}

The COSY--11 cluster target, shown in figure~\ref{target} is installed in front of
one of the COSY dipole magnets.
It can provide streams of clusters of  hydrogen ($H_{2}$) as well as of deuteron ($D_{2}$).
The charged products of the measured reactions are
bent in the magnetic field of the dipole magnet and leave the vacuum
chamber through a thin exit foil,
whereas the beam --- due to the much larger momentum --- remains on its orbit inside the ring.
The charged ejectiles are detected in the drift chambers (D1, D2)~\cite{brauksiepe-nim,smyrski-nim541}
and the scintillator hodoscopes (S1, S3)~\cite{brauksiepe-nim,moskal-phd,wolke-phd}.
Neutrons and gamma quanta are registered in the neutral particle detector~\cite{przerwa-dt}.
The veto detector is used in order to separate
neutrons and gamma quanta from charged particles.
An array of silicon pad detectors {\it (spectator detector)}~\cite{bilger-nim457}
is used for the registration of the spectator protons.

Protons scattered elastically under large angles are measured
in the scintillator detector S4 and in another position sensitive silicon
pad detector $(Si_{mon})$ positioned closely behind the S4 counter.

\section{Measurement method -- general remarks}

The main goal of this experiment is the determination of the total cross section
for the $pn \to pn\eta^{\prime}$ reaction in the excess energy range between
0 and 24~MeV. In order to calculate the total cross section, given in general
by the formula:
\begin{equation}
\sigma(Q_{cm}) = {N(Q_{cm}) \over {L \times E_{eff}(Q_{cm})}},
\end{equation}
one needs to determine: $i)$ the efficiency $E_{eff}$ of the COSY--11 detection
system, $ii)$ the luminosity $L$ integrated over the time of the experiment, $iii)$ the
excess energy $Q_{cm}$ and
$iv)$ the number of $\eta^{\prime}$ meson events $N$ registered for a given excess energy.
In the following the experimental conditions, technique of measurement and method of the
$E_{eff}$, $L$, $Q_{cm}$, $N$ determination will be presented.
\par
Events corresponding to the production of at least one charged particle in coincidence
with a neutron (or gamma quantum) were stored on the tape for later analysis.
Conditions of the main trigger can be written symbolically as:
$$ T_{\eta^{\prime}} = \{N^{\mu\ge1}_{up} \wedge N^{\mu\ge1}_{dw} \} \wedge S1^{\mu\ge1}_{2\ldots6} \wedge S3, $$
which requires that at least a coincidence of one hit (multiplicity $\mu \ge1$) in the neutral particle detector
(in upper and lower photomultiplier), in the S1 detector, and in
the scintillator wall S3 was demanded.
The S1 region was restricted to 5 modules (from 2$^{nd}$ to 6$^{th}$) in which the proton signal from
the $pn \to pn\eta^{\prime}$ reaction was expected.\\
During the experiment an additional trigger referred to
as $T_{elas}$  was set up for the registration of the quasi-free p-p elastic scattering.
Events which gave a signal in the S1 detector in coincidence with the S4
detector~(see figure~\ref{cosy11}) were accepted as a quasi-free proton-proton
scattering or elastic proton-deuteron scattering.\\
For each charged particle, which gave  signals in the drift chambers, the momentum vector
at the reaction point can be determined. For the analysis of the $pn \to pn\eta^{\prime}$
reaction, first the trajectories of the outgoing protons are reconstructed~\cite{sokolowski}
based on signals in the drift chambers D1
and D2 -- and then knowing the magnetic field of the dipole magnet -- its momentum vector is
determined. Independently, the velocity of the particle is derived
based on the time--of--flight measured between the S1 and the S3 detectors.
Knowing the velocity and the momentum of the particle, its mass can be
calculated, and hence the particle can be identified. After the particle
identification the time of the reaction at the target is obtained from the
known trajectory, velocity, and the time measured by the S1 detector.
\par
The neutral particle detector delivers the information about the time at which
the registered neutron or gamma quantum induced a hadronic or electromagnetic
reaction in the detector volume, respectively. The time of the reaction combined with this
information allows to calculate the time--of--flight (TOF$^N$) of the neutron
(or gamma) between the target and the neutral particle detector, and --- in case
of neutrons --- to determine the absolute value of the momentum (p), expressed as:
\begin{equation}
 p = m_n \cdot {l \over {TOF^N}} \cdot  {1 \over {\sqrt{1 - ({l \over {TOF^N}})^2/c^2}}},
\end{equation}
where $m_n$ denotes the mass of the neutron, and
$l$ stands for the distance between the target and the
neutral particle detector.
The direction of the neutron momentum vector is deduced from the angle defined
by the centre of the hit module inside the neutral particle detector.
\par
Similarly, to determine the direction of the momentum
vector of the spectator proton, the angle defined by the centre of the hit segment
inside the spectator detector is used.
In order to calculate the four--momentum of the spectator proton its kinetic energy
$(T_{spec})$ is directly measured as the energy loss in the silicon detector.
Knowing the proton kinetic
energy $(T_{spec})$ one can calculate its momentum using the relationship:
\begin{equation}
     p = \sqrt{{(T_{spec} + m_p)}^2 - {m_p}^2},
\end{equation}
where $m_p$ denotes the proton mass.
\par
To evaluate the luminosity, the quasi-elastic proton-proton scattering is
measured during the data taking of the
$pn \to pn\eta^{\prime}$ reaction,
with one proton detected in the drift chambers and the scintillator hodoscopes
and the other proton registered in the silicon detector $Si_{mon}$. The elastically scattered protons can
be well separated from the multi-particle reactions, due to the two body kinematics.
\par
Based on Monte Carlo studies, the beam momentum value, the position of the
spectator detector and the configuration of the neutral particle detector have
been optimized before the experiment in order to maximize the detection efficiency,
the resolution of the excess energy determination
and in order to achieve a relatively high missing mass resolution~\cite{czyzyk-dt}.
\par
In the next two chapters the method of the detector calibration and evaluation of the data, namely
determination of the excess energy $Q_{cm}$, number of $\eta^{\prime}$ mesons created
in proton--neutron collisions, the luminosity L and the detection efficiency will be presented.


\chapter{Calibration of the detectors}
\markboth{\bf Calibration of the detectors}
         {\bf Calibration of the detectors}

\section{Time-space calibration of the drift chambers}

In this section we give an account on the time-space calibration
of the drift chambers. \\
The drift chamber D1 (see figure~\ref{cosy11}) contains six
detection planes, two planes have vertical wires and four have inclined
wires. The second drift chamber (D2) contains two more planes with vertical wires
(altogether eight). The wires in consecutive pairs of planes
are shifted by  half of the cell width in order to resolve the left-right
position ambiguity with respect to the sense wire. Both chambers were operating
with 50$\%$--50$\%$ argon--ethane gas mixture at atmospheric pressure.
\begin{figure}[H]
\includegraphics[height=.29\textheight]{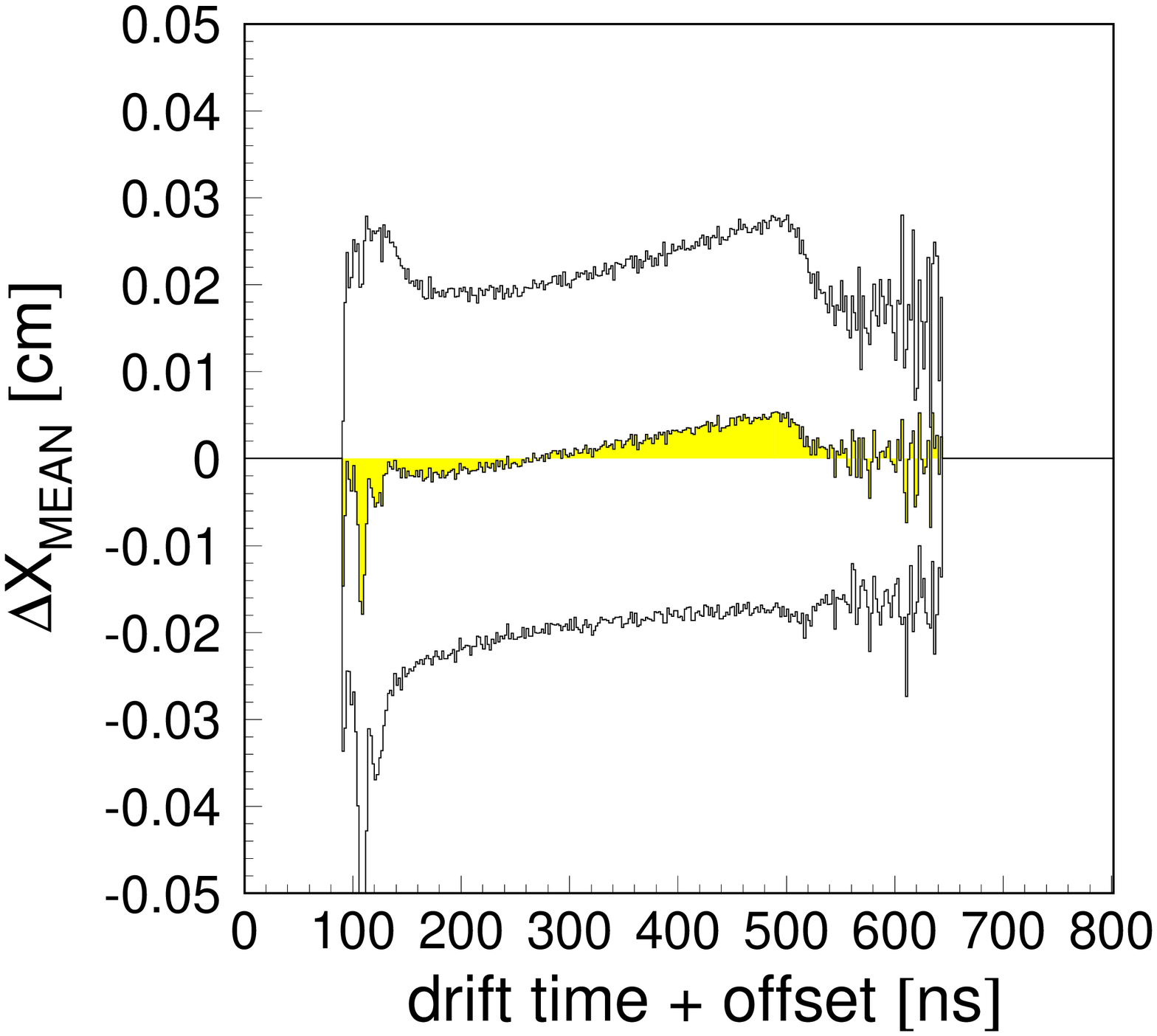}
\includegraphics[height=.29\textheight]{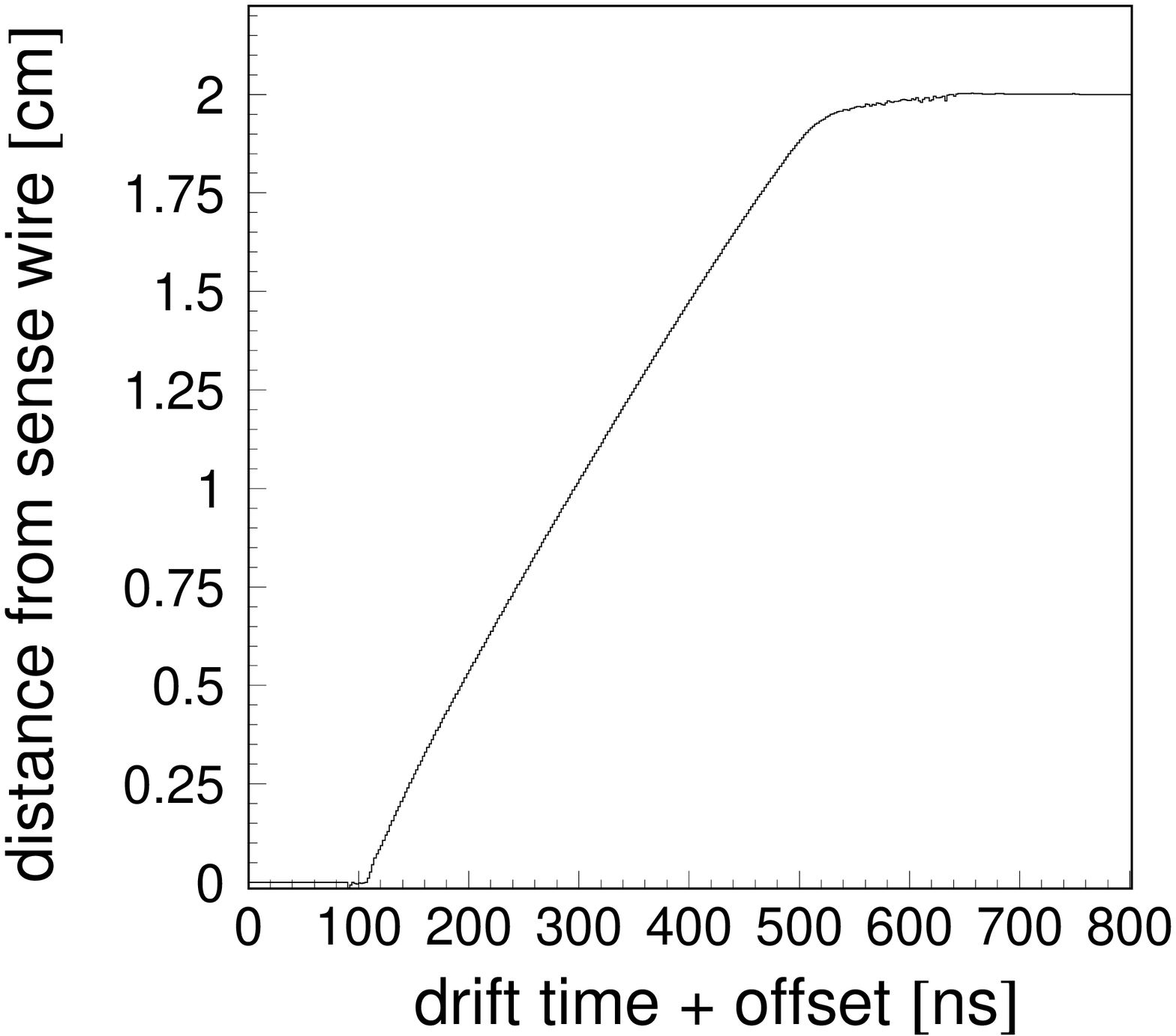}
\caption{ {\bf Left:} Example of the spectrum used for the drift chamber
          calibration. The shaded area presents mean values of
          $\Delta X$ as a function of the drift time.
          Upper and lower histograms indicate a band
          of one standard deviation of the $\Delta X$ distribution.
          The meaning of $\Delta X$ is explained in the text.
          {\bf Right:} Distance from the sense wire as a function
          of the drift time for an arbitrarily chosen plane of the D1 chamber.}
\label{dc_calib}
\end{figure}
When a charged particle passes through the drift cell it
invokes an ionization of the gas mixture. The positively charged ions
drift to the cathode wires with negative potential,
whereas the free electrons drift towards the sense wire with positive potential.
The drift time of electrons is measured
and hence the distance between the particles trajectory and the sense wire
can be determined. A relation between the
drift time and the distance from a sense wire is established from
experimental data.
Due to  variations of the drift velocity caused by
changes of the pressure, temperature and humidity
of the air inside the COSY tunnel,
the drift chamber calibration has to be performed for time intervals
not longer than few hours. The calibration is derived in an iterative way.
First, having an approximate time-space function {\it x(t)} for a given sample
of measured events\footnote{In case of this experiment the first approximation
was taken from the previous COSY--11 runs. It is worth noting that in general
an approximate drift time to drift distance relation can be determined by integration
of the drift time spectra as provided by the uniform irradiation method~\cite{smyrski-ar2004}.},
a distance of particle
trajectory from a sense wire is calculated, and a straight line is fitted to the
obtained points. Next, the deviation $\Delta X$ between
the fitted and measured distances of the particle's trajectory
to the wire is calculated. An example of a mean value of $\Delta X$
as a function of the drift time
is shown in figure~\ref{dc_calib} (left). The shaded area represents the mean
values of $\Delta X$. Upper and lower histograms indicate a band
of one standard deviation of the $\Delta X$ distribution. Afterwards
the relation between the drift time and distance
from the wire is corrected by the determined mean value of $\Delta X$.
The procedure is repeated until corrections
become negligible compared with expected position resolution of the chambers.\\
Figure~\ref{dc_calib}~(right) shows the time-space calibration
of an arbitrarily chosen plane of the D1 chamber. The nearly linear dependence is
seen for the drift time range from 100~ns up to 500~ns.

\section{Time calibration of the S1 and S3 hodoscopes}
In this section a method of the time--of--flight calibration is presented.\\

The time--of--flight measurement on the known distance between the S1
and the S3 hodoscopes enables to calculate the velocity of the particle which
crosses both detectors.
In order to calculate the time--of--flight, relative time offsets between
each detection unit of the S1 and S3 detectors have to be established.
\par
The S1 detector (see figure~\ref{cosy11}) is built out of sixteen
scintillation modules with dimensions 45~cm $\times$ 10~cm $\times$ 0.4~cm.
The modules are arranged
vertically with a 1~mm overlap and are read out at both ends by pairs of photomultipliers.\\
The start time for the time--of--flight measurement is calculated as the mean value of times
measured by the upper and lower photomultiplier of the S1 detector module.
The measured TDC values for a single S1 module may be expressed as:
$$TDC_{S1}^{up} = t_{S1} + {{y}\over{\beta_{L}}} + t^{walk}_{up} + offset_{up} - T_{trigger},$$
\begin{equation}
TDC_{S1}^{dw} = t_{S1} + {{L-y}\over{\beta_{L}}} + t^{walk}_{dw} + offset_{dw} - T_{trigger}.
\label{s1_tdc}
\end{equation}
The $t_{S1}$ is the real time when a particle crosses the detector, $y$ denotes
the distance between the upper edge of the active part of the detector and the hit position,
$t^{walk}_{up}$ and $t^{walk}_{dw}$ stand for the time walk effect, and $offset_{up}$, $offset_{dw}$
comprise all delays due to the utilized electronic circuits.
The common start signal for all TDC modules is denoted
by the $T_{trigger}$ and it is the same for S1 and for S3 detector.
Thus the mean value of the time measured by the S1 module is given by:
\begin{equation}
t_{S1} = {{TDC_{S1}^{up}+TDC_{S1}^{dw}} \over 2} - offset_{S1} - t_{S1}^{walk} + T_{trigger},
\label{ts1real}
\end{equation}
where $offset_{S1}$ comprises all constatnt terms from eq.~\ref{s1_tdc}.\\
The S3 hodoscope is built out of a non-segmented scintillator wall with the dimensions
220~cm~$\times$~100~cm~$\times$~5~cm viewed by a matrix of 217 photomultipliers.
The stop signal for the time--of--flight measurement is calculated as the average
of times obtained from all the photomultipliers that produced a signal.
The TDC value for a single photomultiplier is expressed similarly as in the case
of the S1 detector:
\begin{equation}
TDC_{S3} = t_{S3} + t^{pos}_{S3} + t^{walk}_{S3} + offset_{S3} - T_{trigger},
\label{s3_tdc}
\end{equation}
where $t_{S3}$ denotes the real time when a particle crosses the detector
and the $t^{pos}_{S3}$ is the time needed by the light signal to pass from the
scintillation origin down to the photomultiplier photocathodes.
Applaying equations~\ref{s1_tdc} and~\ref{s3_tdc} one can calculate the
time--of--flight between the S1 and S3 hodoscopes as:
$$TOF^{S3-S1} = t_{S3} - t_{S1},$$
\begin{equation}
TOF^{S3-S1} = TDC_{S3} - t^{pos}_{S3} - t^{walk}_{S3} - offset_{S3}
 - {{TDC_{S1}^{up}+TDC_{S1}^{dw}} \over 2} + t^{walk}_{S1} + offset_{S1}.
\label{tofs1s3}
\end{equation}
Since in both equations~(\ref{s1_tdc}, ~\ref{s3_tdc}) the $T_{trigger}$ time is the same,
the  $TOF^{S3-S1}$ is independent of the triggering time. The $t^{pos}_{S3}$ as well
as $t_{S1}^{walk}$ and $t^{walk}_{S3}$ may be calculated based on the known calibration parameters
and amplitude of the signals~\cite{moskal-phd}. Thus the only
unknown quantities in equation~\ref{tofs1s3} for $TOF^{S3-S1}$ are the time offsets for the individual
photomultipliers of S3 detector ($offset_{S3}$) and time offsets for a single module
of S1 detector ($offset_{S1}$).
\par
In order to establish these effects we have selected events with only one
track reconstructed in the drift chambers, with signals in only one S1  module
and signals in the S3 hodoscope for those photomultipiers which are at the position expected
from the extrapolation of the particle trajectory determined in the drift chambers.\\
In the first approximation the relative offsets of the S1 modules were established comparing
the time between neighbouring modules for signals from particles which
crossed the detector through the overlapping regions.
Assuming, that the time offsets for the S1 detector are approximately
correct, we have established the time offsets for each photomultiplier of S3 detector,
by comparing the time--of--flight values calculated from the time signals in the S1 and S3 detectors
(see eq.~\ref{tofs1s3}) with the time--of--flight determined from the reconstructed momenta
of the particles in the magnetic field. Having the time offsets for the S3 hodoscope adjusted, we have
again re-calculated the offsets for the S1 detector in the same way as it was done for the
S3 detector. This procedure was repeated a few times until the
corrections of the offsets became negligible.\\

\begin{figure}[H]
\centerline{\includegraphics[height=.4\textheight]{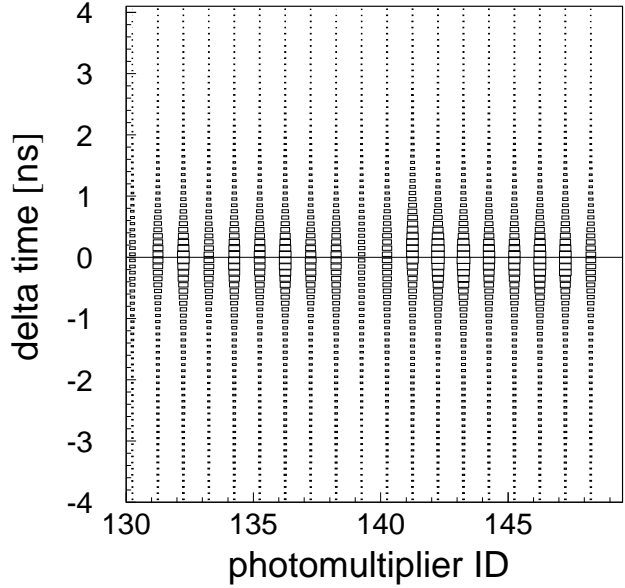}}
\caption{ The difference between the time--of--flight ($TOF_{S3-S1}$)
          determined from signals registered in the S1 and S3 scintillators
          and the time--of--flight calculated from the reconstructed momenta
          of particles versus the photomultiplier ID of S3. The figure shows result after
          the S3 hodoscope calibration. Photomultipliers which are
          positioned at the lower and upper edge of the scintillation wall have registered
          essentially less events (ID~=~130 and 139).}
\label{s3_calib}
\end{figure}
Figure~\ref{s3_calib} presents the distribution of difference between the time--of--flight ($TOF_{S3-S1}$)
on the S1-S3 distance determined from signals registered in S1 and S3 scintillators
and the time--of--flight calculated from the reconstructed momenta of particles for
photomultipliers no. 130--149 as obtained after the third iteration.

\section{The neutral particle detector}

The installation of the neutral particle detector at the COSY--11
facility enabled to study reactions with neutrons in the exit channel.
It allows, for example to investigate quasi-free meson production
in proton-neutron interactions or charged hyperon production like
$\Sigma^+$ via the $pp \to nK^+\Sigma^+$ reaction~\cite{rozek-phd,rozek-plb643}.
This detector~\cite{moskal-ar1997} delivers the time at which
neutron or gamma quantum induces a hadronic or electromagnetic reaction inside the
detector volume, respectively. This information combined with the time of the reaction
at target place --- deduced using other detectors --- enables to calculate
the time--of--flight between the target and the neutral particle detector
and to determine the absolute momentum of the registered neutrons.\\
In this section a method of time calibration will be demonstrated
and results achieved by its application will be presented and
discussed.
\par
Information about the deposited energy is not used in the
data analysis because the smearing of the neutron energy determined
by this manner is by more than an order of magnitude larger than
established from the time--of--flight method.
The experimental precision of the missing mass determination
of the $pn \to pn\eta^{\prime}$ reaction~\cite{moskal-hadron,proposal}
strongly relies on the accuracy of the reconstruction  of the momentum
of neutrons, therefore the time calibration of the neutral particle detector
has to be done with high precision.
\par
\begin{figure}[H]
\centerline{\includegraphics[height=.3\textheight]{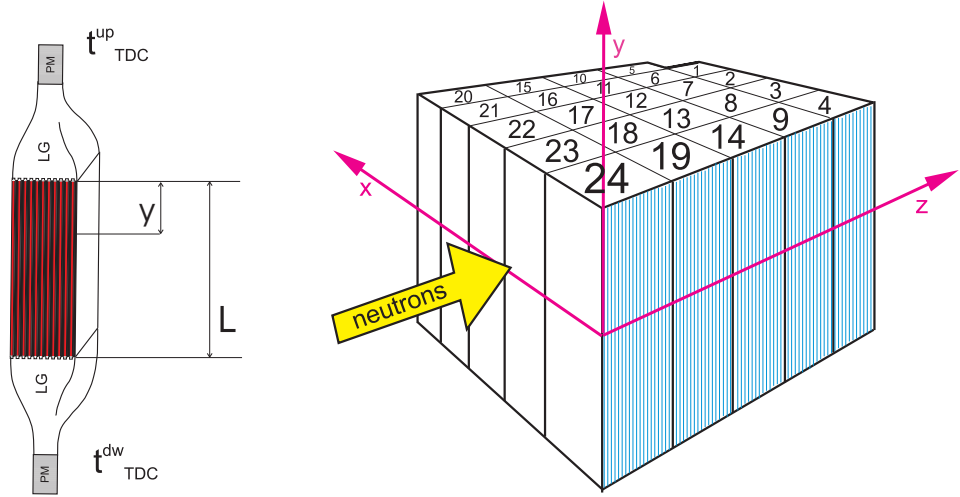}}
\caption{{\bf Left:} Scheme of the single detection unit of the
         COSY--11 neutral particle detector. LG and PM denotes
         light guide and photomultiplier, respectively. The figure shows also
         the definition of variables $y$ and $L$ used in the text.
         {\bf Right:} Configuration of the detection units of
         the neutral particle detector.}
\label{modul_pod_3d}
\end{figure}

Previously, the neutron detector consisted of 12 detection units~\cite{moskal-ar2002},
with light guides and photomultipliers mounted on one side of the module.
In order to improve the time resolution of the detector additional light guides
and photomulipliers were installed, such that the light signals from the scintillation layers
are read out at both sides of the module. The neutral particle detector
used for the experiment described in this thesis consists of 24 modules,
as shown in figure~\ref{modul_pod_3d}.
Each module is built out of eleven plates of
scintillator material with dimensions 240~mm $\times$ 90~mm $\times$ 4~mm interlaced with
eleven plates of lead with the same dimensions. The scintillators are connected
at both ends of a module to light guides --- made of plexiglass --- whose
shape changes from rectangular to cylindrical, in order to accumulate the
produced light on the circular photocathode of a photomultiplier. There
the light pulse is converted into an electrical signal, which is provided to
the ADC and to the TDC converters.\\
The neutron detector is positioned at a distance
of 7.36~m from the target in the configuration schematically depicted
in figure~\ref{modul_pod_3d}. The detector covers the laboratory
angular range of $\pm 1.84^o$ in x and $\pm 1.1^o$ in y direction.

This configuration and the distance to the target has been chosen to optimize
the efficiency for the reconstruction of the $pn \to pn\eta^{\prime}$
reaction~\cite{czyzyk-dt}. The choice was based on results from Monte Carlo
studies of the acceptance and efficiency as a function of the distance and
configuration~\cite{czyzyk-dt}.
The choice for the thickness of the scintillator plates was also based on the
simulation studies~\cite{land}.

\begin{figure}[H]
\centerline{\includegraphics[height=.4\textheight]{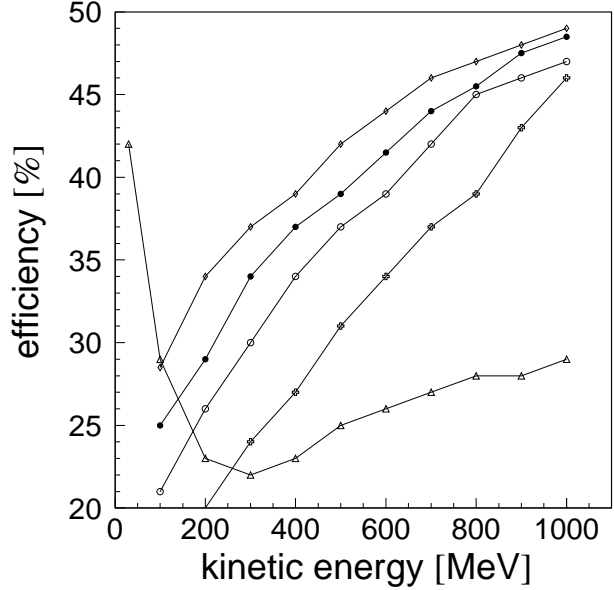}}
\caption{ Efficiency of the 20~cm thick neutron detector as
          a function of neutron kinetic energy simulated for various layer
          thicknesses of the scintillator and iron plates.
          The solid line with triangles is for pure scintillator material.
          Diamonds denote a layer thickness of 0.5~mm, closed circles of 2~mm,
          open circles of 5~mm, and crosses of 25~mm. The figure has been adapted
          from ref.~\cite{land}. Note that for better visualisation the vertical
          axis starts at 20\%.}
\label{land}
\end{figure}

As can be deduced from figure~\ref{land}
the maximum efficiency, for a given total thickness,
for the registration of neutrons --- in the kinetic energy
range of interest for the $pd \to pn\eta^{\prime}p_{sp}$ reaction
($\sim$ 300~MeV -- $\sim$ 700~MeV) ---
would be achieved for a homogeneous mixture of lead and scintillator.
However, in order to optimize the efficiency and the cost
of the detector the plate thickness has been chosen to be 4~mm. This
results in an efficiency which is only of few per cent smaller than the maximum possible.
The functioning of the detector was already tested in previous
experiments~\cite{moskal-prc79,przerwa-dt,rozek-plb643}.
\subsection{Time signals from a single detection unit}

A schematic view of a single detection module of the COSY--11 neutral particle
detector is shown in figure~\ref{modul_pod_3d}~(left).
The time ($T^{exp}$) from a single module is calculated as an average time measured
by the upper and lower photomultiplier. Namely:
\begin{equation}
T^{exp} = {{{TDC}^{up} + {TDC}^{dw}} \over 2 },
\label{t_exp}
\end{equation}
where $TDC^{up,dw}$ denotes the difference between the time of signals arrival from
the photomultiplier and from the trigger unit to the Time--to--Digital--Converter (TDC).
This can be expressed as\footnote{Here we omit the time walk effect which
will be discussed later in section~\ref{walk}}:
\begin{equation}
{TDC}^{up} = {t}^{real} + \mbox{offset}^{up} + {{y} \over c_L} - T_{trigger},
\label{t_up}
\end{equation}

\begin{equation}
{TDC}^{dw} = {t}^{real} + \mbox{offset}^{dw} + {{L-y} \over c_L} - T_{trigger},
\label{t_dw}
\end{equation}
where $L$ stands for the length of a single module, $y$ denotes the distance
between the upper edge of the active part of the detector and the point at which
a neutron induced the hadronic reaction, $t^{real}$ is the time at which the
scintillator light was produced,  $T_{trigger}$
represents the time at which the trigger signal arrives at the TDC converter,
and $c_L$ denotes the velocity of the light signal propagation inside the scintillator
plates. The parameters offset$^{up}$ and offset$^{dw}$ denote the time of propagation
of signals from the upper and lower edge of the scintillator to the TDC unit.
\par
Applying  equations~\ref{t_exp},~\ref{t_up} and~\ref{t_dw} one can calculate
a relation between $T^{exp}$ and $t^{real}$:
\begin{equation}
T^{exp} = t^{real} + {{\mbox{offset}^{up} + \mbox{offset}^{dw} + {{L} \over c_L} } \over 2} -
T_{trigger} = t^{real} + \mbox{offset} - T_{trigger},
\label{t_exp2}
\end{equation}
where the value of "offset" comprises all delays due to the utilized electronic circuits,
and it needs to be established separately for each segment. It is worth noting, that due
to the readout at both ends of the detector the $T^{exp}$ is independent
of the hit position along the module, as it can be deduced from equation~\ref{t_exp2}.

\subsection{Relative timing between modules}

\begin{figure}[H]
\includegraphics[height=.19\textheight]{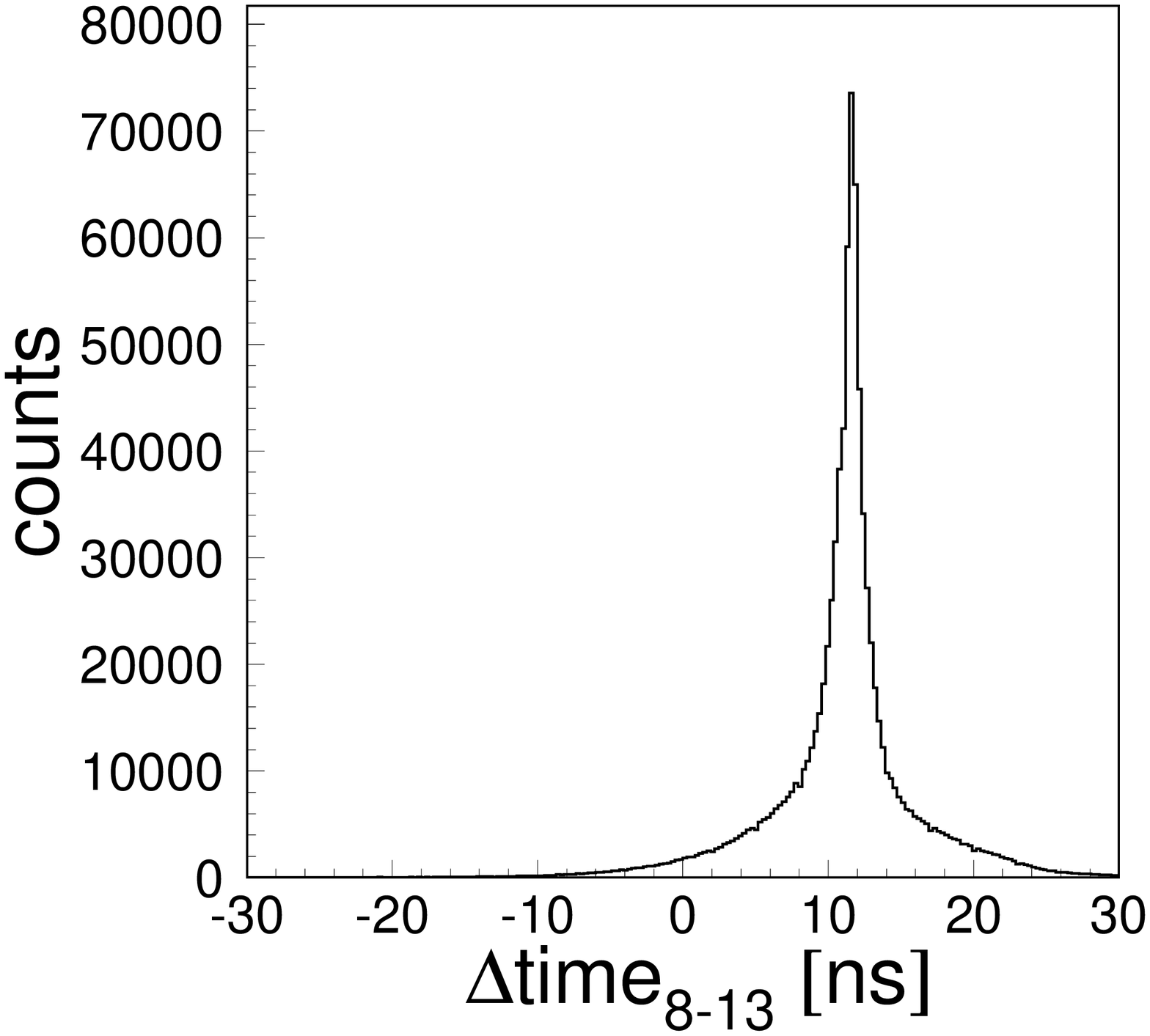}
\includegraphics[height=.19\textheight]{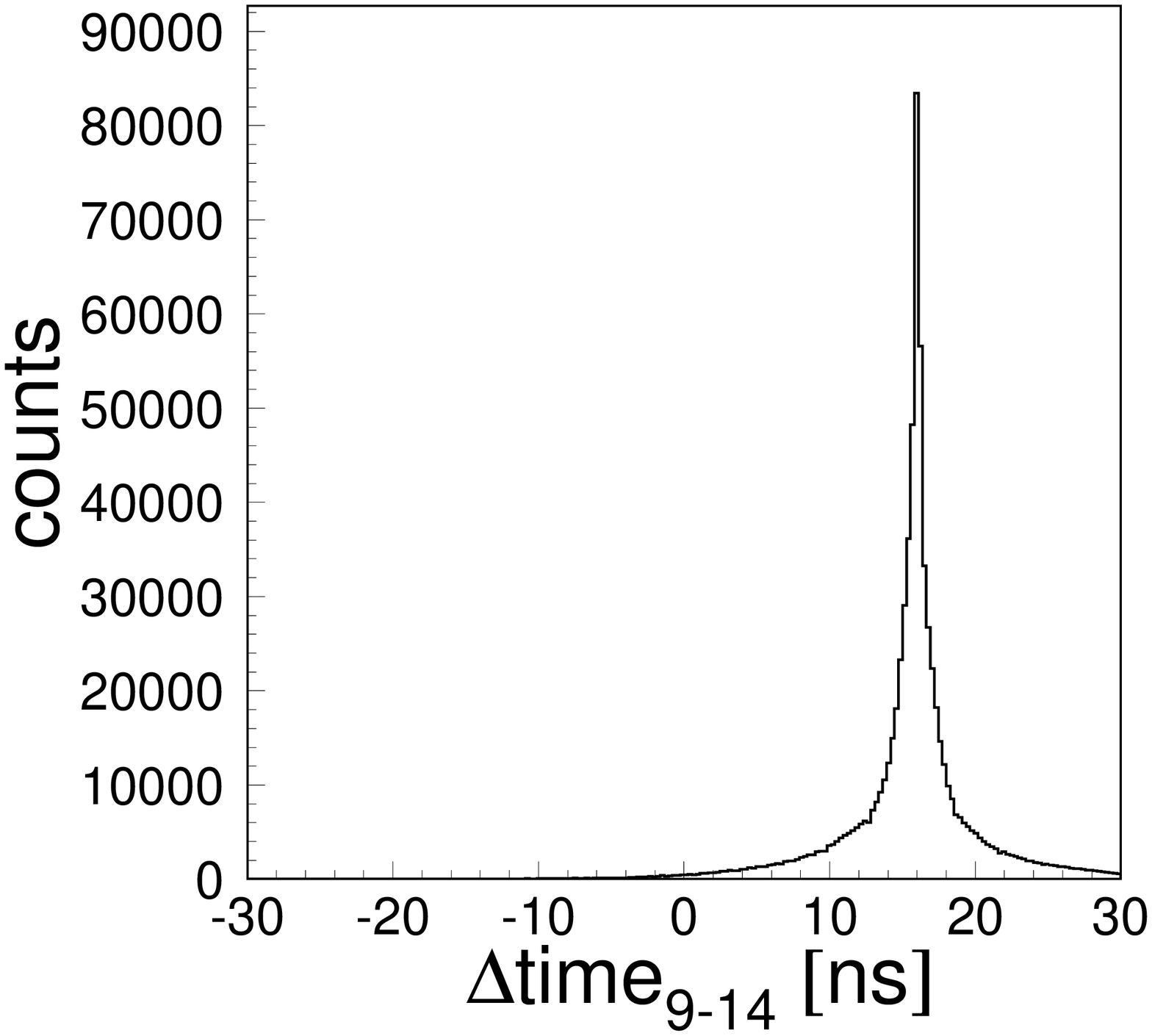}
\includegraphics[height=.19\textheight]{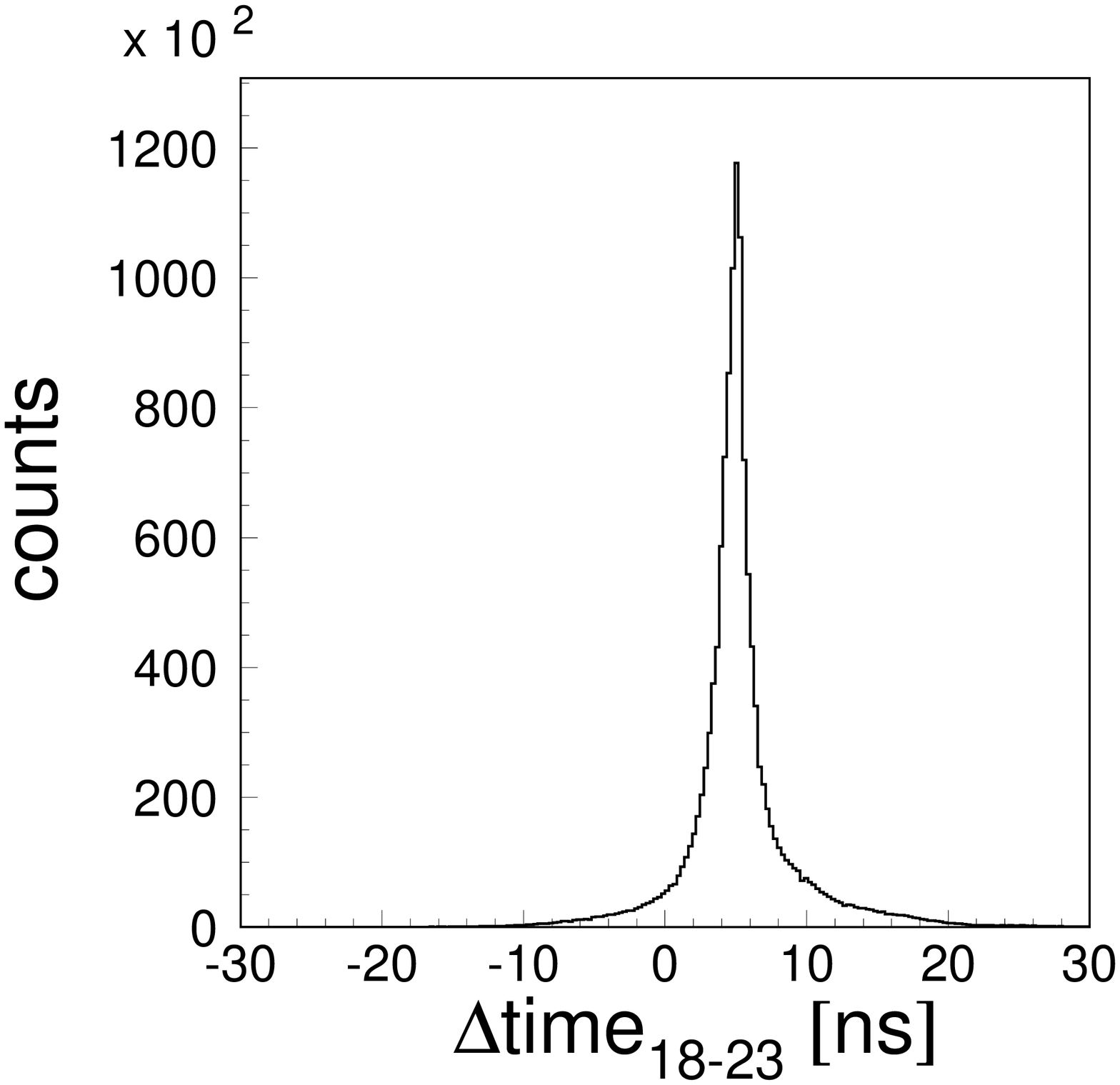}
\caption{ Distributions of the time difference between the $8^{th}$ and the
          $13^{th}$, the $9^{th}$ and the $14^{th}$, and the $18^{th}$ and
          the $23^{th}$ module of the neutron detector, as determined before
          the calibration assuming that offsets are equal to zero.}
\label{offrel_przed}
\end{figure}

Instead of determining the values of "offsets" from equation~\ref{t_exp2}
for each detection unit separately, the relative timing between modules will
be first established and then the general time offset connecting the timing
of all segments with the S1 detector will be found.\\
In order to establish relative time offsets for all
single detection units, distributions of time differences
between neighbouring modules were derived from experimental data
and compared with simulated distributions.
A time difference measured between two modules can be expressed as:
\begin{equation}
   \Delta_{ij} = T^{exp}_{i} - T^{exp}_{j} =
    {t}_{i}^{real} - {t}_{j}^{real} + (\mbox{offset}_{i} - \mbox{offset}_{j}),
\label{delta_ij}
\end{equation}
where $T^{exp}_{i}$ and $T^{exp}_{j}$ stand for the time registered by the $i^{th}$
and $j^{th}$ module, respectively.
Examples of ${\Delta}_{ij}$ spectra determined before the calibration
are presented in figure~\ref{offrel_przed}.

\begin{figure}[H]
\includegraphics[height=.19\textheight]{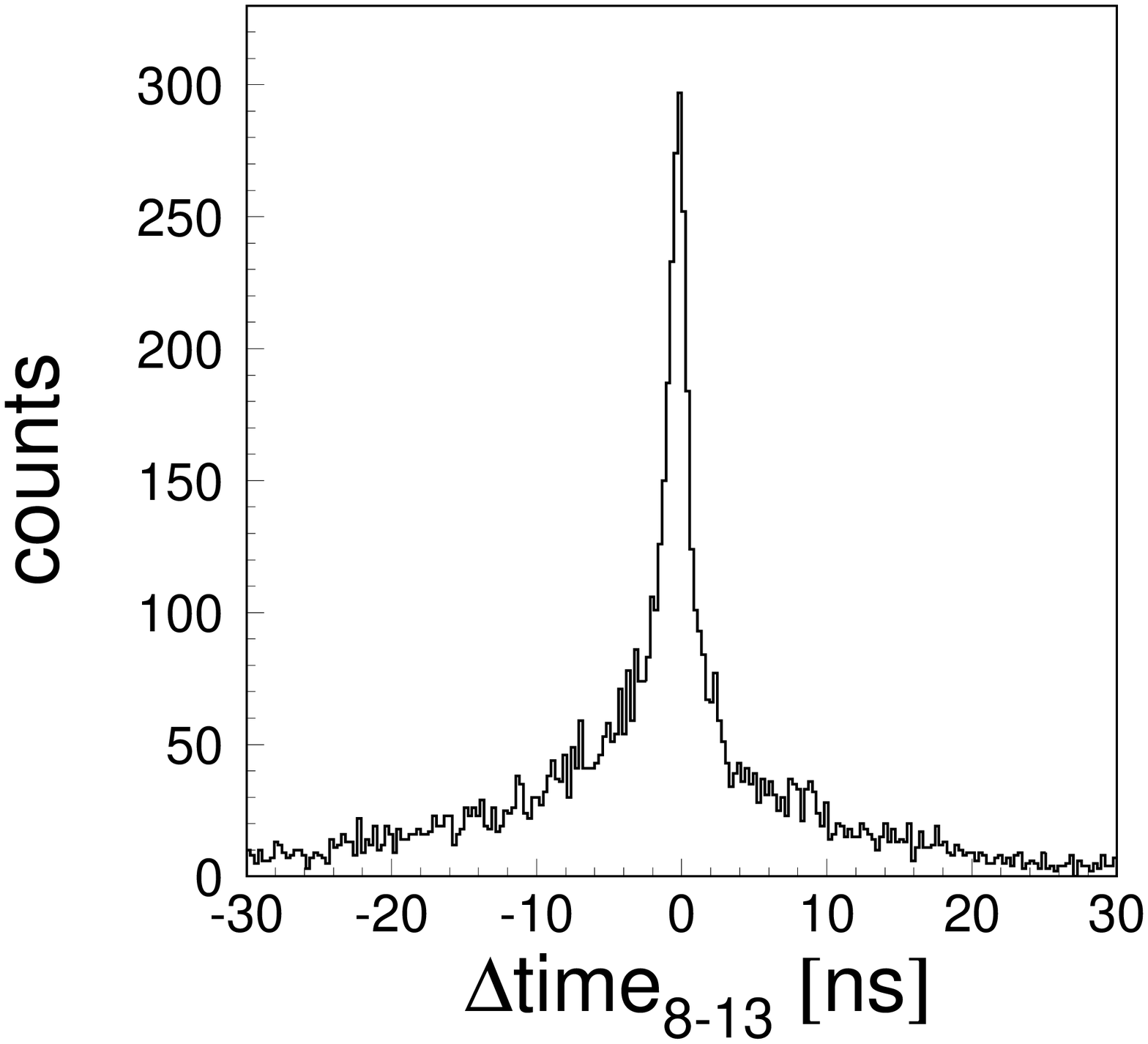}
\includegraphics[height=.19\textheight]{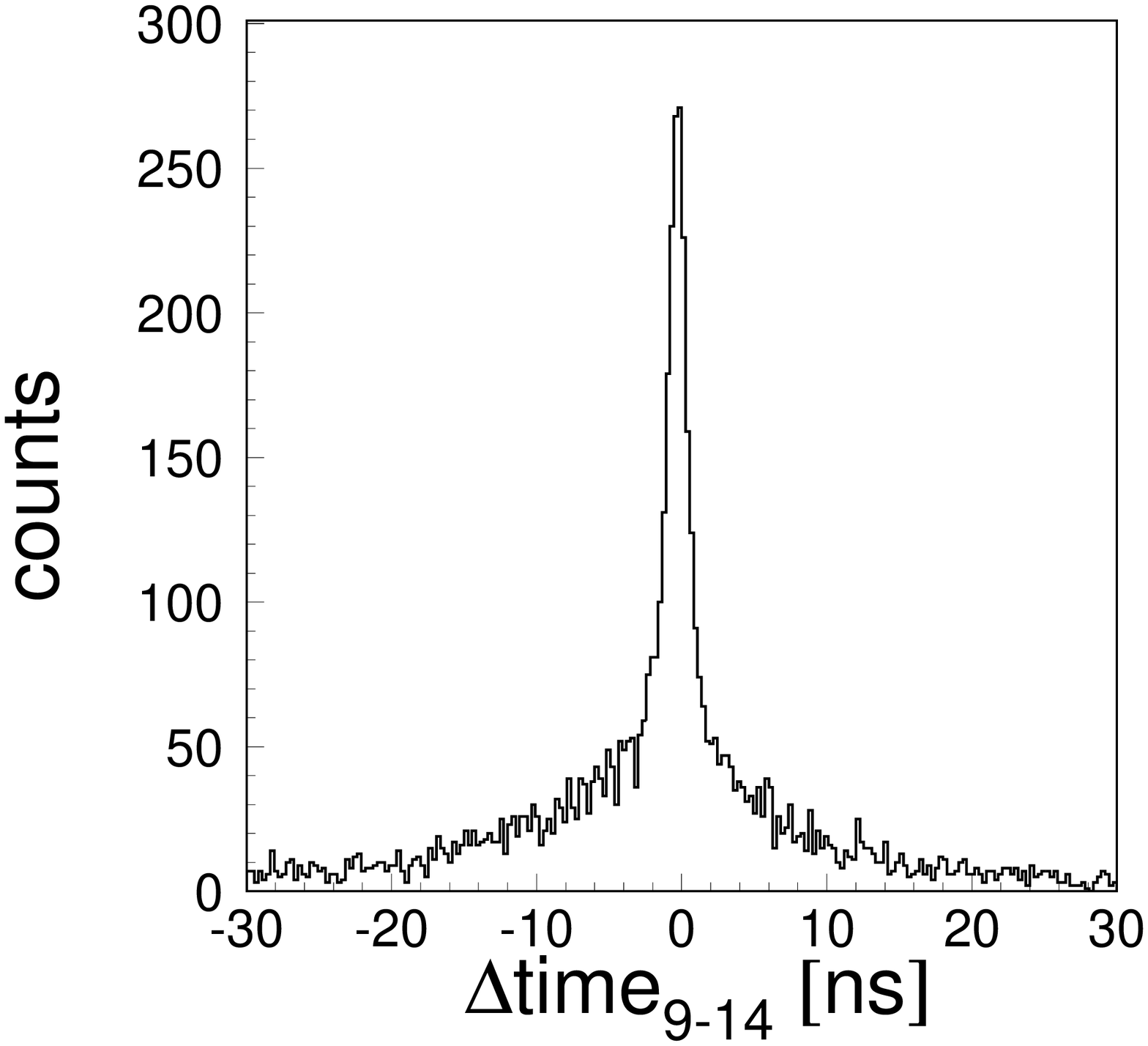}
\includegraphics[height=.19\textheight]{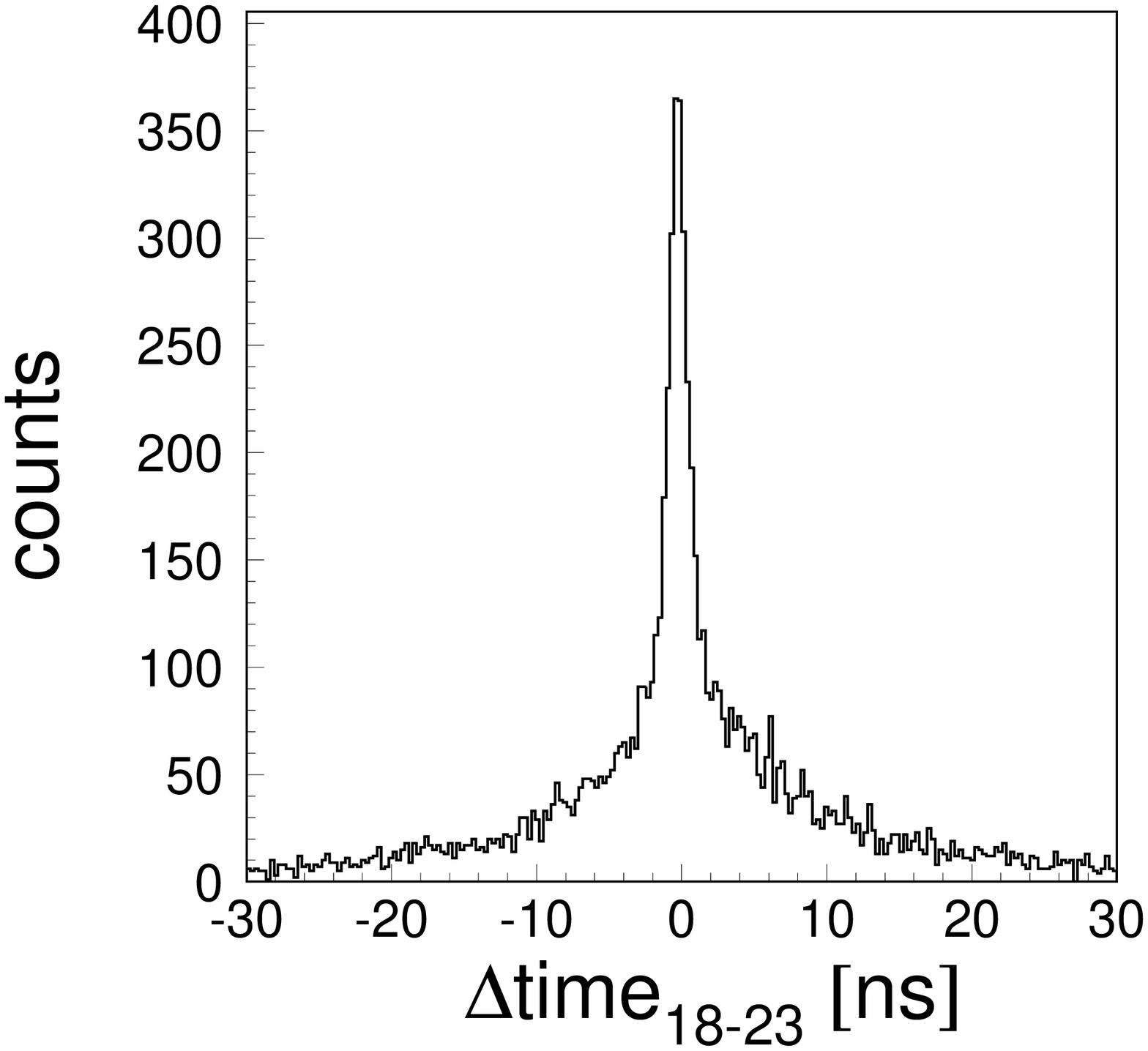}
\caption{ Simulated distributions of the time difference between the $8^{th}$
          and the $13^{th}$, the $9^{th}$ and the $14^{th}$, and the
          $18^{th}$ and the $13^{th}$ module of the neutron detector.}
\label{offrel_sym}
\end{figure}

One can note that the peaks are shifted from the zero value
and additionally the distributions contain tails.
The tails reflect the velocity distribution of the secondary particles.
In order to determine reference spectra
corresponding time differences between the modules were simulated  using the GEANT--3 code.
The result of the simulation is shown in figure~\ref{offrel_sym}.
To produce these spectra the quasi--free $pn~\to~pn\eta^{\prime}$
reaction has been simulated.

The values of the relative time offsets were determined using a dedicated
program written in Fortran 90~\cite{rozek-ar2003,rozek-priv}.
It adjusts values of offsets such that the time difference obtained from the experimental data
and from simulations equals to each other for each  pair of detection units.
Furthermore, from the width of the spectra one can obtain the information
about the time resolution of a single module, which was extracted to be
0.4~ns~\cite{rozek-ar2003}.

\begin{figure}[H]
\includegraphics[height=.19\textheight]{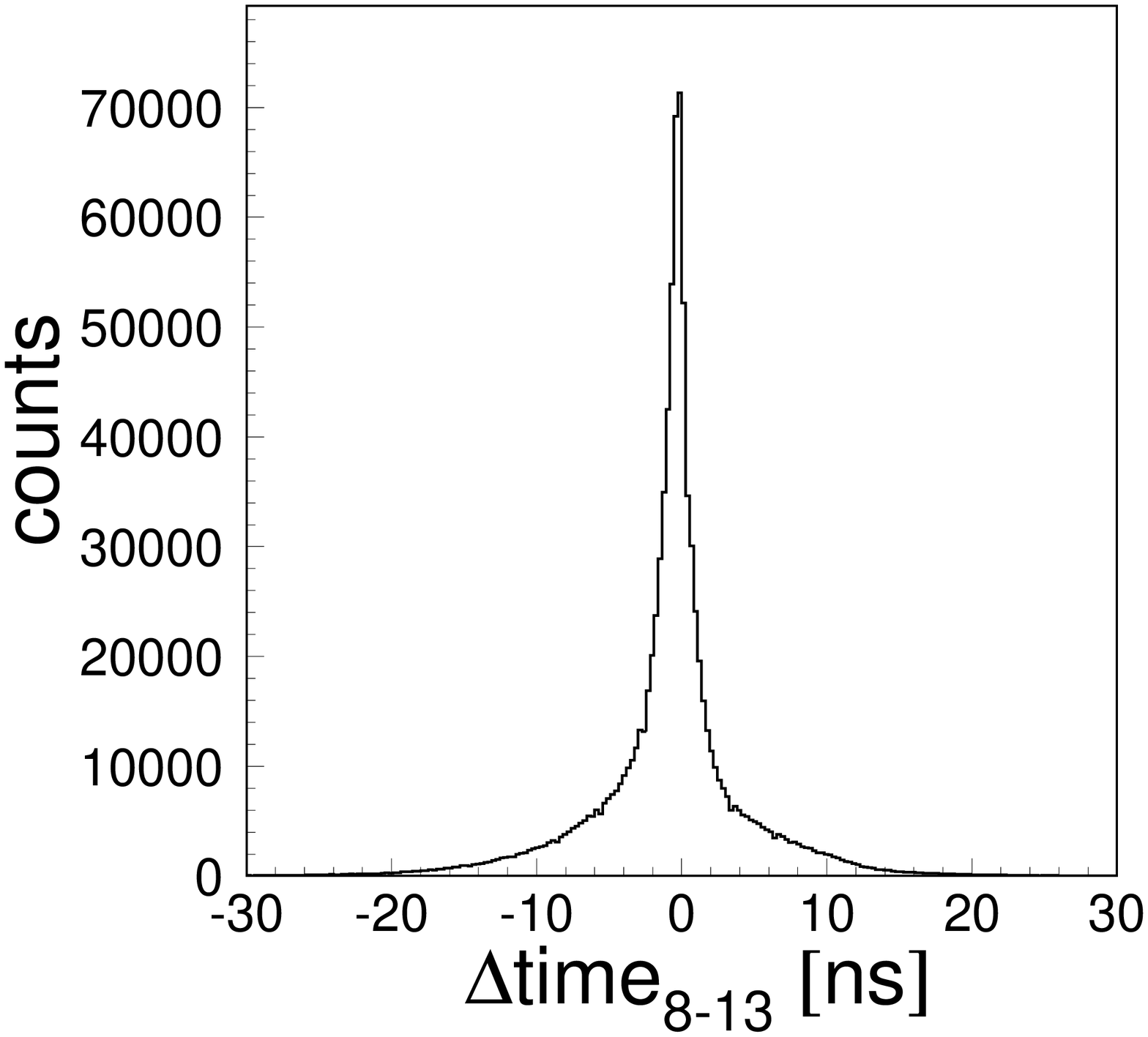}
\includegraphics[height=.19\textheight]{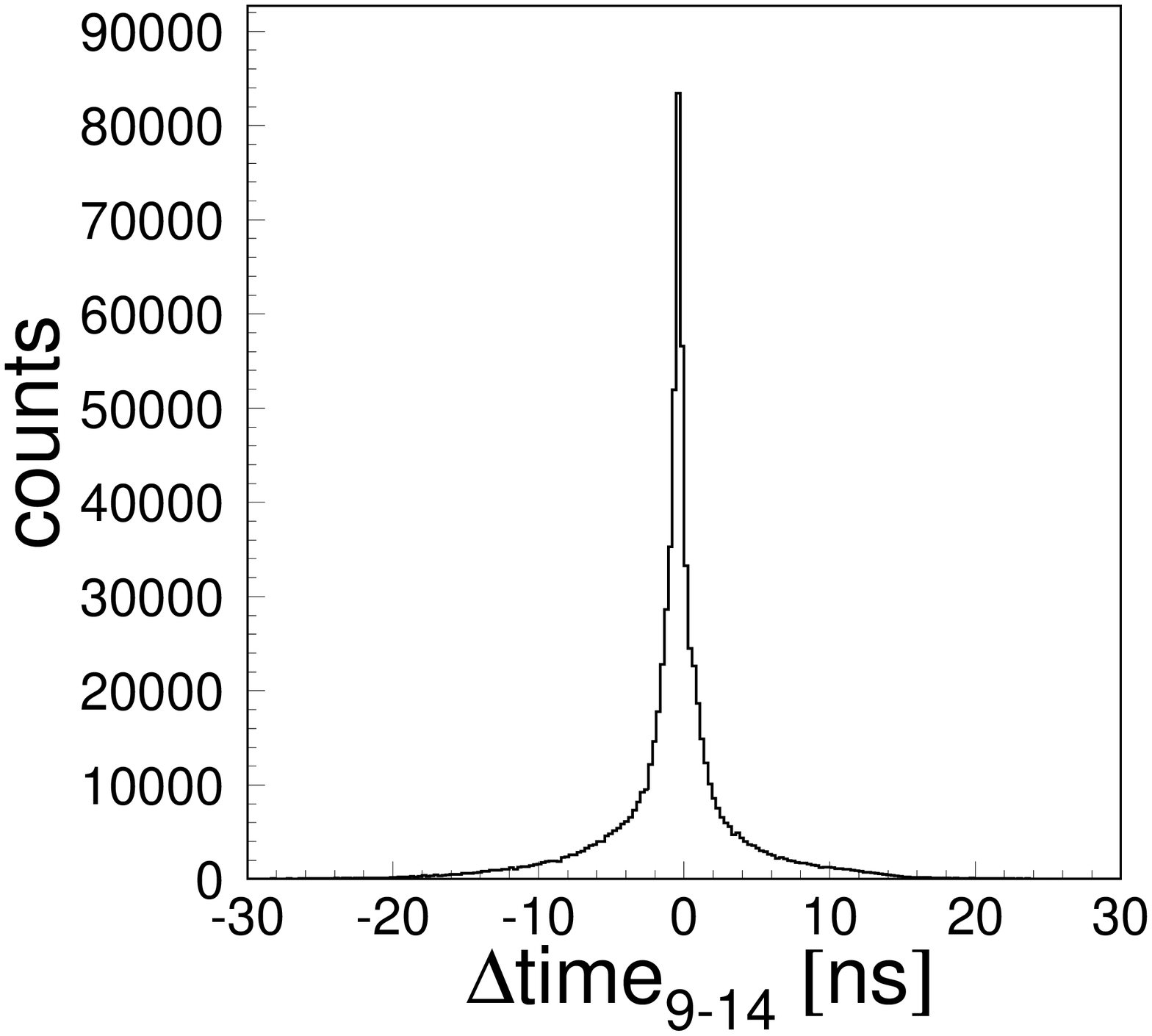}
\includegraphics[height=.19\textheight]{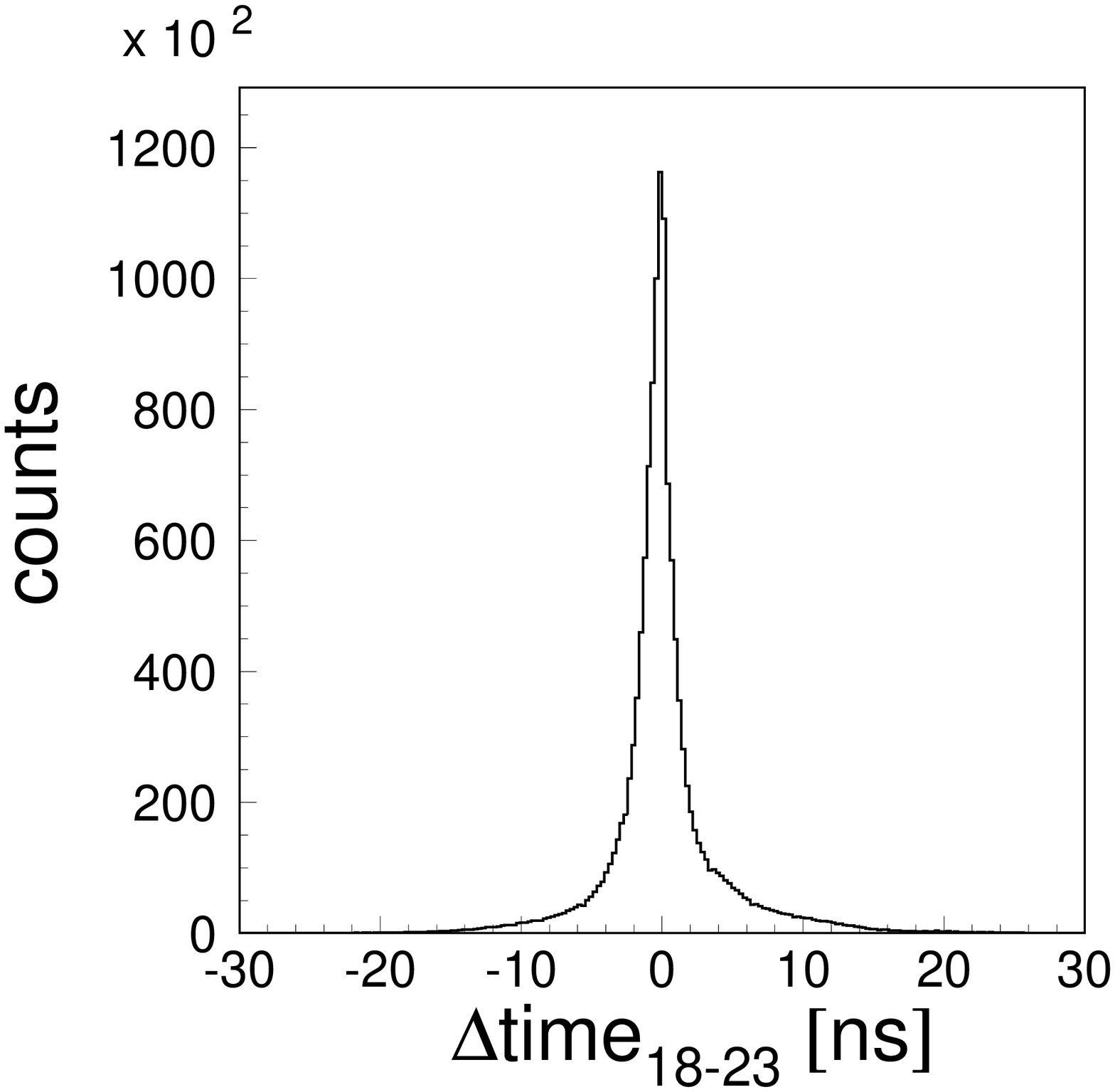}
\caption{ Distributions of the time difference between the $8^{th}$ and the
          $13^{th}$, the $9^{th}$ and the $14^{th}$, and the $18^{th}$ and
          the $23^{th}$ module of the neutron detector, as obtained after
          the calibration.}
\label{offrel_po}
\end{figure}

Figure~\ref{offrel_po} presents experimental distributions of time differences between
neighbouring modules as determined after the calibration.
Examining figures~\ref{offrel_po} and~\ref{offrel_sym}  it is evident that now
the peaks are positioned as expected from simulation.

\subsection{General time offset and the time walk correction}
\label{walk}

The momenta of neutrons are determined on the basis of the time between the reaction
and the hit time in the neutral particle detector.
The time of the reaction can be deduced from the time when the exit proton from
the $pn \to pn\eta^{\prime}$ reaction
crosses the S1 detector (see fig.~\ref{cosy11}) because the
trajectory and the velocity of this proton can be reconstructed.
Therefore, to perform the calculation of the  time--of--flight between the target and
the neutral particle detector,
a general time offset of this detector with respect to the
S1 counter has to be established. This can be done by measuring the gamma quanta
for which the velocity is constant.
\par
\begin{figure}[H]
\includegraphics[height=.29\textheight]{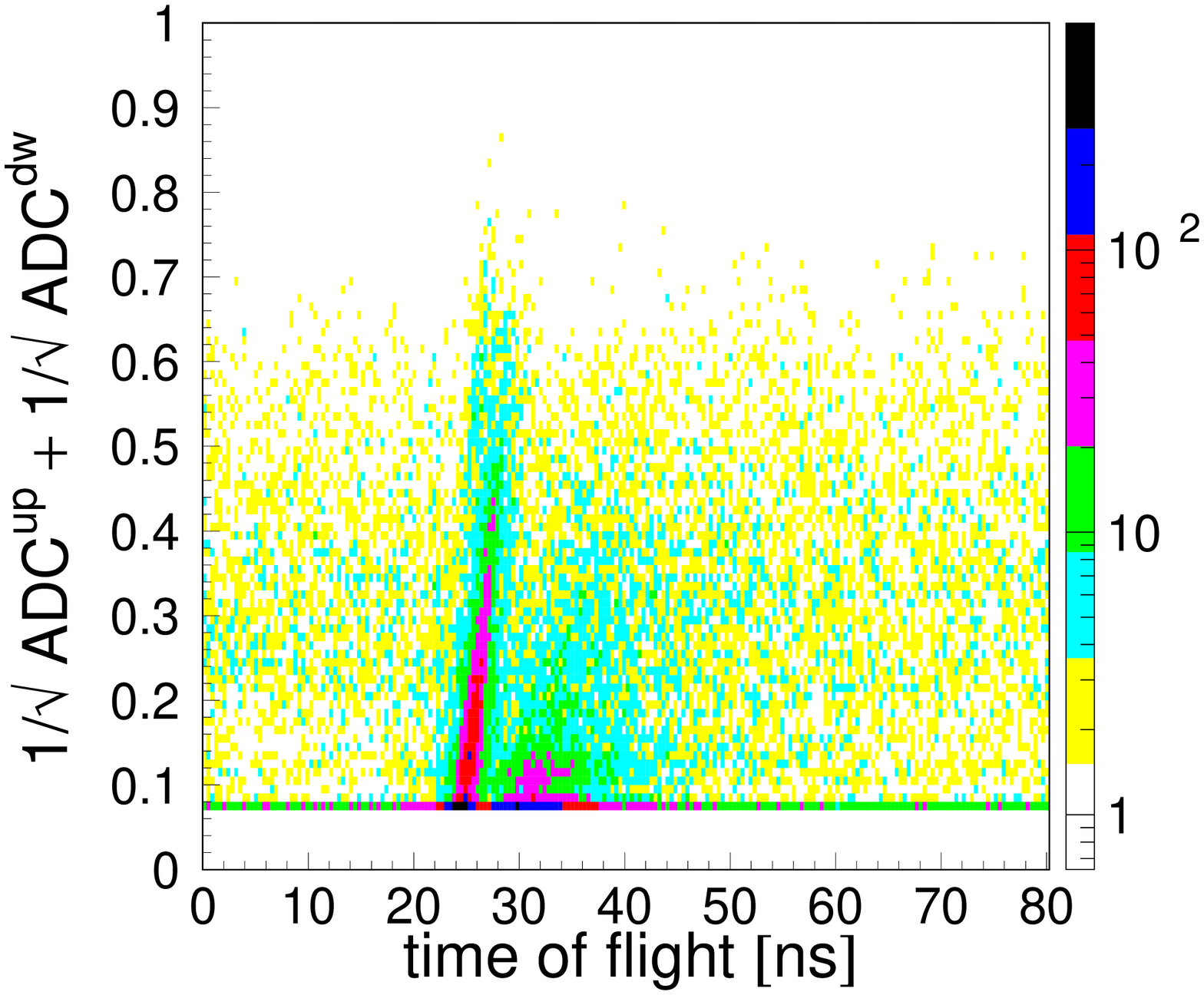}
\includegraphics[height=.29\textheight]{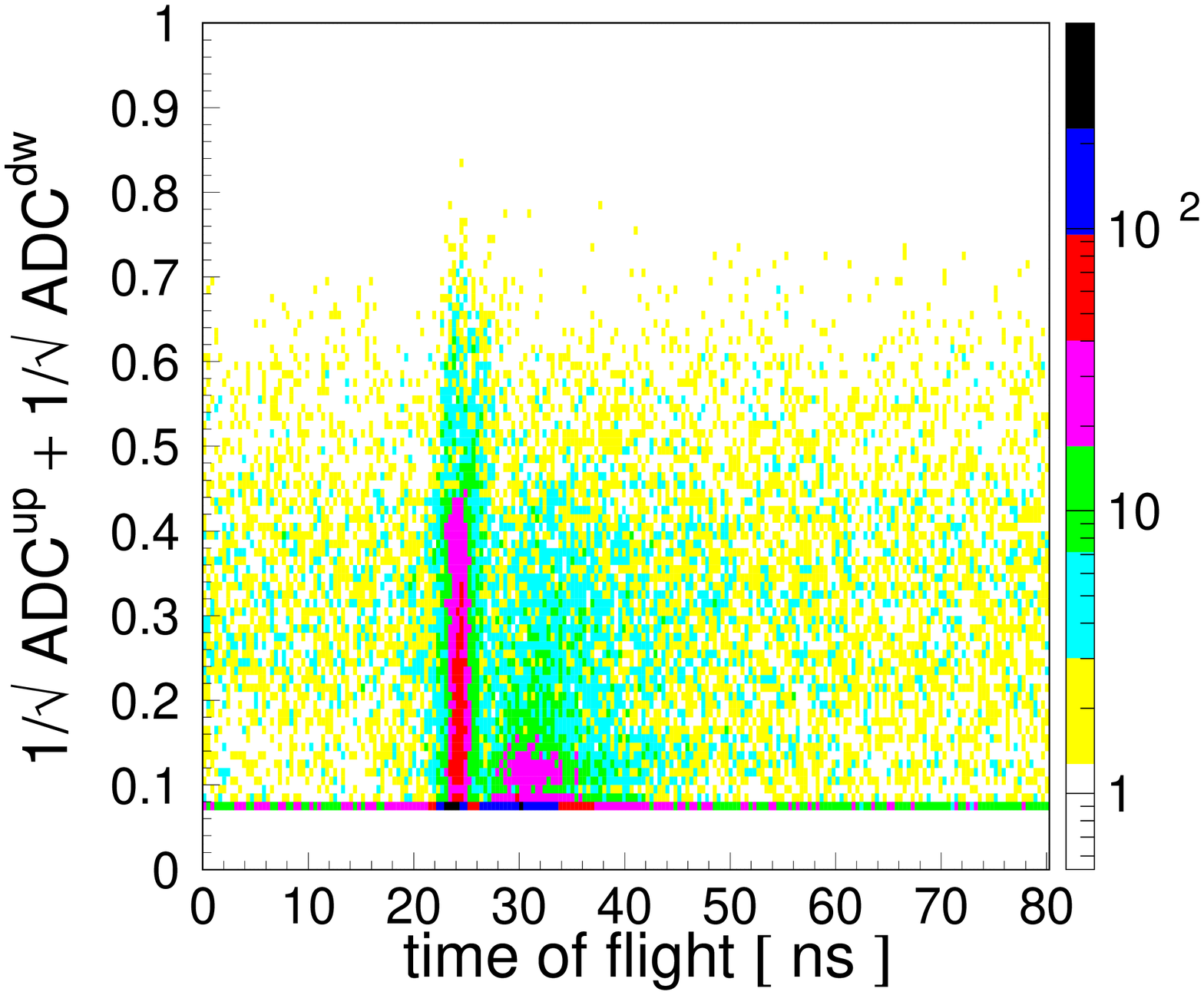}
\caption{ Time walk effect. A value of
          ${ 1 \over{ \sqrt{ADC^{up}}} } + { 1 \over{ \sqrt{ADC^{dw}}} }$
          measured in the neutral particle detector
          as a function of the time--of--flight ($TOF^{N}$) between the target and
          the neutral particle detector as obtained before {\bf (left)} and after {\bf (right)}
          time walk correction.}
\label{9955}
\end{figure}

The time--of--flight between the target and the neutral particle detector $({TOF}^{N})$
is calculated as the difference between the time of the module
in the neutral particle detector which fired as the first one ($t_{N}^{real}$)
and the time of the reaction in the target $(t^{real}_{r})$:
\begin{equation}
    {TOF}^{N} =  t_{N}^{real} - t^{real}_{r}
\label{TOF_neut}
\end{equation}
The time of the reaction is obtained from backtracking the protons through the known
magnetic field and the time measured by the S1 detector ($t_{S1}$). Thus $t^{real}_{r}$
can be expressed as:
\begin{equation}
t^{real}_{r} = t_{S1} - TOF^{S1} = T_{S1}^{exp} - offset^{S1} + T_{trigger} - TOF^{S1},
\label{t_real_r}
\end{equation}
where $TOF^{S1}$ denotes the time--of--flight between the target and the S1 counter,
$offset^{S1}$ denotes all delays of signals from the S1 detector, and analogically
as for equation~\ref{t_exp2} the $t_{S1}$ and $T_{S1}^{exp}$ denote the time
at which the scintillator light in the S1 detector was produced and the averaged time registered
by the TDC unit~(see eq.~\ref{t_exp}), respectively.
Since the time in the neutron detector reads:
\begin{equation}
 t_{N}^{real} = T_{N}^{exp} - offset^{N} + T_{trigger}
\label{t_n_real}
\end{equation}
we have:
\begin{equation}
{TOF}^{N} = T_{N}^{exp} + TOF^{S1} - T_{S1}^{exp} - offset^{N} + offset^{S1}
      = T_{N}^{exp} + TOF^{S1} - T_{S1}^{exp} + offset^{G}
\label{TOF_neut1}
\end{equation}
By $offset^{G}$ the general time offset of the neutron detector with respect to S1 is denoted.
In order to determine the value of $offset^{G}$ the $pd \to A^{+}B^{+}\gamma X$
reaction is used, where $\gamma$ is measured by the
neutral particle detector and for which the ${TOF}^{N}$ between the target
and the neutral particle detector is known.
$A^{+}$ and $B^{+}$ represent any two positively charged particles~($\pi^+\pi^+,~p\pi^+,~pd,~d\pi^+)$.
Events corresponding to the $pd \to A^{+}B^{+}\gamma X$ reaction
have been identified by measuring the outgoing charged as well as neutral ejectiles.
Positively charged particles (pions, protons and deuterons) were detected by means of
the drift chambers (D1, D2) and scintillator hodoscopes (S1, S3).
Gamma quanta originated predominately from the $\pi^0$ meson decay
are registered in the neutral particle detector.
Knowing the time--of--flight for gamma quanta $T_{\gamma}$ on the distance
between the target and the neutral particle detector the general time offset
was determined by comparing the
measured time with the nominal value of $T_{\gamma}$. The measured time--of--flight
spectrum is shown in figure~\ref{9955}. $T_{\gamma}$ should be equal to 24.5~ns.
As can be inspected from the figure~\ref{9955}, the time depends on the
amplitude. This phenomenon is occuring for measurements using a
leading-edge discriminator and is referred to as the time-walk effect~\cite{leo}.
In the case of a constant threshold value, two signals of different
pulse hights but exactly coincident in time may trigger the discriminator
at different times. An offline correction can be applied
to minimize this effect assuming a linear dependence between the time walk
and the inverse of the square root of the signal charge~\cite{tanimori-nim216}.
The function used for offline time-walk correction was
\begin{equation}
 t_c = t_m - \alpha + \beta ({ 1 \over{ \sqrt{ADC^{up}}} } + { 1 \over{ \sqrt{ADC^{dw}}} }),
\label{t_walk}
\end{equation}
where $t_c$ and $t_m$  are corrected and measured times,
and $\alpha$ and $\beta$ are
coefficients determined from the data,
and $ADC^{up}$ and $ADC^{dw}$ denote the charge of the signal
measured at the upper and the lower end of the detection module.

Figure~\ref{9955} shows the dependence of the time--of--flight for neutral particles
as a function of values of ADC signals in the neutral particle detector.
The time--of--flight shown in the figure was measured between the target and the neutral particle
detector before (left) and after (right) the time walk correction.
The spectrum was obtained under the condition that in coincidence
with a signal in the neutral particle two charged particles were registered in
the drift chambers. The time in the neutral particle detector was corrected in
such a way as if the interaction point of particles was in the first row of the detector
modules only. A clear signal originating from the gamma rays is seen
over a broad enhancement from neutrons. This histogram  shows that a discrimination
between the signals originating from neutrons and gamma quanta can be done by a cut
on the time of flight.
For the gamma quanta the time--of--flight value
is independent of their energy. This fact was used to determine the $\alpha$ and $\beta$
coefficients. The right panel of figure~\ref{9955} presents the analogous distribution for
which  a corrected time $t_c$ was used for the time--of--flight calculation
instead of the $t_m$.
\begin{figure}[H]
\centerline{\includegraphics[height=.32\textheight]{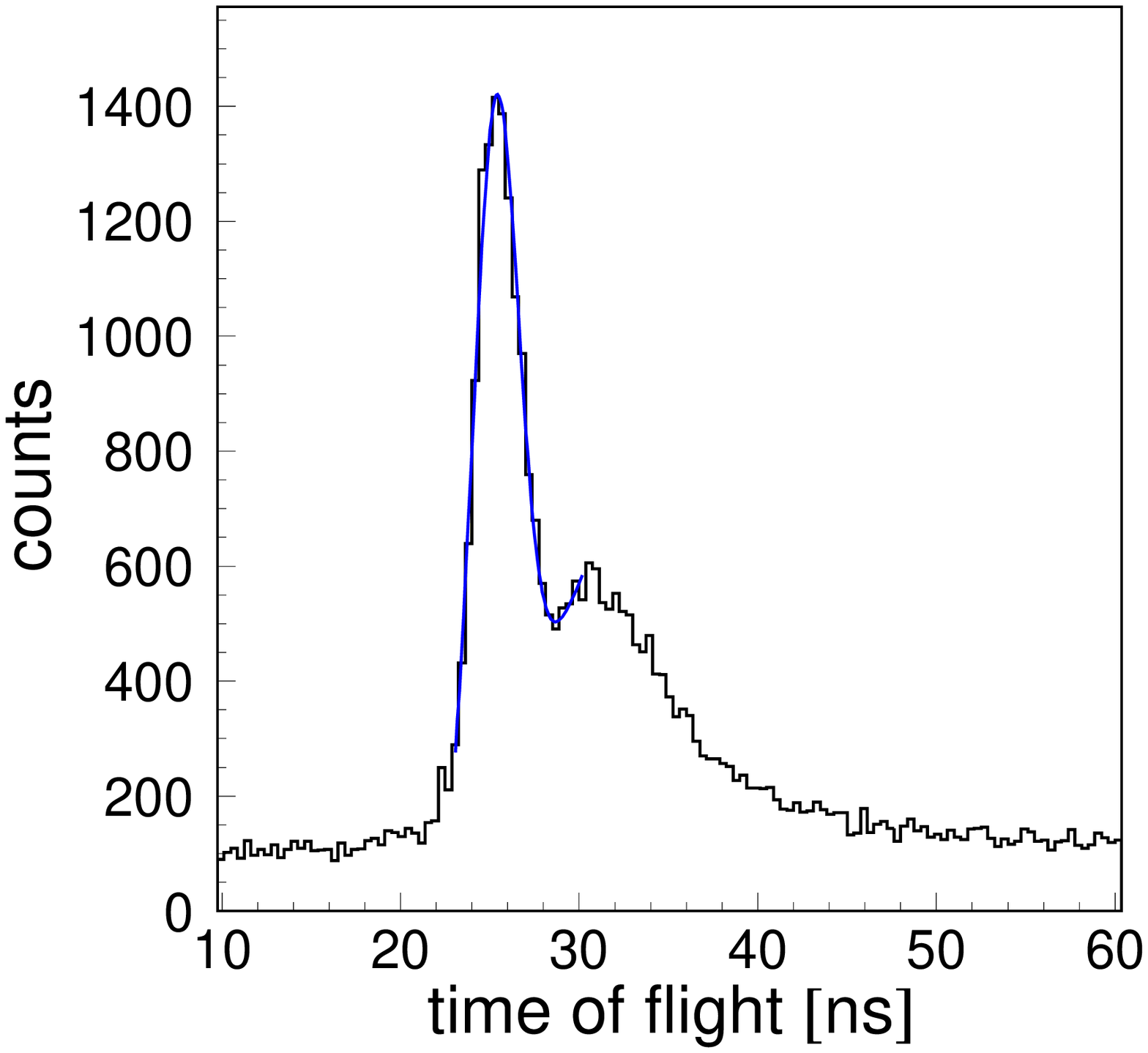}
\includegraphics[height=.32\textheight]{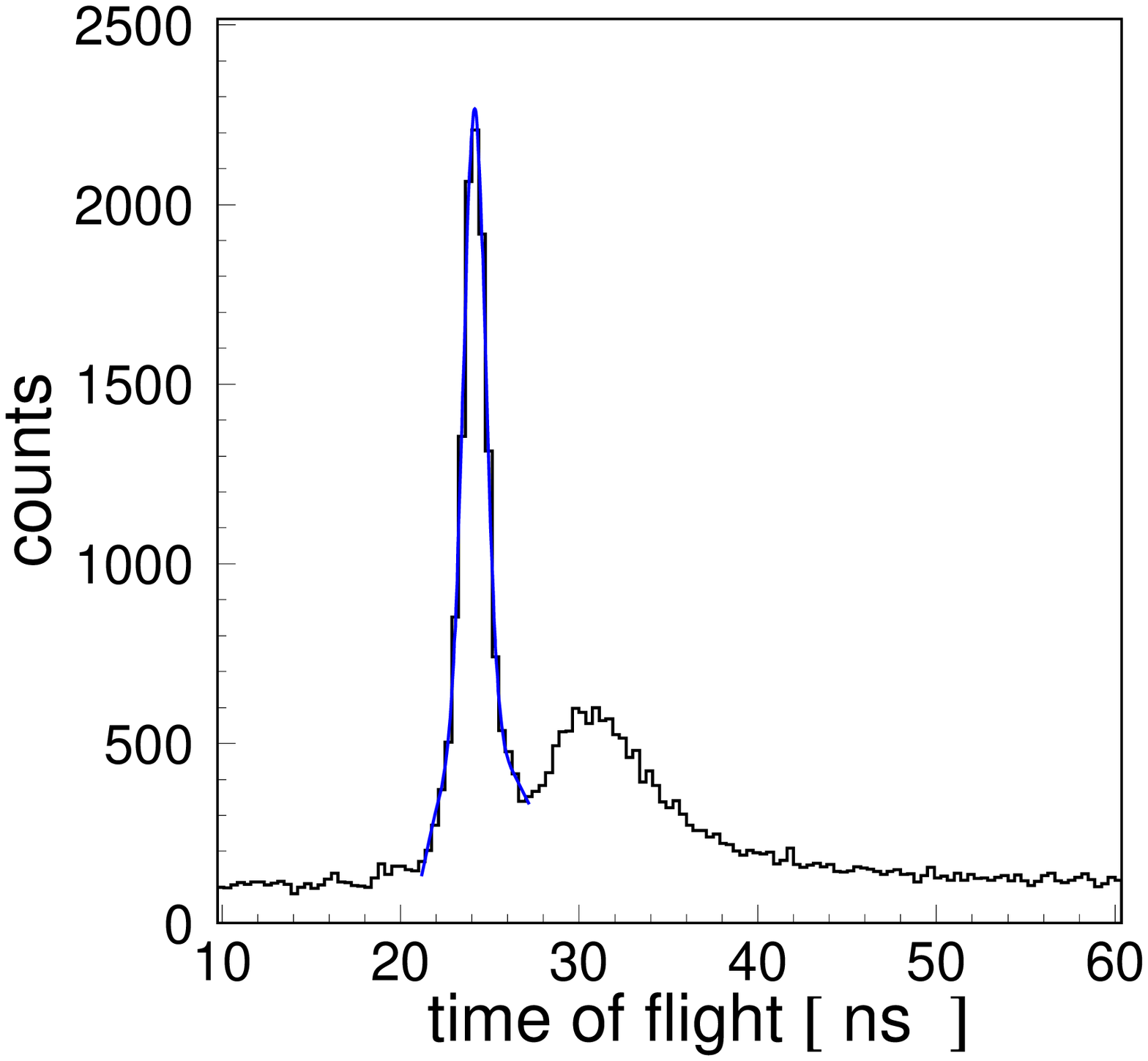}}
\caption{ Distribution of the time--of--flight between the target and the neutral particle
          detector before {\bf (left)} and after {\bf (right)} time walk correction.}
\label{tof_walk}
\end{figure}
Figure~\ref{tof_walk} presents the projection of the distributions from figure~\ref{9955}
on the time--of--flight axis. Both distributions of the time--of--flight were fitted by a
Gaussian function in the range where the clear peak originating from gamma quanta
appears. The area around the peak was fitted by a second order polynomial.
The left panel of figure~\ref{tof_walk} presents the distribution of the time--of--flight
before the time walk correction, and the fit was resulting in a time--of--flight resolution
of $\sigma~=~1.2$~ns. The time--of--flight for which the corrected time in the neutral particle detector
was used is shown in the right panel of figure~\ref{tof_walk}.
The application of the time walk correction improved the time resolution to $\sigma~=~0.6$~ns.
It is worth noting that this is an overall time--of--flight resolution resulting from
the time resolution of the neutron and S1 detectors and the accuracy of the momentum reconstruction
of charged particles, needed for the determination of the time of the reaction
in the target.
\par
\begin{figure}[H]
\includegraphics[height=.29\textheight]{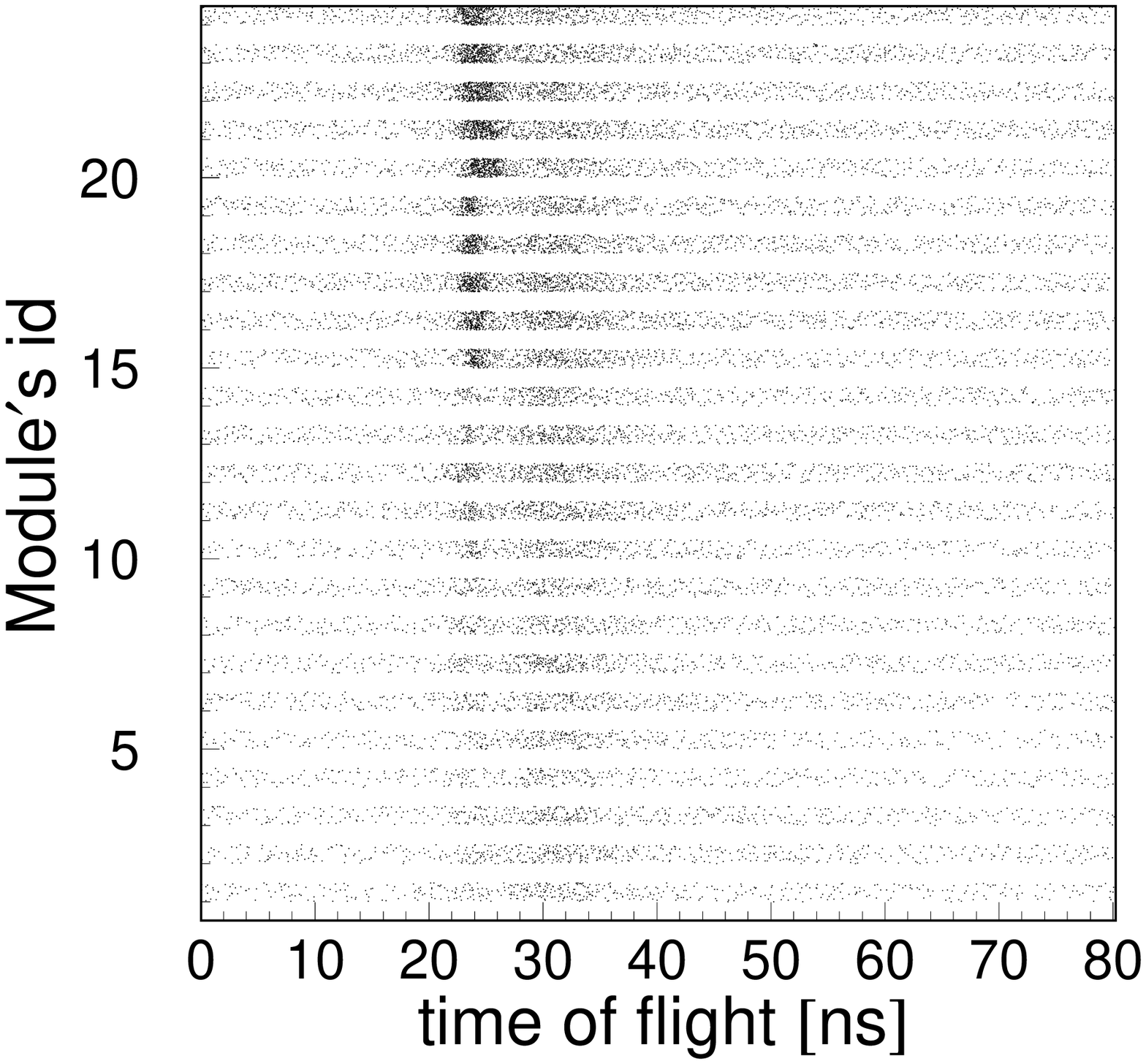}
\includegraphics[height=.29\textheight]{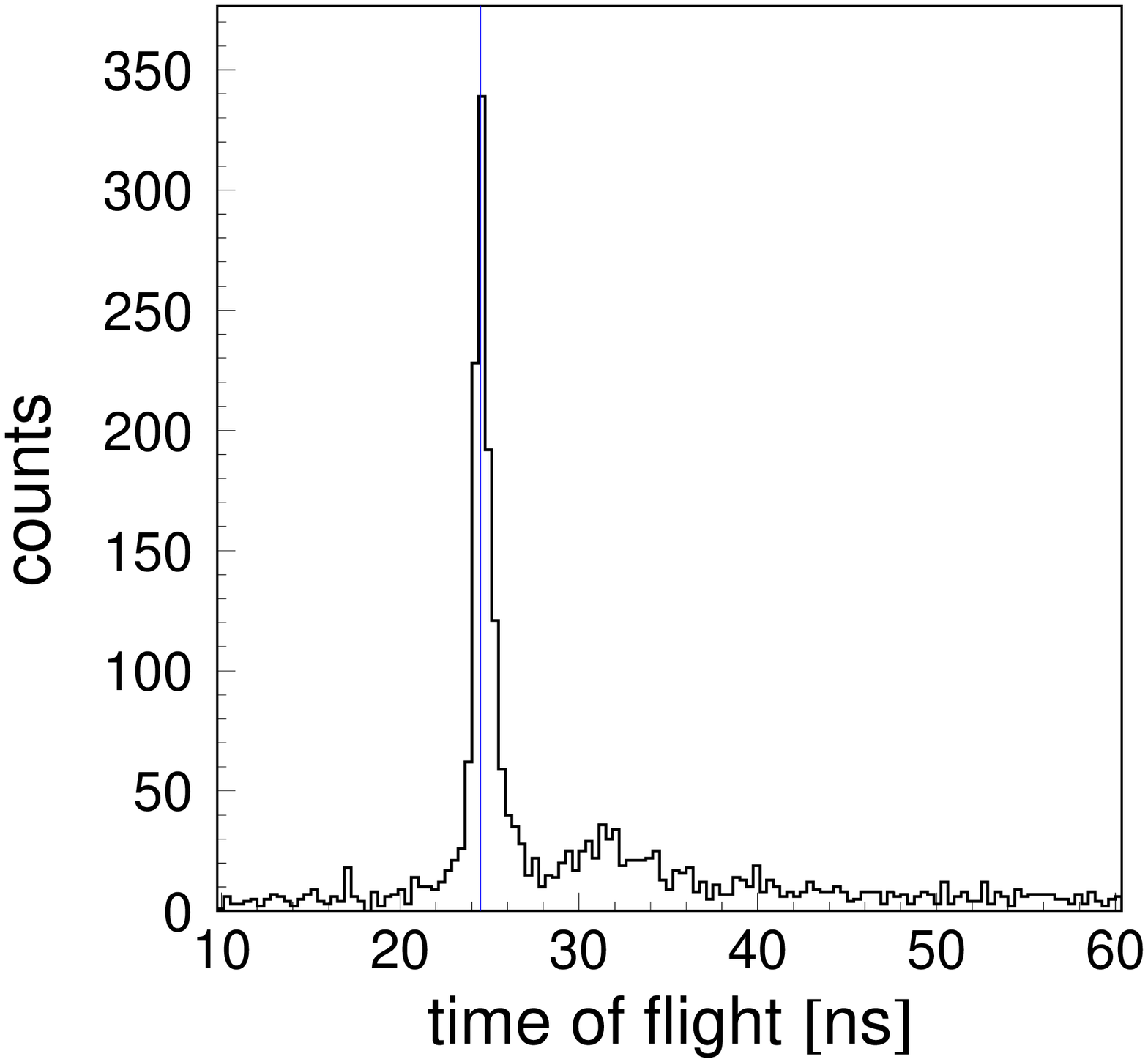}
\caption{{\bf Left:} Time--of--flight determined between the target and the neutron
          detector as a function of modules's number. {\bf Right:} ${TOF}^{N}$ distribution
          determined taking signals from first row of the neutral particle detector only.}
\label{tof_mod_id}
\end{figure}

Figure~\ref{tof_mod_id} (left) shows the TOF distribution as a function of modules number.
As expected from the known absorption coefficients~\cite{hagiwara-prd66},
the gamma quanta are predominantly registered in the first row of the
detector (see fig.~\ref{modul_pod_3d}) whereas the interaction points of the
neutrons are distributed more homogeneously.
Indeed, a clear signal originating from the gamma quanta is seen on the time--of--flight
distribution, when taking into account signals from the first row of the neutron
detector only, as it is shown in figure~\ref{tof_mod_id} (right).
Therefore, in order to raise the confidence to the determination of the general offset
we have taken signals in the first row of the neutral particle detector.
The data have been restricted only for the $pd \to pd\gamma~X$ channel. In this case due to the
baryon number conservation there is only one possible source of a signal in the
neutral particle detector, namely
a gamma quantum, which originates predominantly from the decay of $\pi^0$ mesons in the target.

\begin{figure}[H]
\centerline{\includegraphics[height=.4\textheight]{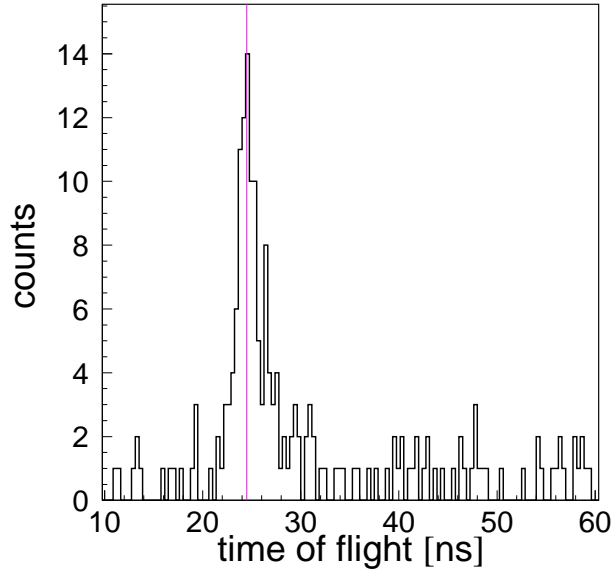}}
\caption{Time--of--flight distribution determined between the target and the neutron
         detector obtained under the assumption that additionally to a signal in the
         neutron detector one proton and  one deuteron were identified from the
         signals in the drift chambers. }
\label{tof_pd}
\end{figure}

Figure~\ref{tof_pd} presents the experimental distribution of the time--of--flight
between the target and the neutron detector for neutral particles,
with the requirement that  additionally to a signal in the
neutron detector two charged
particles were registered and that one of them was identified as a proton
and the other as a deuteron\footnote{The method of identifying the particles
registered by the drift chambers and scintillator hodoscopes is described
in Chapter~6.}.
As expected, in this spectrum only a signal from the gamma quanta is seen.
The clear peak is positioned at the value equal to 24.5 ns, which corresponds to the
time--of--flight for the light on the target--neutral particle detector distance.

\subsection{Efficiency determination}
\label{efficiency_sim}

The efficiency of the COSY--11 neutral particle detector --
which is an important factor for determining the absolute
values of cross sections -- was determined using two independend
simulation programs. For the first case, a procedure based on the
GEANT-3 ({\bf GE}ometry {\bf AN}d {\bf T}racking) code~\cite{geant} was
used for the simulation of the hadronic cascades induced in matter by neutrons.
The same procedure was repeated using the FLUKA\footnote{The simulations were
performed with the 2008 version.} ({\bf FLU}ktuierende {\bf KA}skade)~\cite{fluka,zdebik-dt}
simulation program.
\par
\begin{figure}[H]
\includegraphics[height=.29\textheight]{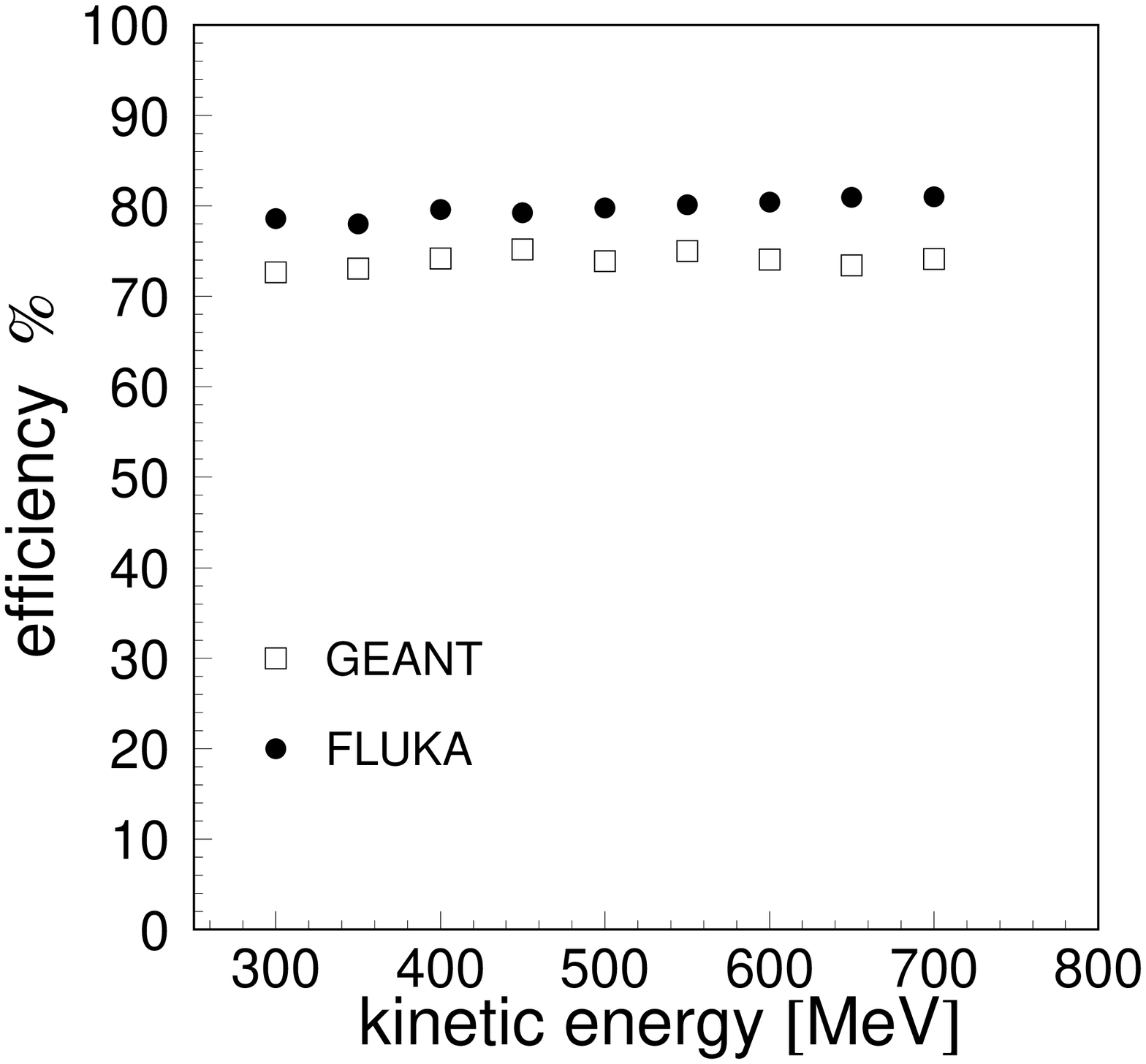}
\includegraphics[height=.29\textheight]{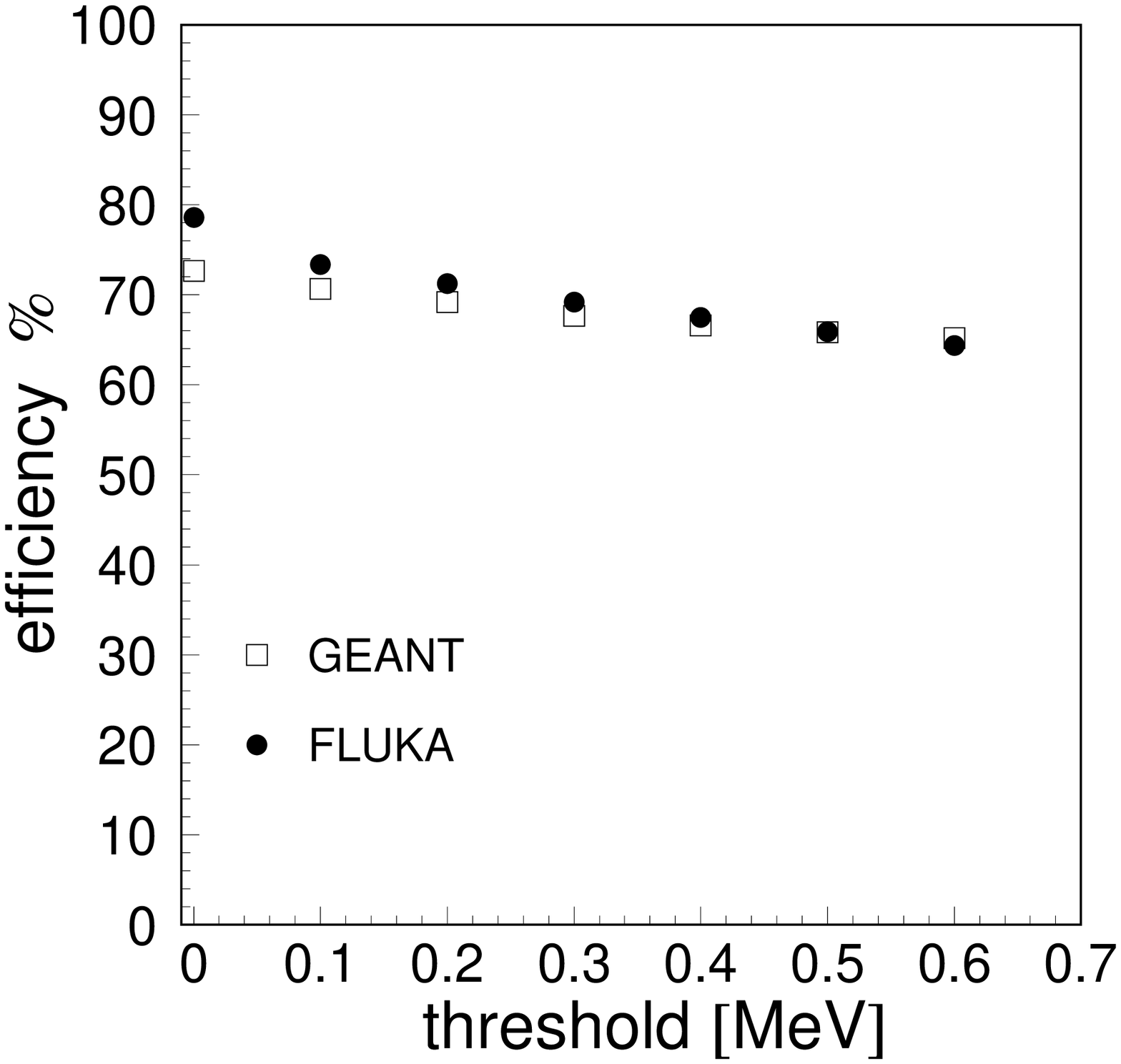}
\caption{ {\bf Left}: The efficiency distribution as a function of the kinetic energy
                      of the neutrons determined assuming that the threshold is equal to 0 MeV.
         {\bf Right}: The relation between the threshold value and the efficiency for
                      neutrons with an energy of 300 MeV.}
\label{neut_eff}
\end{figure}

The efficiency of the neutron detector is given by the ratio of the number
of events, for which an energy deposited in the scintillator material was
larger than the threshold value at least in one of 24 detection units
to the number of generated neutrons.
The value of the calculated efficiency as a function of the kinetic energy of the neutrons
is shown in fig.~\ref{neut_eff} (left). Open squares denote result obtained
using the GEANT-3 package and the outcome of the simulation using FLUKA-2008 is presented as black
circles~\cite{jklaja-app31}.
The kinetic energy of the neutrons from the $pn \to pn \eta^{\prime}$
reaction varies from 300~MeV up to 700~MeV for the 3.35~GeV/c beam
momentum, and as can be inspected from fig.~\ref{neut_eff} (left) the
efficiency is fairly constant in this range. It is important to
mention that the results determined with the different programs are in agreement
within $\pm3\%$, independently of the threshold value.\\
Also studies of the variation of the efficiency
depending on the threshold have been conducted. In the experiment the
threshold was set to about 0.1 MeV and therefore
the values from 0 up to 0.6~MeV were scanned.
The result is presented in fig.~\ref{neut_eff} (right).
For both, the GEANT and FLUKA-2008 simulation the efficiency
changes by about 10\% over the 0.6~MeV range of the threshold.

\section{Veto detector}
\begin{figure}[H]
\centerline{\includegraphics[height=.3\textheight]{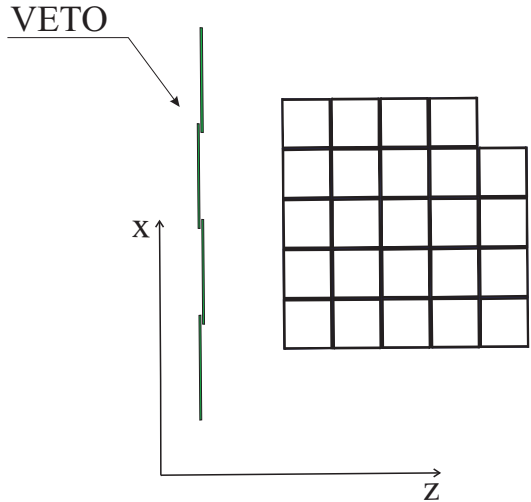}}
\caption{Layout of veto detector and neutral particle detector.}
\label{veto_cosy11}
\end{figure}

In order to distinguish between charged and neutral particles, in front of the
first layer of the neutral particle detector an additional scintillator detector
was installed as it is shown in figure~\ref{veto_cosy11}.
This detector -- referred to as veto detector  -- is built out of four
overlapping modules with dimensions of 400~mm~$\times$~200~mm~$\times$~4~mm.
The lightguides and photomultipiers are mounted at the upper and lower edge of
the modules such that the light signals are read out at both sides of the module.\\
It permits to reject the background originating from charged particles hitting
the neutral particle detector. The discrimination between signals arising from
the only possible remaining
neutrons and gamma quanta is done by a cut on the time--of--flight~(see chapter 6.).
\begin{figure}[H]
\includegraphics[height=.29\textheight]{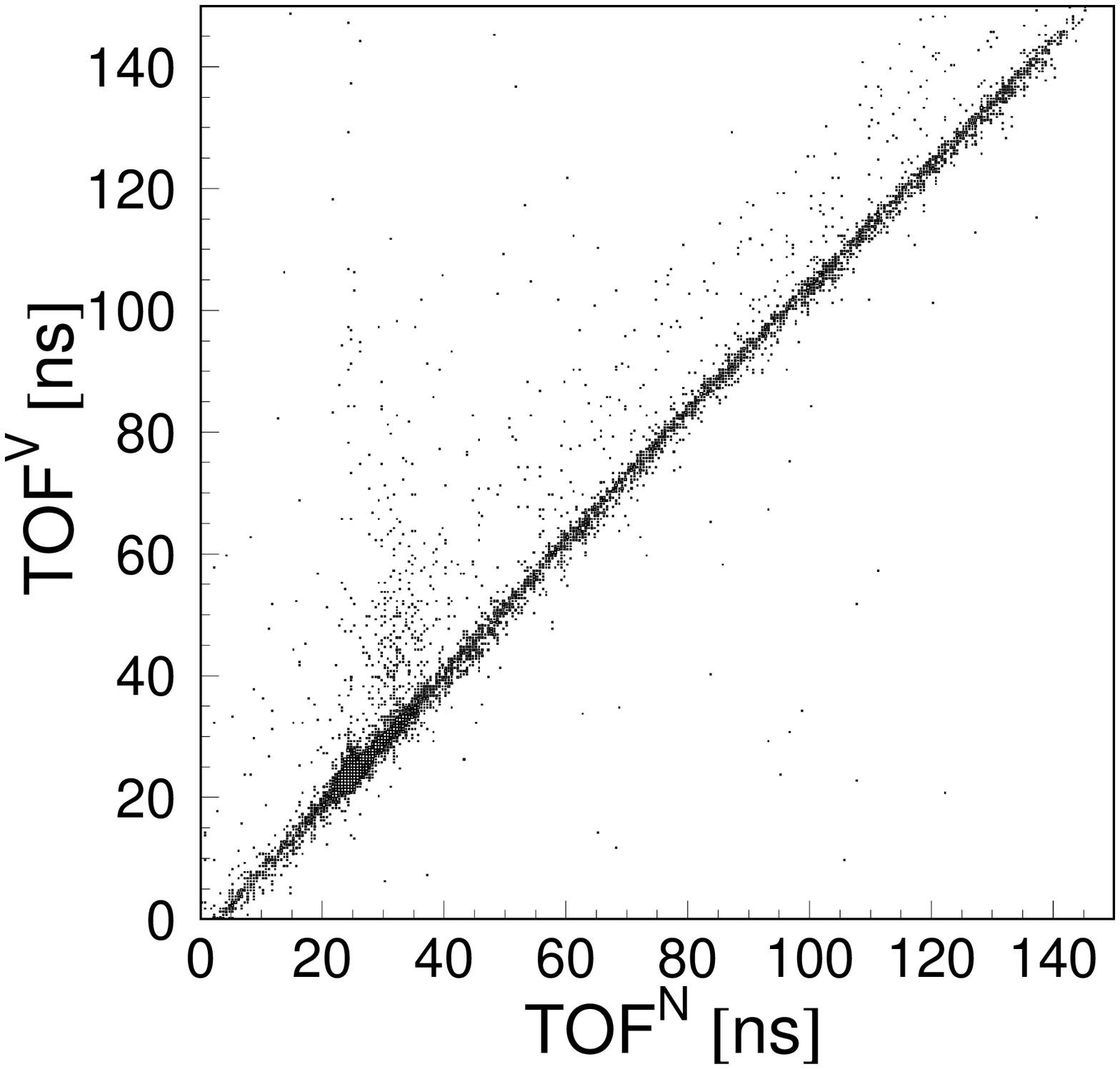}
\includegraphics[height=.29\textheight]{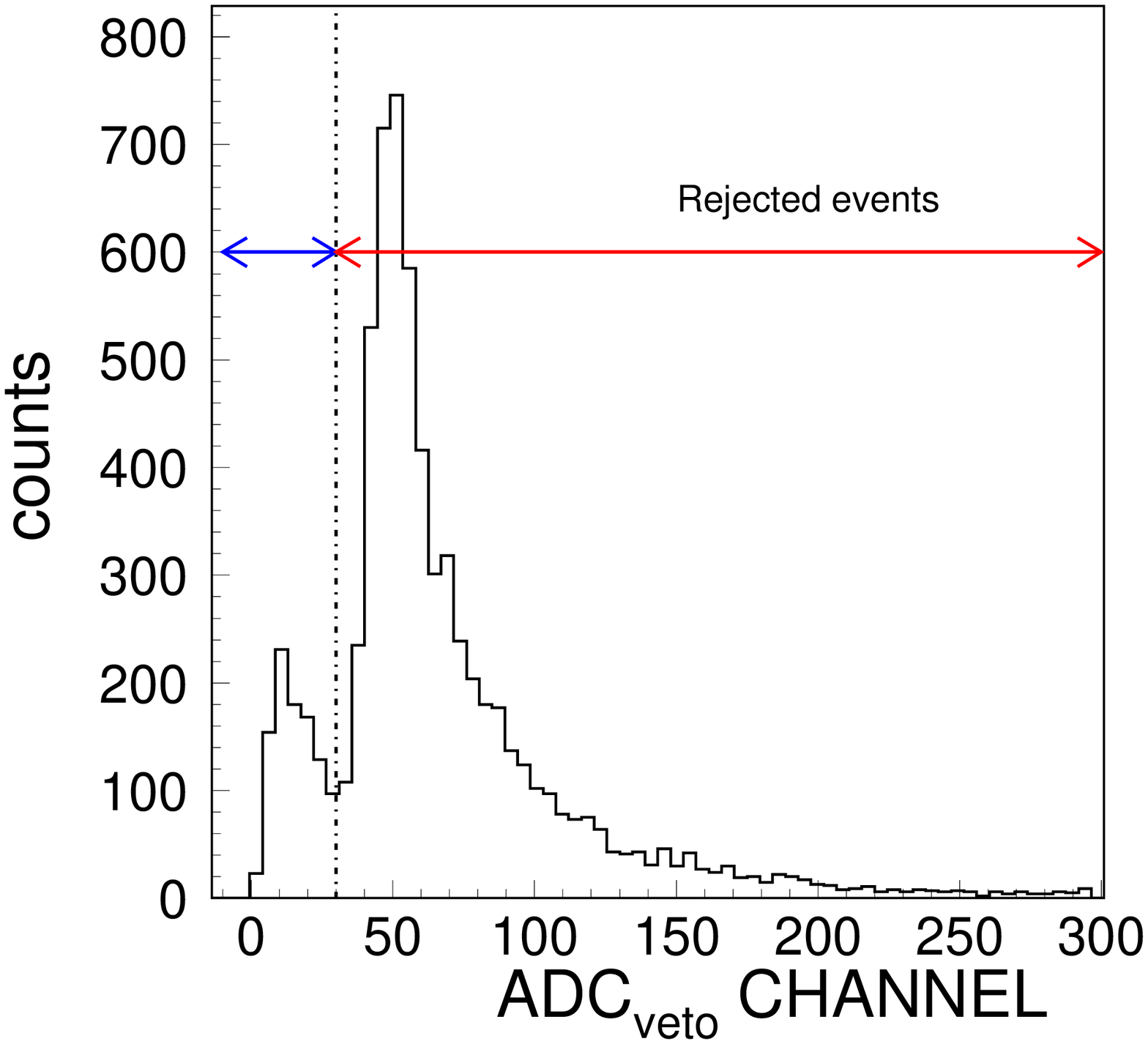}
\caption{{\bf Left:} The relation between the time--of--flight between the target
                     and the neutral particle  detector ($TOF^N$) and the
                     time--of--flight between the target and veto detector ($TOF^V$).
         {\bf Right:} ADC signals in the scintillator veto detector. The red arrow shows the
                      ADC range for charged particles, which are rejected in the further
                      analysis, whereas the blue arrow point to the ADC range for which events
                      are interpreted as neutral particles. Noise signals are extinguished
                      by the coincidence between the upper and lower photomultiplier.}
\label{veto}
\end{figure}
As a first step of the time calibration of the veto detector
the time--of--flight between the target and the veto detector has been calculated
for each module separately assuming that all time offsets are equal to zero. Next, the values of the
calibration constants (offsets) were adjusted such that the time--of--flight
value is equal for each module.
Figure~\ref{veto}~(left) shows the relation of the
time--of--flight between the target and the neutral particle  detector ($TOF^N$)
and the time--of--flight between the target and the veto detector ($TOF^V$).
In principle this relation should be linear, however we observe events
beyond the correlation line with time signals in the veto detector larger than in the neutral
particle detector. This may happened when a neutral particle (which does not give a
signal in the veto detector) induces electromagnetic or hadronic reactions in the neutral
particle detector, and charged products of these reactions scattered backwards resulting in
signals in the veto detector. This effect is reflected in figure~\ref{veto}~(right)
which shows the ADC signals in one of the veto detector module. Two
well separated peaks are observed. The peak in the range from 30 to 300 channels originates from
events with charged particles which crossed the veto detector and then entered the neutral particle
detector. When taking into account events which are above the correlation line only,
the peak in the range between channel numbers 30 and 300 vanishes and only events with
small ADC value up to 30 remain. Therefore, to discriminate between charged and neutral particles
a cut on the veto detector ADC was applied, which is denoted in the figure ~\ref{veto}~(right)
by a dotted line.

\section{Spectator detector}

In order to identify the unobserved $\eta^{\prime}$ meson in the quasi-free
$pn \to pnX$ process by means of the missing mass technique,
one has to measure not only the momenta of outgoing
proton and neutron, but also a determination of the $pn$ reaction energy
$(\sqrt{s})$ for all measured events is needed. The latter was realised by measuring
the four-momentum vector of the spectator proton.

\subsection{Scheme of the spectator detector}

Before installation at the COSY-11 detection setup, the spectator detector has
been used by the PROMICE/WASA collaboration at the CELCIUS storage ring of The
Svedberg Laboratory~\cite{bilger-nim457}.
Detectors which are located inside the scattering chamber, close to the
interaction point, must be compatible with high vacuum and resistant to radiation
environment. Moreover, a very good energy resolution is highly required in order to determine
the four-momentum of the spectator protons. These factors exclude the use of plastic scintillator.
Solid state devices are well suited
for this purpose, therefore silicon was chosen as an active material in the
spectator detector.\\
\begin{figure}[H]
\centerline{\includegraphics[height=.23\textheight]{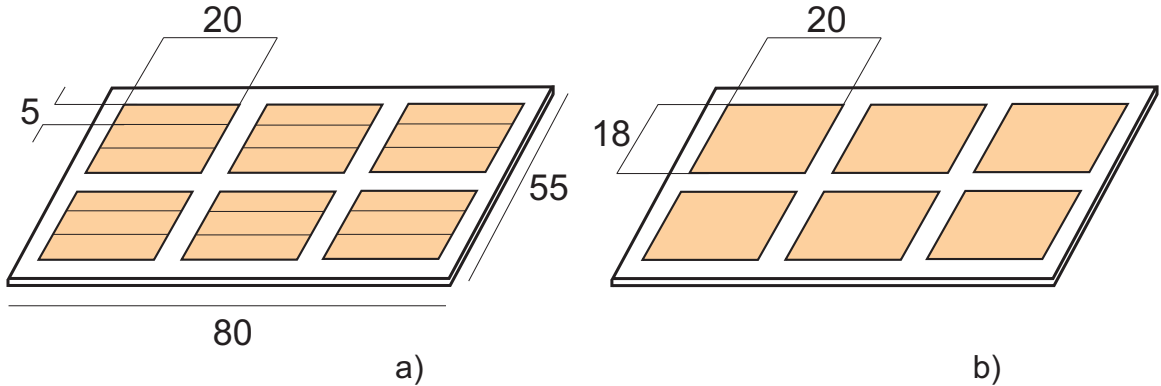}}
\caption{Schematic view of the spectator detector's single modules.
         {\bf Left:} front layer (dE). {\bf Right:} back layer (E).
         Indicated dimensions are given in mm.}
\label{spec_scheme}
\end{figure}
The spectator detector consists of four double--layered modules. The layers are
produced of 300~$\mu$m thick silicon. Each of the front layers, which are the planes
closer to the beam, contain eighteen silicon pads. The active area of a single pad amounts to
20~mm~$\times$~5~mm. The back layer contains six silicon pads, with an active area of
20~mm~$\times$~18~mm. The schematic picture of the front~(dE) and the back~(E) layers is
shown in figure~\ref{spec_scheme}.\\
\begin{figure}[H]
\centerline{\includegraphics[height=.25\textheight]{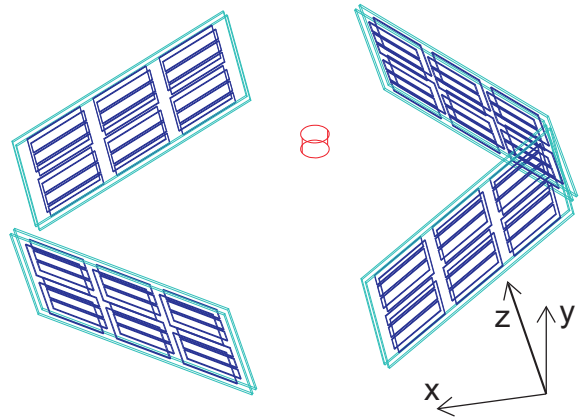}}
\caption{Schematic view of the spectator detector as installed at the COSY-11
         facility. The z-axis corresponds to the direction of the beam, x is
         chosen parallel to the acceleration plane and y is vertical.}
\label{spec_3d}
\end{figure}
The silicon detector
system is placed in the high vacuum 
at the COSY storage
ring at a distance of about 5~cm upstream from the interaction point
as it is schematically shown in figure~\ref{spec_3d}.
The active area covers about 22\% of the full solid angle. The arrangement was a
compromise between the technical needs for the installation and the angular
resolution.

\subsection{Energy calibration and position optimization}

During the experiment, the charge of the signal
induced in the detector by the particles, which is proportional
to the energy deposited in the detection unit, has been measured. Denoting by $ADC_{1}$
and $ADC_{2}$ the measured charge in the first and second layer, respectively,
the real energy loss can be expressed as: $\Delta~E_{1} = \alpha_{1} ADC_{1}$ and
$\Delta~E_{2} = \alpha_{2} ADC_{2}$, where $\alpha_{1}$ and $\alpha_{2}$
are the calibration constants for two layers.
The aim of the spectator detector calibration was to find the
exact position of the detector and then to establish the
calibration constants for each of the 96 silicon pads.
The position of the spectator detector inside the beam pipe
is described by only three parameters~\cite{czyzyk-dt}.
Variable {\it dz} denotes the translation along the z-axis
in the main reference system (MARS). Parameters {\it dl} and {\it alpha}
account for translation and rotation of the flange used for supporting the
spectator detector.
\par
\begin{figure}[H]
\centerline{\includegraphics[height=.4\textheight]{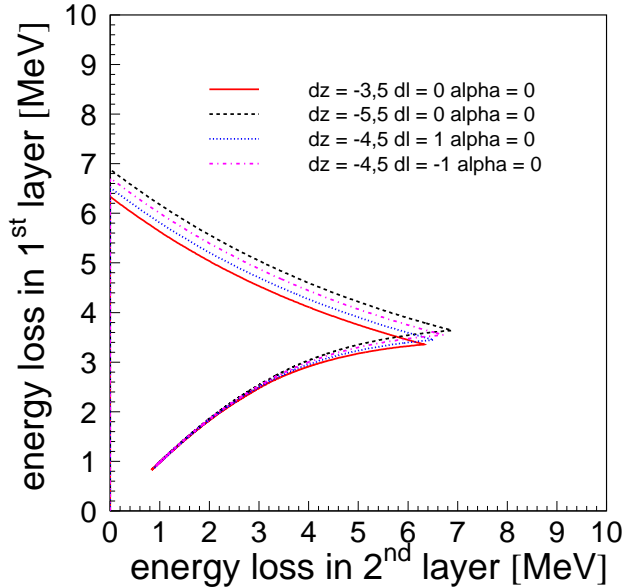}}
\caption{Energy losses in the first layer versus energy losses in the second layer
         as calculated for different positions of the spectator detector.
         The values of {\it dz}, {\it dl} are given in cm and {\it alpha} is given in rad.}
\label{parametry_spec}
\end{figure}

Differences in shapes of the $dE_1(dE_2)$ distributions are
due to the fact that the effective thickness of the silicon modules
of the detector "seen" by particles outgoing from the interaction point
depends on the particles' incident angle on the detector surface.
These variations allowed to find the exact position and orientation of the
spectator detector inside the COSY--ring.
Figure~\ref{parametry_spec}
presents an example of the calculated $dE_1(dE_2)$ distributions for a few
arbitrarily chosen sets of parameters {\it dz, dl,} and {\it alpha}.
Assuming that the set of
calibration constants ($\alpha_{1}$, $\alpha_{2}$) are correct, experimental
distributions of the $dE_{1}(dE_{2})$ have been fitted to the calculated energy losses,
and examined by varying  the parameters
describing the position of the spectator detector.
After the determination of the new position of the detector
the experimental data points on the
$dE_{1}(dE_{2})$ plot were fitted to the expected $dE_{1}(dE_{2})$ function,
with calibration constans
$\alpha_{1}$ and $\alpha_{2}$ as free parameters for each detection pair.
The procedure was repeated until the changes become negligible.
\begin{figure}[H]
\includegraphics[height=.29\textheight]{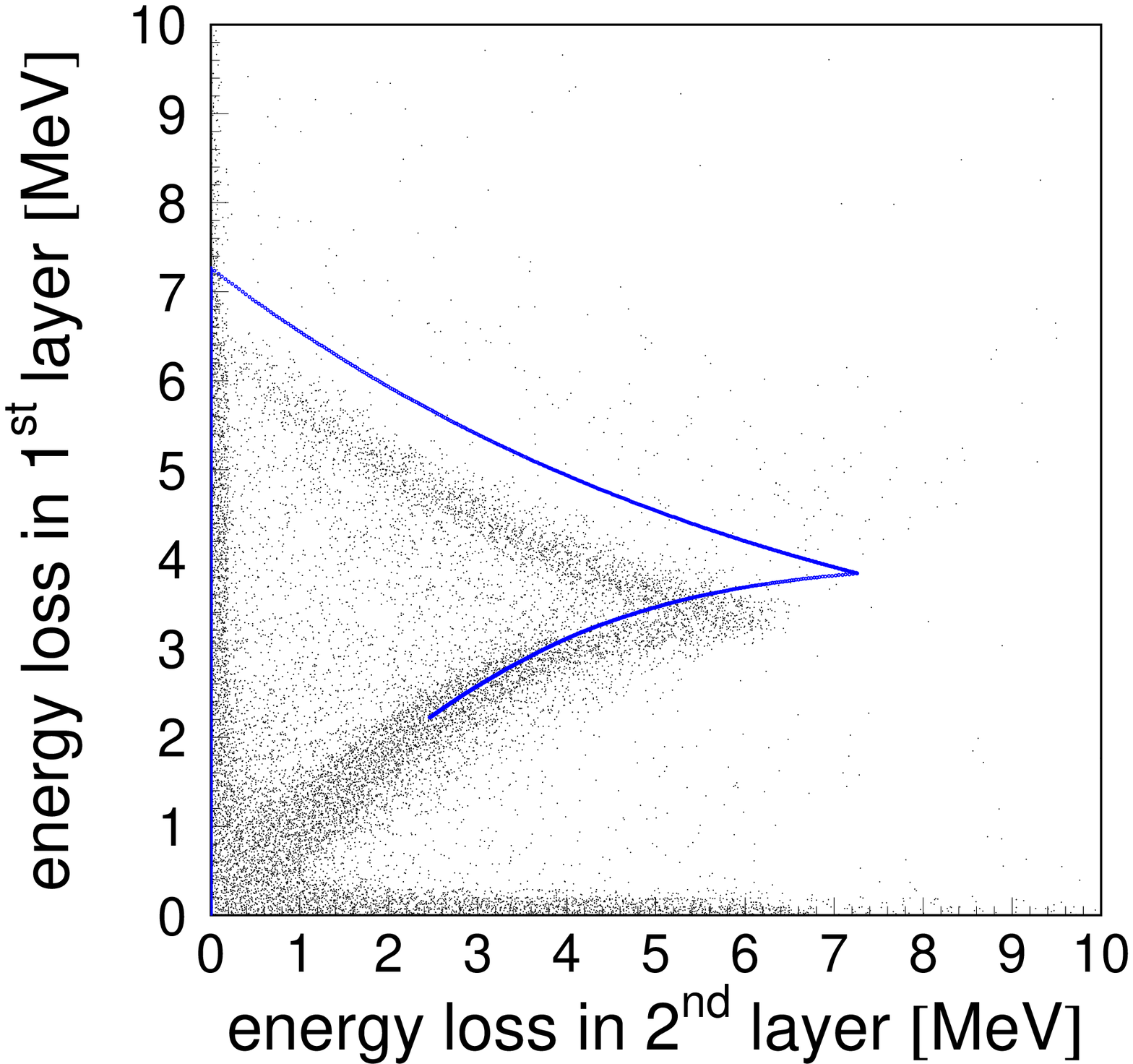}
\includegraphics[height=.29\textheight]{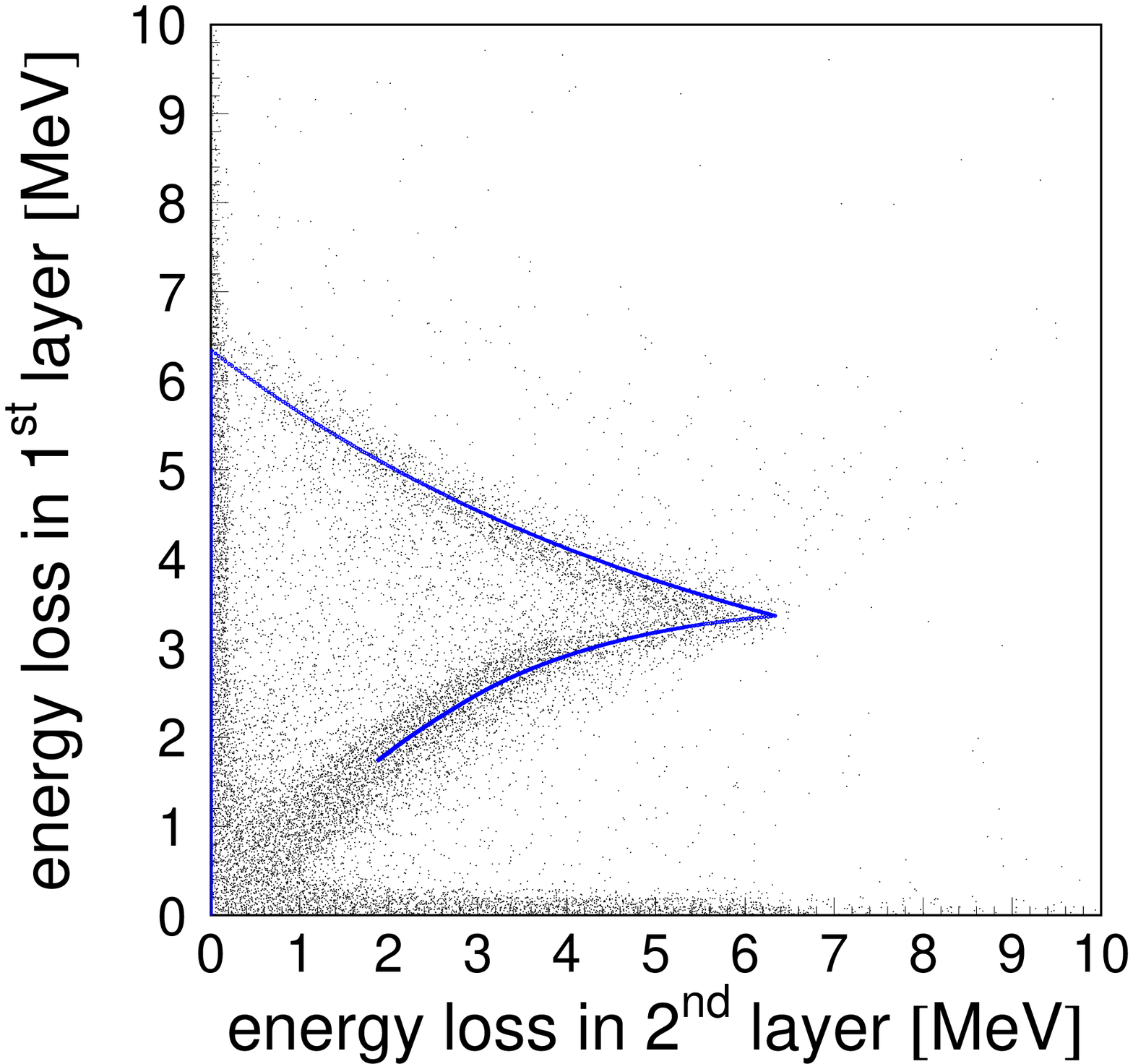}
\caption{ Energy losses in the first layer versus the second
          layer as measured at COSY--11 with a deuteron target
          and a proton beam with a momentum of 3.35 GeV/c before
          {\bf (left)} and after {\bf (right)} the calibration. Points represent the
          experimental data, whereas the theoretical calculations
          are indicated by the solid curve.}
\label{banan}
\end{figure}
Figure~\ref{banan} (left) shows the energy losses in the first layer versus
the second layer before the calibration constants $\alpha_{1}$ and $\alpha_{2}$
were determined. The theoretical calculations are indicated by the solid curve.
Fig.~\ref{banan} (right) shows the corresponding plot after the calibration.
The position of the
spectator detector has been determined with an accuracy of $\pm$1~mm, consistent
with the nominal values based on the geometrical designs of the setup and with
the values obtained for the analysis of the $pn \to pn\eta$ test reaction~\cite{moskal-prc79}.
Taking into account the size of the
stream of the deuteron target with a diameter of 9~mm~\cite{dombrowski-nim386} the
accuracy of the position determination within $\pm$1~mm is satisfactory.\\
Figure~\ref{par1_par4} shows the distribution of the excess energy with respect to the
$pn\eta^{\prime}$ system determined for the $pn$ reaction.
The solid histogram was obtained with the position
of the spectator detector determined in the above described procedure.

\begin{figure}[H]
\centerline{\includegraphics[height=.4\textheight]{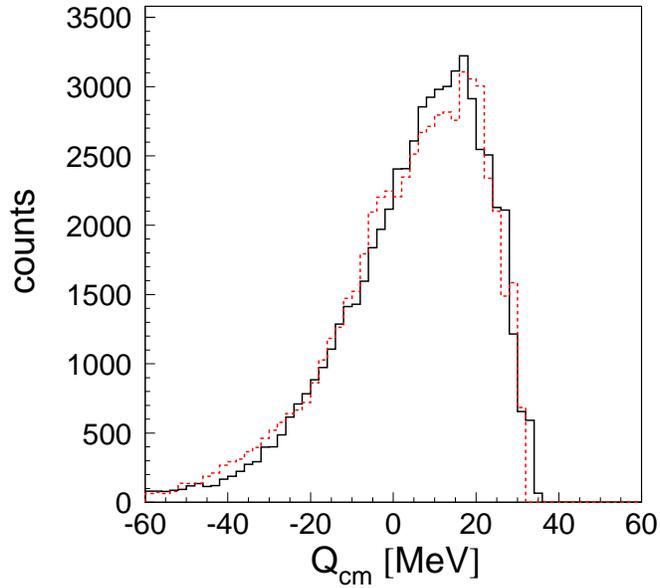}}
\caption{The distribution of the
         excess energy $Q_{cm}$ determined with respect to the $\eta^{\prime}$
         meson production threshold for the $pn$
         reaction as obtained with sets of parameters given by the present
         calibration (solid histogram) and with the orientation of the
         spectator detector determined previously from the analysis of the $pn \to pn\eta$
         reaction~\cite{moskal-prc79} (dashed histogram).}
\label{par1_par4}
\end{figure}
As a cross check,
the same data were analysed with the set of parameters describing the orientation of the
spectator detector used previously in the  analysis of the $pn \to pn\eta$ reaction.
The result is presented as a dashed histogram.

 
\chapter{Analysis of the experimental data}
\markboth{\bf Analysis of the experimental data}
         {\bf Analysis of the experimental data}

In this chapter the determination of the four-momentum vectors of the
outgoing nucleons will be presented, and the missing
mass technique as well as the method of background subtraction will be described.

\section{Event selection}

The first step of the off-line analysis of the experimental data was the software
selection of events with one track reconstructed in the drift chambers
and a signal in the neutral particle detector. As a next step particles registered
in drift chambers were identified.

\subsection{Proton identification}

All charged particles originating from the proton-neutron reaction are separated from
the proton beam in the magnetic field of the dipole magnet due to the smaller momenta
and some of them are registered in the drift chambers
and scintillator detectors (S1, S3). For each event the time--of--flight of particle
between the S1 and S3 hodoscopes
$(\Delta~t~=~t_{S3} - t_{S1})$ is measured.
\begin{figure}[H]
\centerline{\includegraphics[height=.4\textheight]{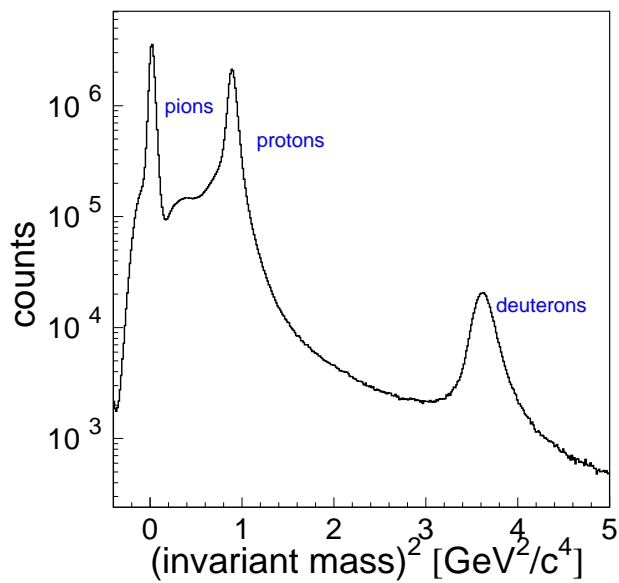}}
\caption{Distribution of the squared mass of charged particles
         originating from the proton-deuteron reaction performed
         at a beam momentum of 3.35~GeV/c.}
\label{inv_mass}
\end{figure}
Knowing the relative distances
between these detectors, the velocity of the particles is calculated.
Based on the signals in the drift chambers D1 and D2 the trajectories
of particles are reconstructed. Furthermore the particles momenta are reconstructed
by tracking the trajectories back through the
known magnetic field of the dipole to the centre of the interaction region
(the overlap between the target and the beam).
The precision of the momentum reconstruction is equal to 6~MeV/c~\cite{moskal-prc69}.
Having the velocity of the particle and its total momentum the squared
invariant mass is determined. \\
Charged ejectiles can well be identified as shown in figure~\ref{inv_mass},
where the distribution of the squared mass of charged particles
originating from the proton-deuteron reaction performed at a beam momentum
of 3.35~GeV/c is presented. Three clear peaks are evidently visible and correspond
to the squared mass of pions, protons and deuterons. For the further analysis
events with invariant mass corresponding to the proton mass are selected
(0.2~GeV$^2$/c$^4$ $\le$ m$^2$ $\ge$ 1.6~GeV$^2$/c$^4$).\\
After the particle identification, the time of the reaction $(t^{real}_r)$ at
the target point is obtained from the known velocity, trajectory and the time
measured by the S1 detector.

\subsection{Neutron identification}

\begin{figure}[H]
\centerline{\includegraphics[height=.4\textheight]{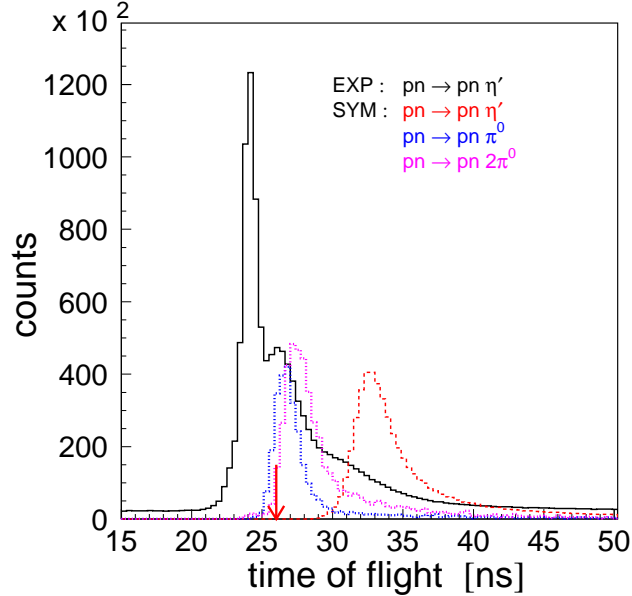}}
\caption{Distributions of the time--of--flight between the target and the neutral
         particle detector ($T^{N}$) for neutral particles. The experimental
         distribution depicted as a black solid histogram is compared with $i)$
         a simulated spectrum of the $pn \to pn\eta^{\prime}$ reaction (dashed histogram)
         and $ii)$ simulated exemplary background contributions (dotted histograms).}
\label{sym_tof_all}
\end{figure}

The determination of the four-momentum of the neutron is based on the
time--of--flight measurement. The absolute value of the neutron momentum
can be expressed as:
\begin{equation}
p = m_n \cdot {l\over{TOF^{N}}} \cdot {1\over{\sqrt{1 - ({l\over{TOF^{N}}})^2/c^2}}}.
\label{ped_neut}
\end{equation}
The $m_n$ denotes the mass of the neutron, $l$ is a distance from the centre of
the interaction region to the central part of the module
which fired as the first one. Thus the time--of--flight $TOF^{N}$ is taken as a difference
between the time of the reaction $t^{reac}_r$ (known from the backtracking protons
to the interaction point) and the shortest time of the neutral particle detector.
The individual momentum components are determined from the angle
defined by the centre of the hit segment. The momentum component along the y-axis
is assumed to be zero since the sensivity of the neutral particle detector is insufficient  in the
up-down direction. The granularity of the detector allows to determine the horizontal
hit position with an accuracy of $\pm 4.5$~cm. \\
Figure~\ref{sym_tof_all} presents the time--of--flight distribution between the target and the
neutral particle detector as measured for neutral particles. The veto detector installed
in front of the neutral particle detector discriminates signals originating from
charged particles, however, it does not discern gamma quanta, which are detected
in the neutral particle detector as well. Therefore, a clear signal originating from
the gamma rays is seen at 24.5~ns over a broad yield from neutrons. This histogram shows
that a discrimination between signals originating from neutrons and gamma quanta can
be done by a cut on the time--of--flight. From the Monte Carlo simulations
of the $pn \to pn\eta^{\prime}$ reaction
the largest expected momentum value of the neutron is eqaul to 1.4~GeV/c
which corresponds to the time--of--flight value of $28$~ns
(see dashed histogram in fig.~\ref{sym_tof_all}). This allows to select events
for which the time--of--flight is higher than $26~ns$ as it is indicated by an arrow in
fig.~\ref{sym_tof_all}. This condition excludes  background of gamma quanta.
Background reactions with neutrons in the exit channel and their contribution to the
time--of--flight distribution have been  investigated also. The simulated reactions
were analysed in the same way as the experimental data. The distributions of the
time--of--flight for neutrons originating from quasi-free $pn \to pn~pions$ reactions
are shown in figure~\ref{sym_tof_all} with dotted histograms. The {\it time--of--flight cut}
does not exclude neutrons from different quasi-free reactions,
and this will constitute the main background which must be subtracted
(see section~6.3.1).

\section{Determination of the excess energy}

Having proton and neutron identified, the consecutive software selection of the
experimental data was applied. In addition to one proton registered in the drift chambers and
a neutron in the neutral particle detector, at least a signal in one layer of the spectator
detector was demanded.\\
As it was disscused in chapter~3, to establish the exces energy $Q_{cm}$ with respect
to the $pn \to pn\eta^{\prime}$ reaction, the measurement of the proton spectator momentum,
and the determination of the total energy $\sqrt{s}$ is indispensable.
\begin{figure}[H]
\centerline{\includegraphics[height=.4\textheight]{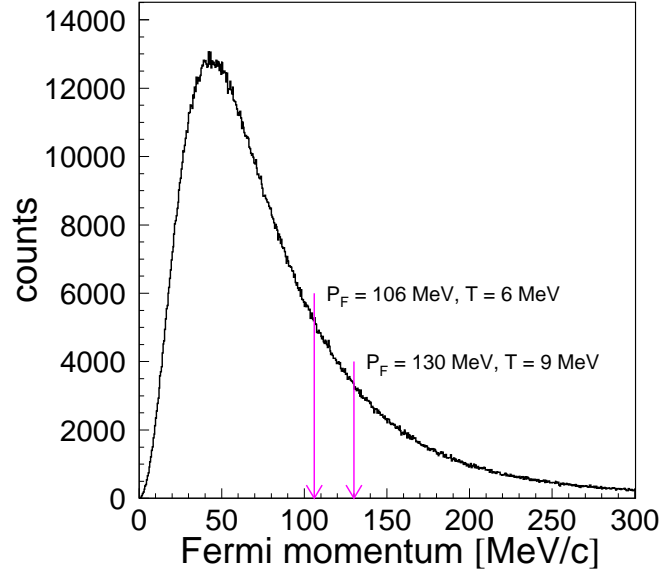}}
\caption{Simulated distribution of the Fermi momentum of nucleons inside the deuteron
         based on the Paris potential~\cite{lacombe-plb101}.
         Arrows depict values of momenta for which the spectator protons are stopped
         in the first or in the second layer of the spectator detector.}
\label{spec_ped_detect}
\end{figure}
The spectator detector is designed such, that protons impinging perpendicular to the
detector surface and having a kinetic energy of
less than 6~MeV are stopped in the first layer (dE). Protons with kinetic energy
in the range between 6~MeV and 9~MeV are stopped in the second layer (E).
All protons with $T_{spec} > 9$~MeV are passing through
both layers of the spectator detector. As it is seen in figure~\ref{spec_ped_detect}
a large number of spectator protons (with maximum Fermi momentum of 130~MeV) will be
stopped in the detector. In the experiment we registered the spectator protons with
momenta larger than 35~MeV/c since it is not possible to separate the energy losses
for protons with lower momenta from the noise level. The {\it spectator noise cut} has been
performed for each pad of the detector separately, using the spectra of energy loss
triggered by a pulser with a frequency of 1~Hz additionally to other
experimental triggers.
Figure~\ref{banan_all_pads}~(left)
shows the energy losses in the $1^{st}$ layer of the spectator detector versus
the $2^{nd}$ layer.
Slow spectator protons are stopped in the first or second
layer of the detector whereas fast particles (mainly charged pions) cross
both detection layers. Signals from deuterons are not observed since deuterons
cannot be emitted backward due to the kinematics.
Having the deposited energy $(T_{spec})$ and the emission angles  $(\theta_{spec}, \phi_{spec})$
we calculate the energy of the spectator proton $(E_{spec} = T_{spec} + m_p)$ and its momentum
$(\vec{p}_{spec})$.
Figure~\ref{banan_all_pads}~(right) presents the momentum distribution
of protons considered as spectator (points) compared with simulations
taking into account a Fermi motion of nucleons inside the deuteron (solid line).
\begin{figure}[H]
\includegraphics[height=.29\textheight]{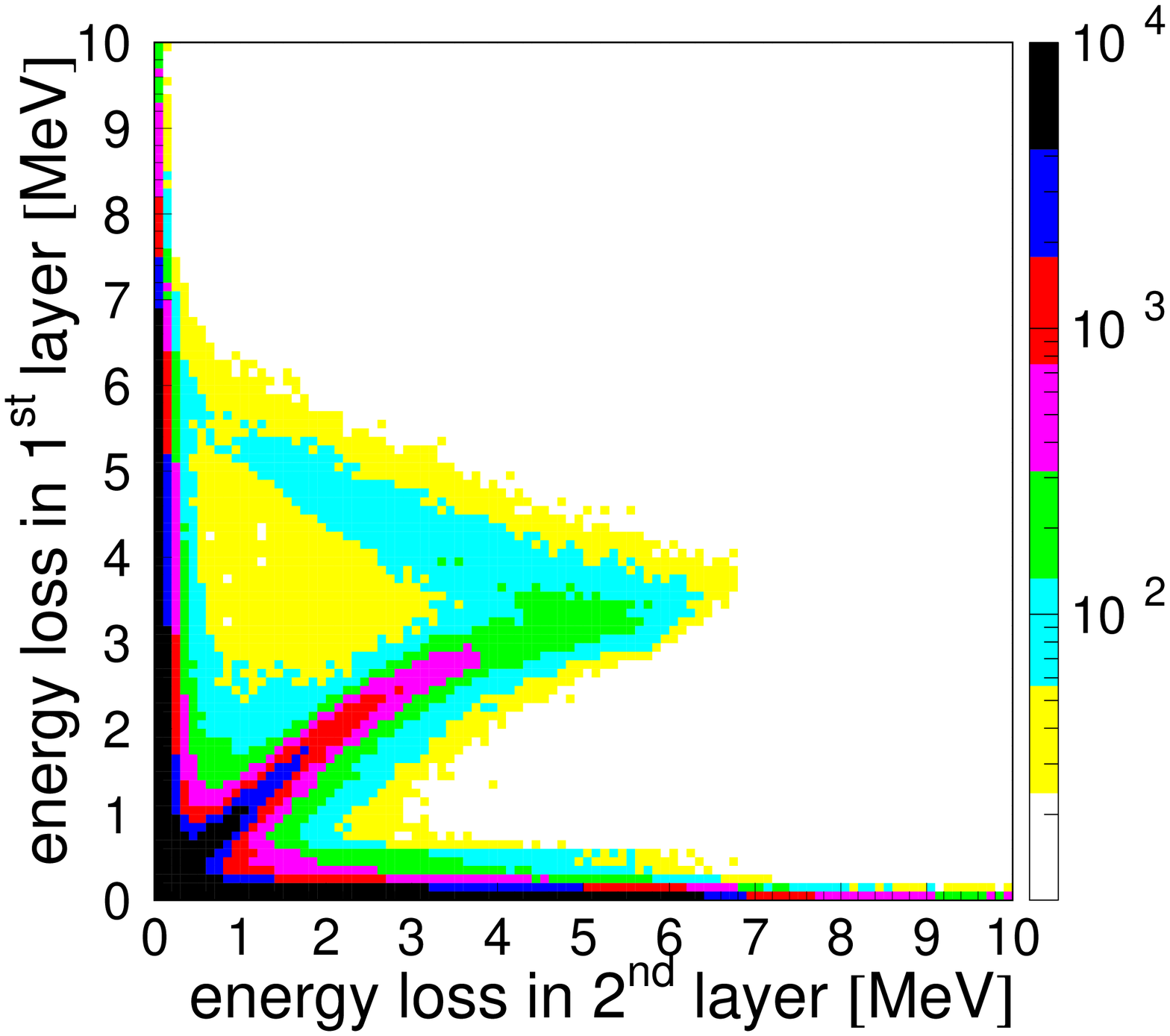}
\includegraphics[height=.29\textheight]{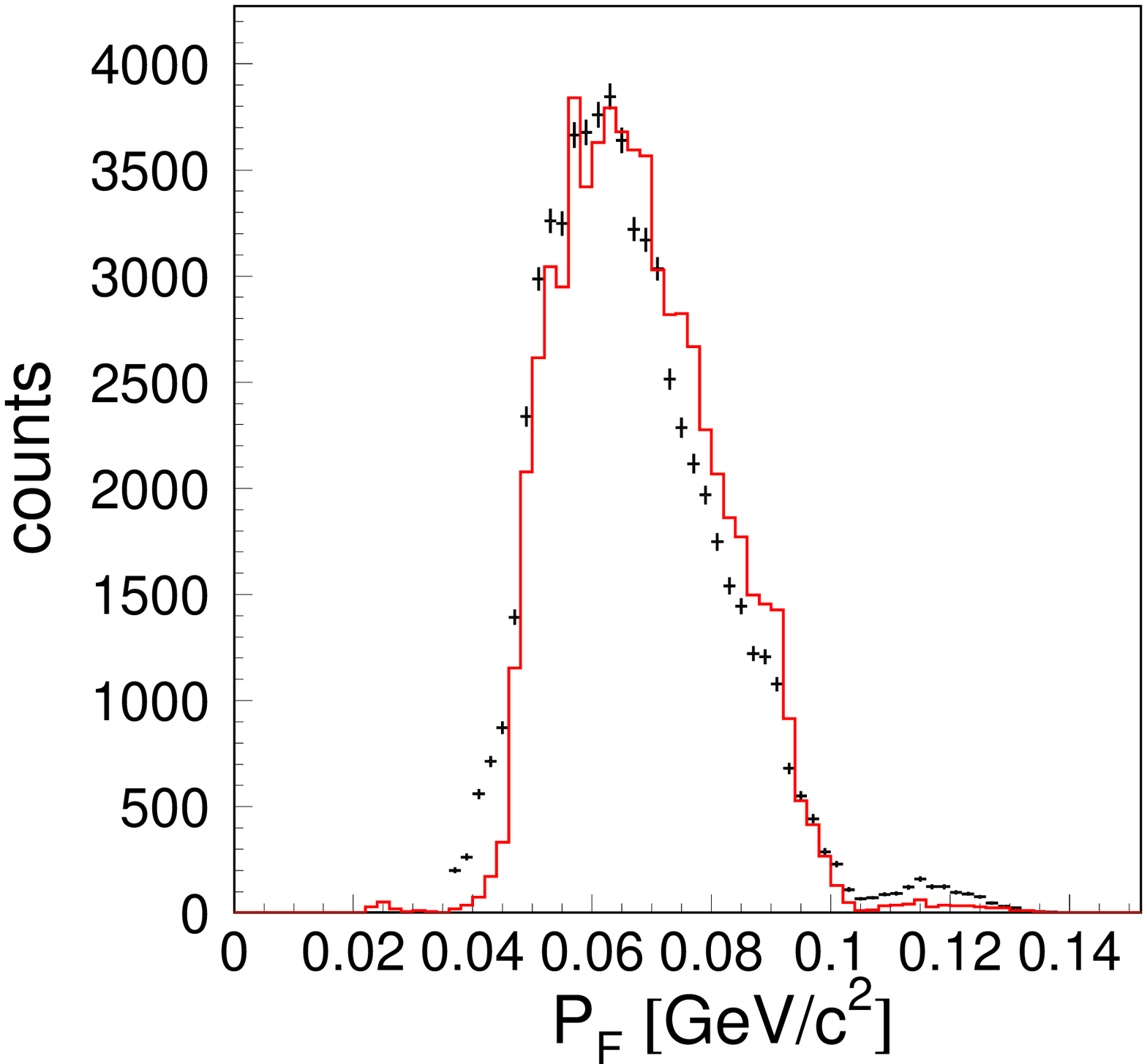}
\caption{{\bf Left:} Energy losses in the first layer versus the second
         layer as measured with a deuteron target and a proton beam with
         a momentum of 3.35 GeV/c. {\bf Right:} Momentum distribution of
         the spectator proton as reconstructed in the experiment (points)
         in comparison with simulation calculations taking into account
         Fermi momentum distribution of nucleons inside the deuteron
         according to the Paris potential~\cite{lacombe-plb101}
         (solid histogram). The simulated events were analysed in the same
         way as the experimental data.}
\label{banan_all_pads}
\end{figure}
The total energy $(\sqrt{s})$ available for the quasi-free proton-neutron
reaction can be calculated for each event from the vectors of the spectator
momentum and beam momentum.
$$ s~=~ (E_b~+~E_{t})^2~-~(\vec{p}_b~+~\vec{p}_t)^2, $$
$$ E_t = m_d - E_{spec}, $$
\begin{equation}
\vec{p}_t = -\vec{p}_{spec}
\label{s}
\end{equation}
$E_b$ and $\vec{p}_b$ are the energy and momentum vector of the proton beam.
$E_{t}$ and $\vec{p}_t$ are the energy and momentum vector of the neutron target,
calculated from the energy and the momentum of spectator proton ($E_{spec}$, $\vec{p}_{spec}$);
$m_d$ and $m_p$ denote the deuteron and proton masses, respectively.
The excess energy with respect to the
$pn \to pn\eta^{\prime}$ process is equal to:
\begin{equation}
Q_{cm} = \sqrt{s} - (m_p + m_n + m_{\eta^{\prime}}),
\end{equation}
where $m_p$, $m_n$ and $m_{\eta^{\prime}}$ are the mass of proton, neutron and
the $\eta^{\prime}$ meson, respectively.
The experimental distribution of the excess energy is shown in the left
panel of figure~\ref{excess_ene_exp_sym}. As was previously discussed, with one fixed
value of the beam momentum -- due to the Fermi motion of nucleons inside the
deuteron -- one can scan a broad range of the excess energy,
from 0 up to 32~MeV for $p_{beam}~=~3.35$~GeV/c. Negative values of the excess energy
correspond to events, when the total energy in the centre-of-mass frame $(\sqrt{s})$ was
insufficient for the production of a $\eta^{\prime}$ meson and only the production of multi-pion
background events could occur. For excess energies larger than zero, additionally to the pion
production also the $\eta^{\prime}$ meson is created. A simulated
distribution of the excess energy for quasi-free $pn \to pn\eta^{\prime}$ reaction, calculated
with respect to the $pn\eta^{\prime}$ system is depicted in
figure~\ref{excess_ene_exp_sym}~(right).\\

\begin{figure}[H]
\includegraphics[height=.29\textheight]{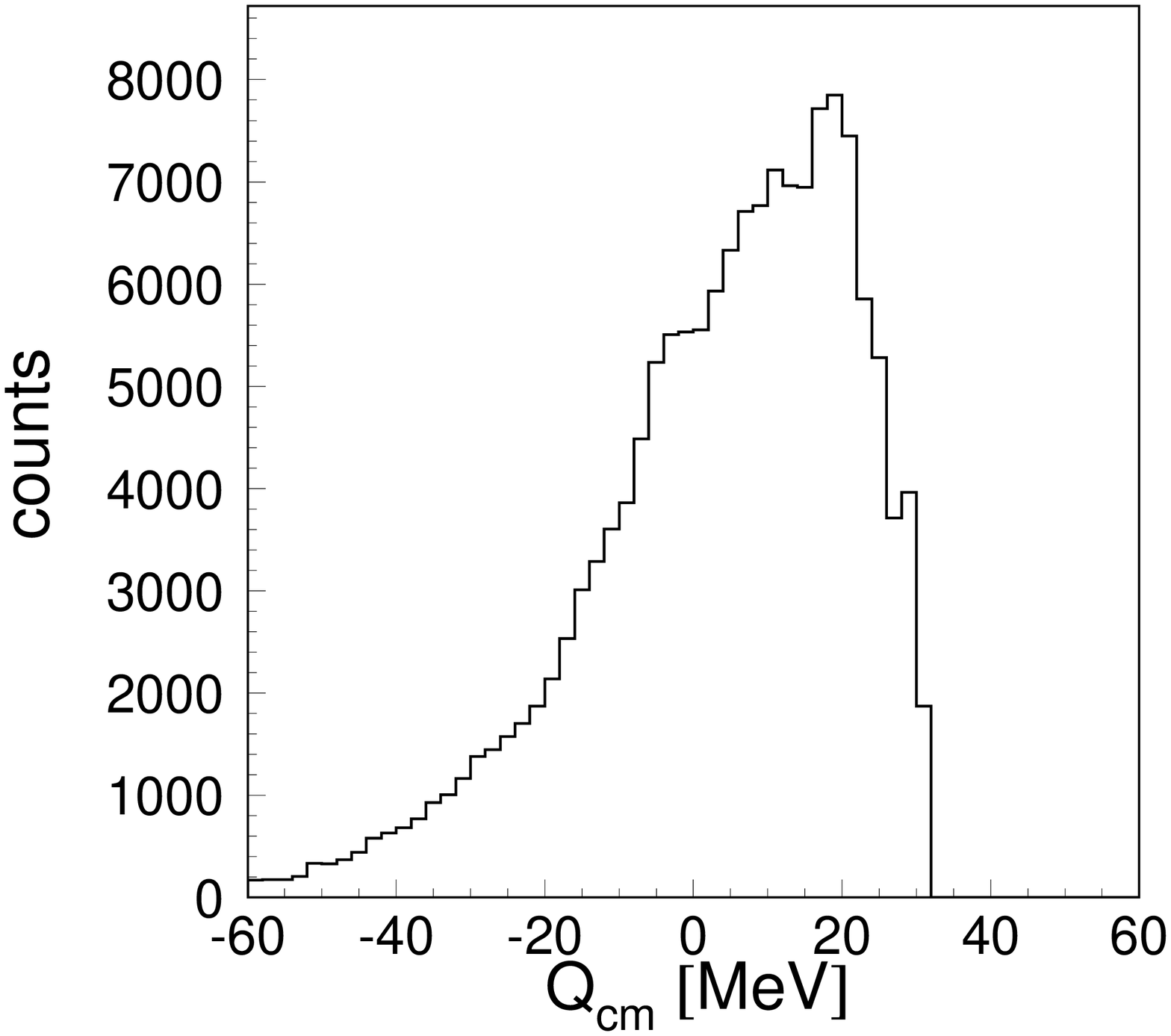}
\includegraphics[height=.29\textheight]{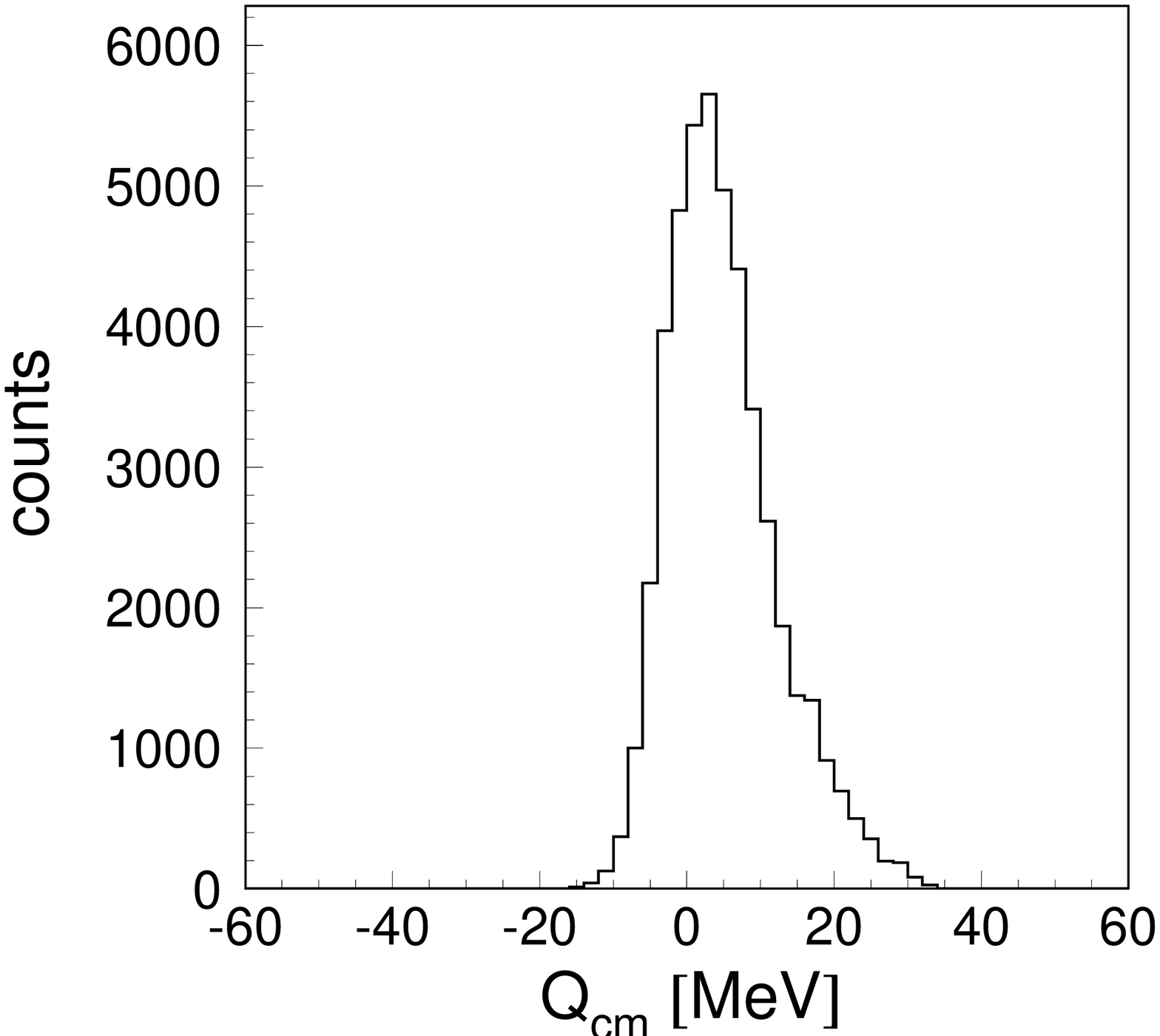}
\caption{Distributions of the excess energy $Q_{cm}$ for the quasi-free
         $pn \to pnX$ reaction determined with respect to the $pn\eta^{\prime}$
         threshold as obtained experimentally {\bf (left)} and for Monte
         Carlo simulation {\bf (right)}. In the simulation all known detector
         features were taken into account (for details see text).}
\label{excess_ene_exp_sym}
\end{figure}

A number of factors influence the registered yield of the $pn\to pn\eta^{\prime}$
reaction as a function of the excess energy: $i)$ the Fermi momentum
distribution of nucleons in the deuteron target, $ii)$ the relative settings
of the spectator detector and the target, $iii)$ the absolute momentum
of the proton beam and $iv)$ the acceptance of the COSY--11 detection setup.
Due to the rapid growth of the total cross section close to
the production threshold the accuracy of the excess energy determination
is one of the most important parameters and it is mandatory to estimate
correctly its uncertainty. The accuracy of the excess energy determination
was estimated based on Monte Carlo simulations. For this purpose we have
simulated $N_0~=~10^9$ events of the $pn\to pn\eta^{\prime}$ reaction taking
into account the Fermi motion of the nucleons inside the deuteron,
the size of the target (diameter~$\sim~9$~mm), the spread of the beam momentum
($\sigma p_b~=~1.5$~MeV/c) and the horizontal ($\sigma(x)~\sim~2$~mm)
and vertical ($\sigma(y)~\sim~4$~mm) beam size~\cite{moskal-nim466}.
Simulated data have been analysed in the same way as the
experimental data, taking into consideration the energy- and time-resolution
of the detectors.
\begin{figure}[H]
\centerline{\includegraphics[height=.4\textheight]{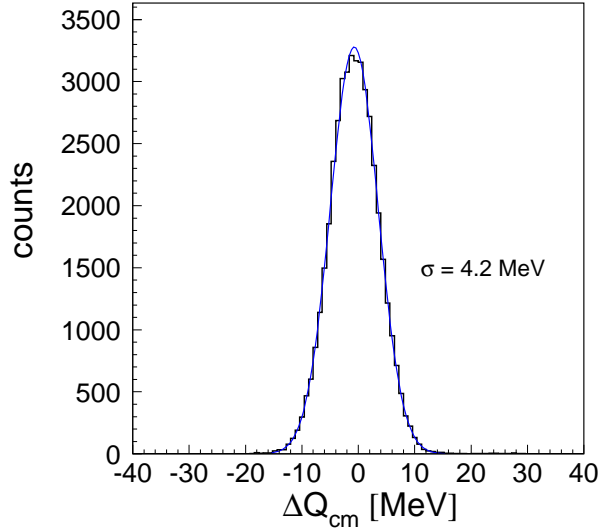}}
\caption{ Difference between the generated $(Q_{cm}^{gen})$ and reconstructed
          $(Q_{cm}^{rec})$ excess energy for the $pn \to pn\eta^{\prime}$ reaction
          simulated for a proton beam momentum of 3.35~GeV/c and a deuteron target.}
\label{excess_ene_resol}
\end{figure}

Figure~\ref{excess_ene_resol} presents the distribution of differences between
the generated excess energy $(Q_{cm}^{gen})$ and the reconstructed excess energy from signals
simulated in the detectors $(Q_{cm}^{rec})$. The distribution of
$\Delta~Q_{cm}~=~Q_{cm}^{gen}~-~Q_{cm}^{rec}$ was fitted by a Gaussian function
resulting in an excess energy resolution of $\sigma(Q_{cm})~=~4.2$~MeV.
For the $pn \to pn\eta$ reaction near threshold a standard deviation of
the excess energy was derived to be $\sigma(Q_{cm})~=~2.2$~MeV~\cite{moskal-prc79},
which is comparable with $\sigma(Q_{cm})~=~1.8$~MeV obtained at similar conditions
at the PROMICE/WASA setup~\cite{bilger-nim457}.

\section{Identification of the $pn \to pn\eta^{\prime}$ reaction}

\begin{figure}[H]
\centerline{\includegraphics[height=.4\textheight]{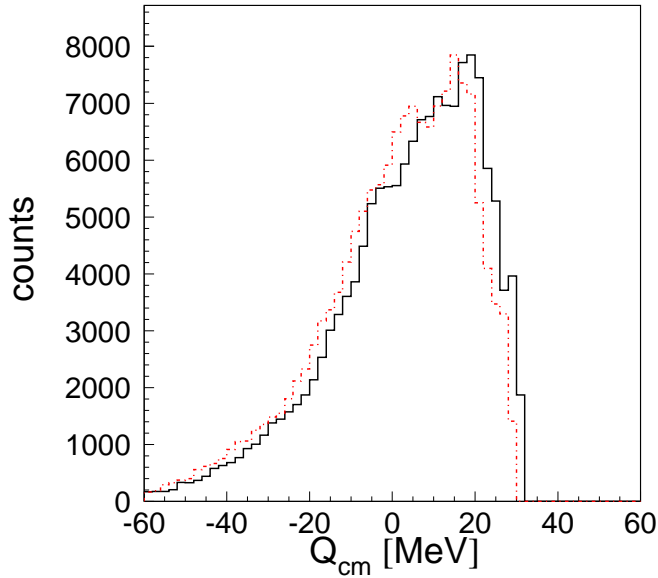}}
\caption{The distribution of the
         excess energy $Q_{cm}$ determined for the $pn \to pn\eta^{\prime}$
         reaction as obtained with the position of the spectator detector
         fixed based on the data (solid histogram) and with the detector
         shifted from this position in z-axis direction
         by 3~mm (dashed histogram). The shift of 3~mm is done for demonstration
         purpose. Note that in chapter~5.5.2 the accuracy of the position
         determination of the spectator detector was established to be $\pm$1~mm.}
\label{excess_ene_spec_p}
\end{figure}

The $\eta^{\prime}$ mesons produced in the $pn \to pn\eta^{\prime}$ reaction
are identified using the missing mass technique.\\
Knowing the four-momentum vectors of a proton and a neutron in the initial and final
state, and employing the principle of momentum and energy conservation
one can calculate the squared mass of the unmeasured particle or group of particles:
\begin{equation}
 m_{x}^{2} = E_{x}^{2} - {\vec {p_{x}}}^{2} = {(E_{b} + E_{t} - E_{p} - E_{n})}^2 -
  {( \vec p_{b} + \vec p_{t} - \vec p_{p} - \vec p_{n} )}^2
\end{equation}
where,\\
$E_{b}, \vec p_{b}$ is the energy and momentum of proton beam,\\
$E_{t}, \vec p_{t}$ is the energy and momentum of neutron target,\\
$E_{p}, \vec p_{p}$ is the energy and momentum of outgoing proton, and\\
$E_{n}, \vec p_{n}$ is the energy and momentum of outgoing neutron.\\
The method of the proton identification and of the neutron selection was described
in sections 6.1 and 6.2.\\
Before the missing mass was calculated, the collected data were grouped
according to the excess energy. The available range of the excess energy $Q_{cm}$
above the $\eta^{\prime}$ meson production threshold (from 0 to 32 MeV) has been
divided into four bins, each 8~MeV wide. The choice of the interval width is a
compromise between the statistic and the resolution and reflects the accuracy
(FWHM) of the determination of the excess energy. Due to the very low statistic
and small acceptance, events corresponding to the excess energy interval between
24 and 32~MeV are not taken for further analysis. Moreover, in this range the excess
energy distribution is very sensitive to changes of the position of the
spectator detector, what can be seen from figure~\ref{excess_ene_spec_p}
showing the distribution of the excess energy $Q_{cm}$ determined for the
$pn \to pn\eta^{\prime}$ reaction as obtained with the position of
the spectator detector fixed based on the data (solid histogram)
and with the detector shifted from this position
in the z-axis direction by 3~mm (dashed histogram). A significant reduction of number of
counts in the interval between 24 and 32~MeV is clearly observed.

\begin{figure}[H]
\parbox{0.45\textwidth}\centering{\includegraphics[width=.45\textwidth]{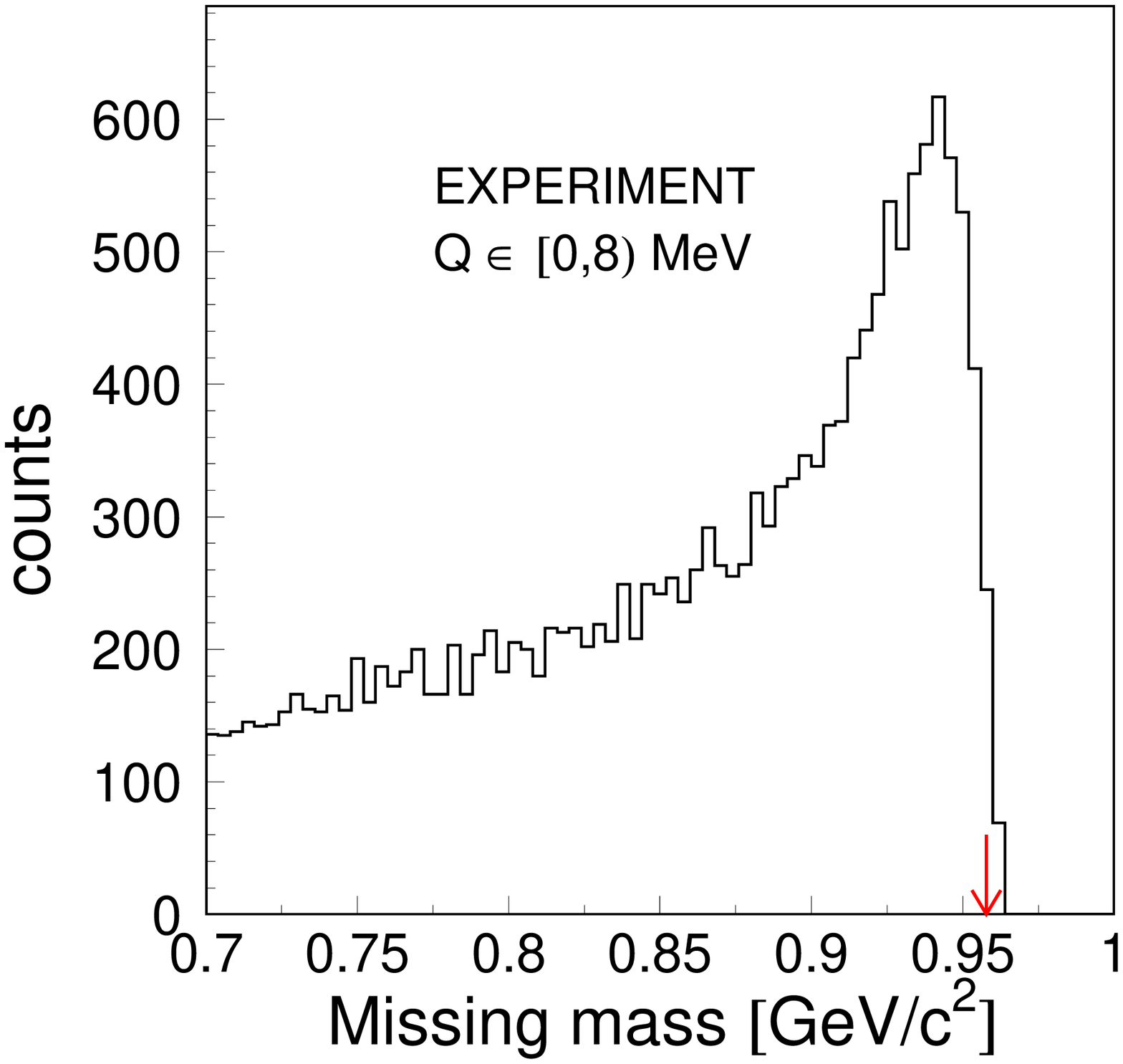}}
\parbox{0.45\textwidth}\centering{\includegraphics[width=.45\textwidth]{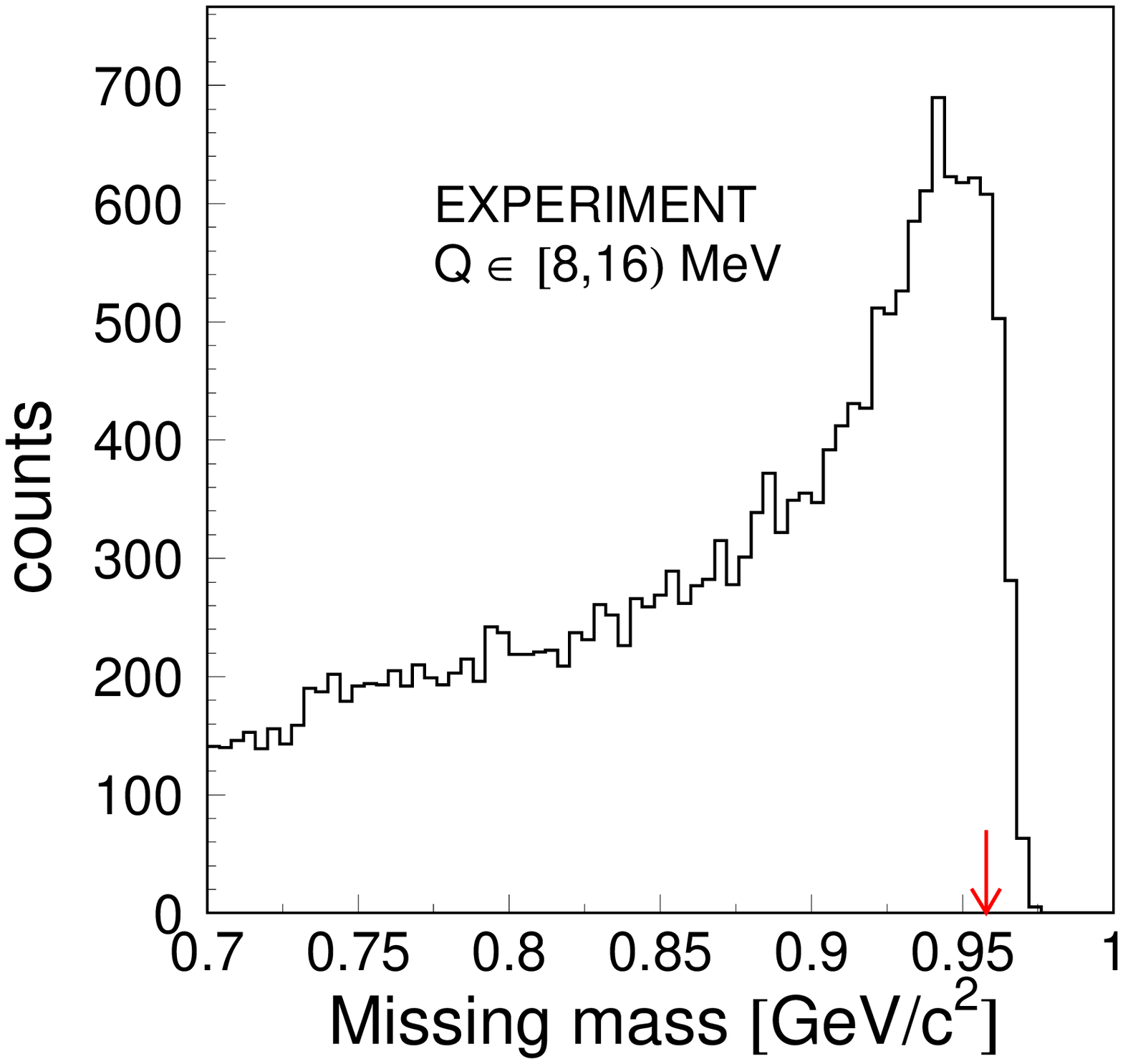}}
\parbox[c]{0.45\textwidth}{\centering\includegraphics[width=.45\textwidth]{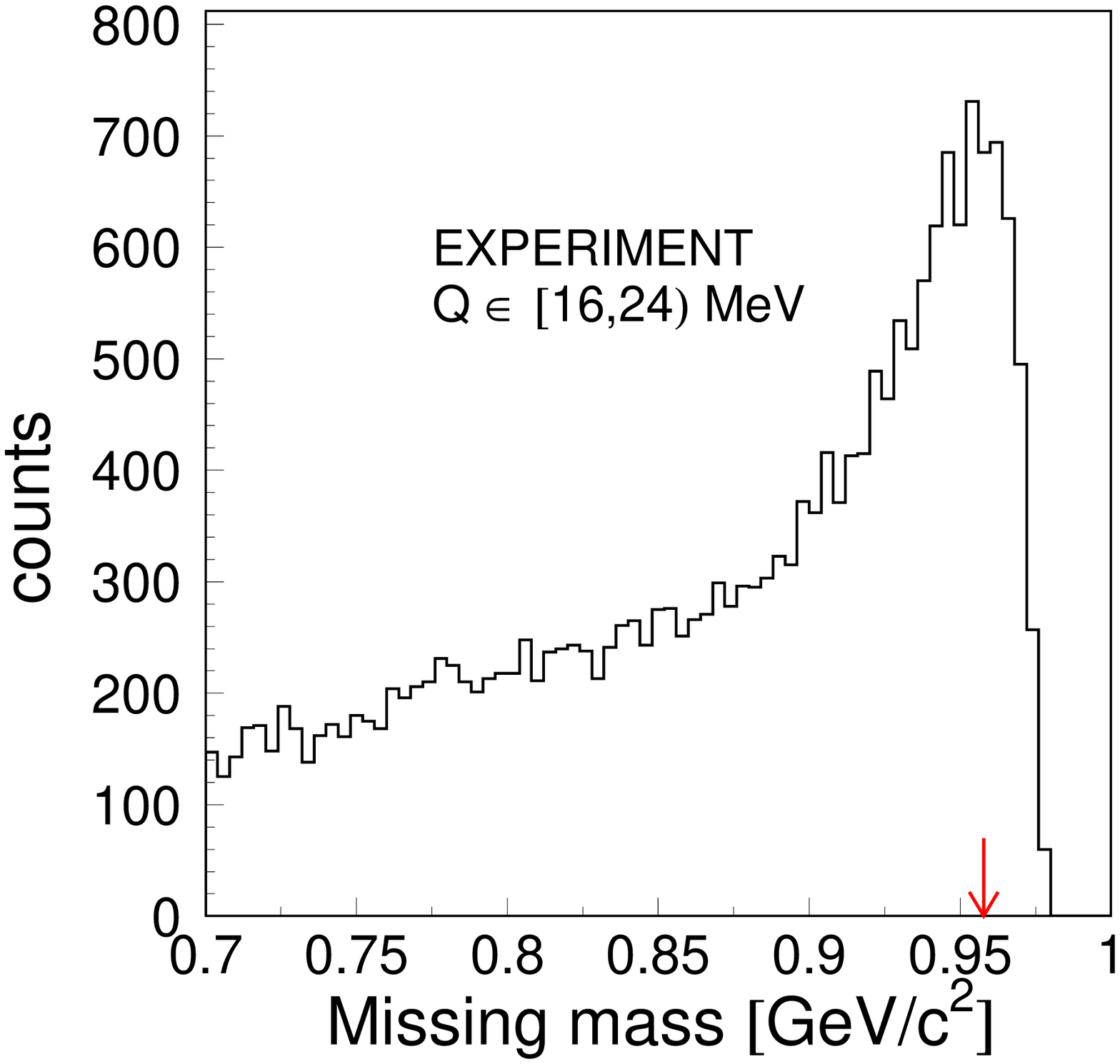}}\hfill
\parbox{.5\textwidth}{\caption{Experimental missing mass distributions of the $pn \to pnX$
         reaction for ranges of the excess energy $Q_{cm}$, from 0 to 8~MeV
         {\bf (upper panel left)}, from 8~MeV to 16~MeV {\bf (upper panel right)} and from
         16~MeV to 24~MeV {\bf (lower panel)}. Arrows depict
         the nominal value of the $\eta^{\prime}$ mass.\label{mm_ztlem}}}
\end{figure}

Figure~\ref{mm_ztlem} presents the experimental missing mass spectra of the
$pn \to pnX$ process calculated for the excess energy ranges $Q_{cm}~=~[0,8)$~MeV
$Q_{cm}~=~[8,16)$~MeV and $Q_{cm}~=~[16,24)$~MeV. For all missing mass distributions
we observe an increase of counting rate towards the higher masses ending at
the kinematical limit. The shape is mostly due to the detector acceptance.
A signal from $\eta^{\prime}$ meson production is expected on top
of the multi-pion mass distribution at the position corresponding to the $\eta^{\prime}$
mass -- denoted by the arrow -- however no enhancement around this region is seen.
Therefore, to extract number of $\eta^{\prime}$ events,
the multi-pion background must be disentangled from the missing mass distributions.

\subsection{Method of background subtraction}

The background investigation in the missing mass spectrum of the quasi-free $pn \to pnX$
reaction is a challenging task, especially close to threshold. This is even more difficult
when the resolution of the mass determination is comparable with the excess energy.\\
The method of separating contributions from the multi-pion production in the missing
mass spectrum of the quasi-free $pn \to pnX$ reaction is described in details
in a dedicated article~\cite{moskal-jpg32}. Therefore, here only the general principle will be pointed out.\\
One can disentangle the number of $pn \to pn\eta^{\prime}$ events from single-
and multi-pion background without conducting model-dependent simulations by comparison
of the missing mass distributions for the negative values of $Q_{cm}$, when only pions
can be created and for $Q_{cm}$ larger than zero.
\begin{figure}[H]
\centerline{\includegraphics[height=.45\textheight]{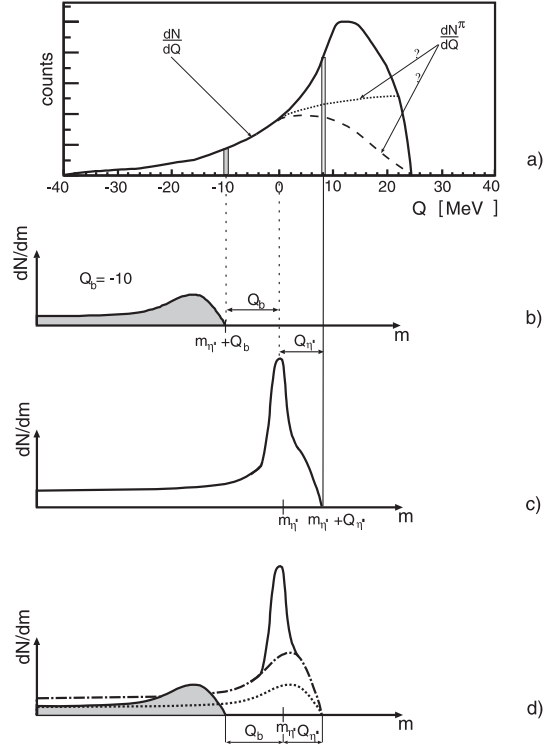}}
\caption{Schematic presentation of the background construction and subtraction.
         $a)$ Distribution of the excess energy with respect to the $pn \to pn\eta^{\prime}$
         reaction. $b)$ and $c)$ Form of the missing mass spectra as derived for negative
         $(Q_{b})$ and positive $(Q_{\eta^{\prime}})$ values of Q. $d)$ The shape of the background
         constructed from the shape of the multi-pion mass distribution (shaded histogram),
         shifted to the kinematical limit (dotted line) and normalised at low masses
         (dash-dotted line). The picture has been adapted from reference~\cite{moskal-jpg32}}
\label{opis_metody_tlo}
\end{figure}
Let us assume that the number of registered events is very large and that the shape
of the reconstructed invariant mass of the multi-pion background is independent
of the excess energy $Q$. The form of the background could then be determined from
the missing mass spectrum $dN^{\pi}/dm$ from an infinitesimal range at any negative
value of $Q$. The shape of the background can be expressed in terms of a normalised
function $B(m_{\eta^{\prime}}~+~Q~-~m)$ of the difference between the kinematical limit
$(m_{\eta^{\prime}}~+~Q)$ and the given mass $m$. Assuming infinite statistics, one could
divide the positive $Q$ range into very narrow subranges within which the resultant
missing mass spectrum would be a sum of a signal from the meson ($\eta^{\prime}$)
and the form $B$. This situation is schematically depicted in figure~\ref{opis_metody_tlo}.
In order to  derive a signal of the meson from the missing mass spectrum
for positive $Q$ it would be sufficient to subtract from this spectrum a missing mass
distribution determined for negative $Q$ values after the shift of the latter to the kinematical
limit (dotted line) and normalisation at the very low mass values where no events
from the $\eta^{\prime}$ meson are expected (dashed-dotted line)~\cite{moskal-jpg32}.
At real experimental conditions, due to the finite statistics, the missing mass spectrum
is studied in finite intervals of $Q$, what alters the form of the multi-pion background.
Therefore one has to take as narrow intervals of $Q$ as reasonable, taking into account
the experimental resolution in $Q$ and assuming that within this range the multi-pion background
is constant. An natural choice is to take the width of the $Q$ subranges equal to the
FWHM of the resolution of the $Q$ determination.

\begin{figure}[H]
\centerline{\includegraphics[height=.4\textheight]{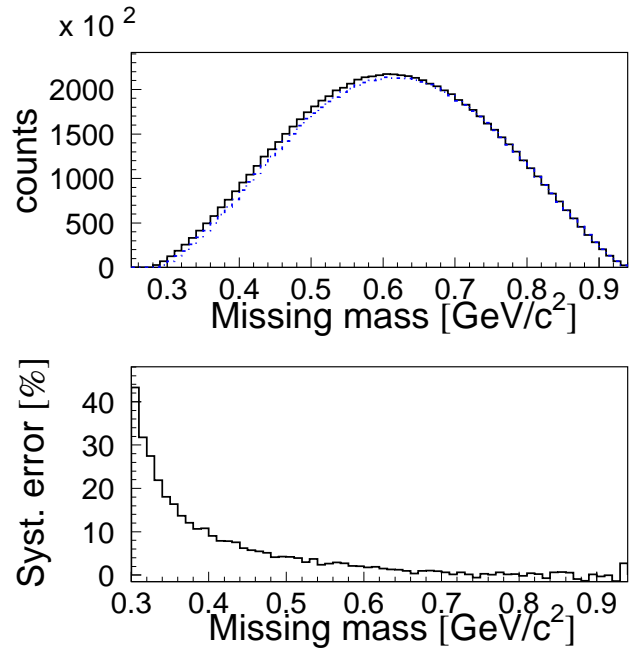}}
\caption{{\bf Upper:} Missing mass distribution determined for the $pn \to pn\pi\pi$
         reaction at the excess energy range $Q_{cm}$~=~(-30,-20)~MeV (dotted line)
         and for the excess energy range $Q_{cm}$~=~(-20,-10)~MeV (solid line).
         The excess energy was calculated with respect to the $pn\eta^{\prime}$ threshold.
         The dotted histogram was shifted to the kinematical limit of the solid histogram.
         {\bf Lower:} Difference between the spectra of the upper panel normalised to the
         solid line.}
\label{mm_genbod_testy}
\end{figure}

In order to gain more certainty that the measured shape of the multi-pion background
does not alter significantly with changes of the excess energy, we have performed
simulations of the background originating from two pion creation in proton-neutron collisions.
Figure~\ref{mm_genbod_testy} presents the missing mass distribution of the two pion system
produced via the $pn \to pnX$ reaction. Events were simulated at the excess energy intervals
of $Q_{cm}$~=~(-30,-20)~MeV and $Q_{cm}$~=~(-20,-10)~MeV with respect to the $pn\eta^{\prime}$
system. From the comparison of the form of both spectra one realises that there
is no noticeable difference in the shape at the missing mass range of the $\eta^{\prime}$ mass.
The systematic error caused by the subtracion of the
shifted and normalised histogram is shown in the lower panel of figure~\ref{mm_genbod_testy}.
It is demonstrated that the fractional systematic error due to the multi-pion background
subtraction is in order of one per cent of the background value within the range of 200~MeV
in the vicinity of the kinematical limit.

\subsection{Background subtracted missing mass distribution}
\label{background}

As was already discussed, events with excess energies larger than zero were
divided into intervals of $\Delta Q_{cm}$~=~8~MeV. For each interval
the missing mass was calculated separately.
Next, for each distribution of the missing mass spectra, a corresponding
background spectrum has been constructed from events with $Q_{cm}$ belonging to the
$(~-~30,~-~10)$~MeV
range applying the method described in the previous section.
As a cross check, to gain more confidence about the applied procedure
we have constructed alternative background distrubutions taking events from a $Q_{cm}$ range
from -20~MeV to 0~MeV. The result of this investigation has shown a consistency of
both background shapes in the order of 1\%.

\begin{figure}[H]
\parbox{0.45\textwidth}\centering{\includegraphics[width=.45\textwidth]{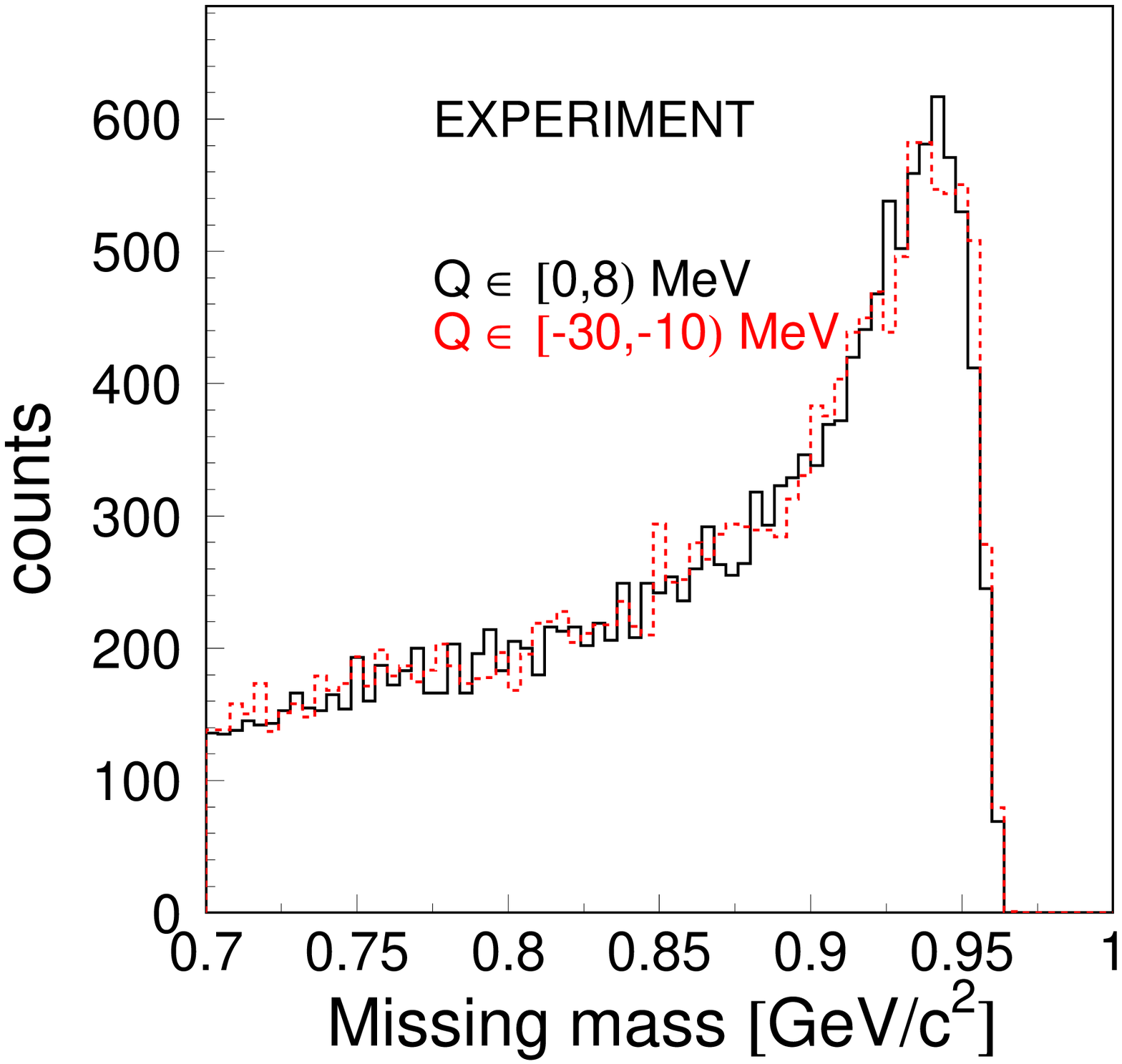}}
\parbox{0.45\textwidth}\centering{\includegraphics[width=.45\textwidth]{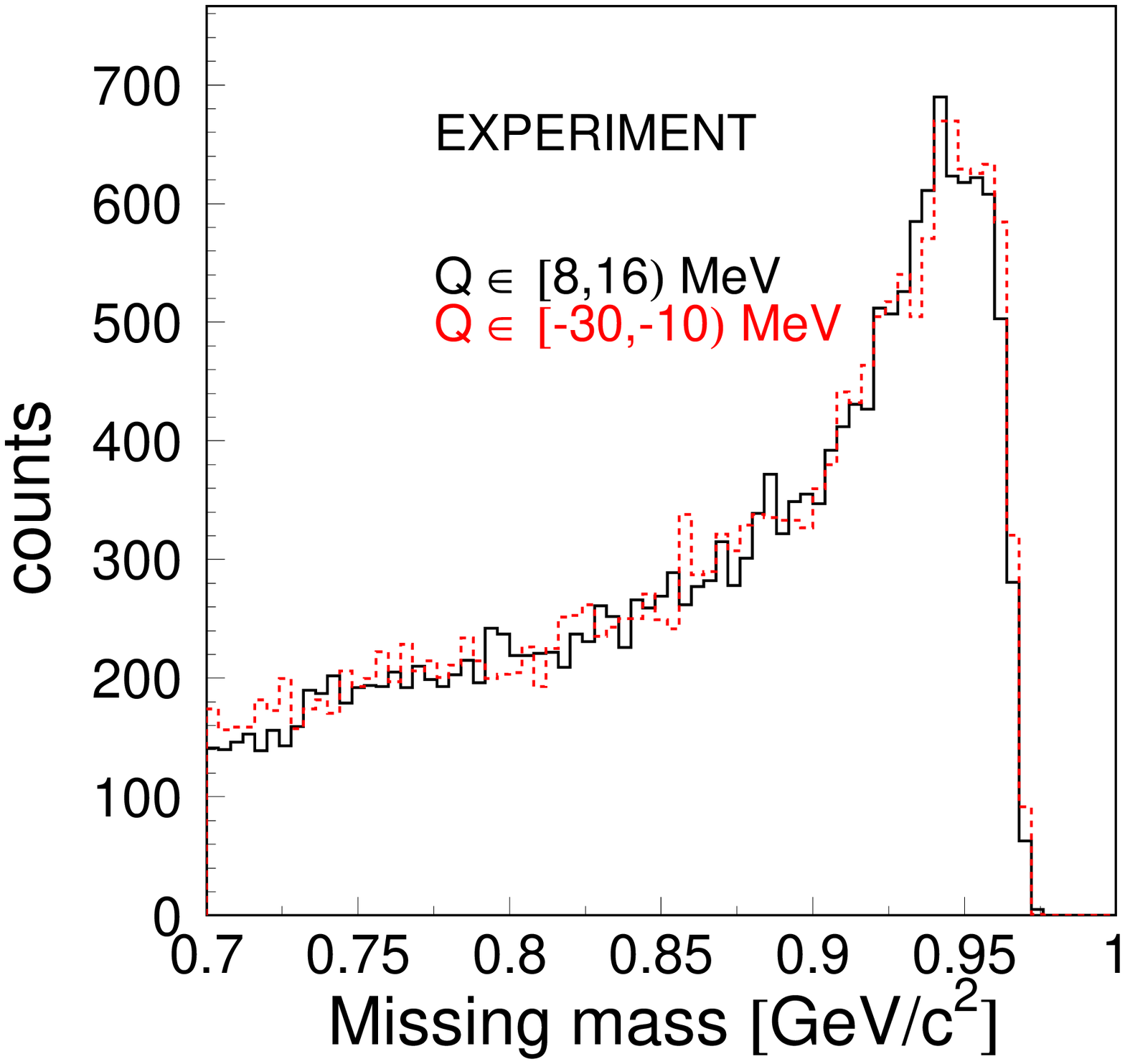}}
\parbox[c]{0.45\textwidth}{\centering\includegraphics[width=.45\textwidth]{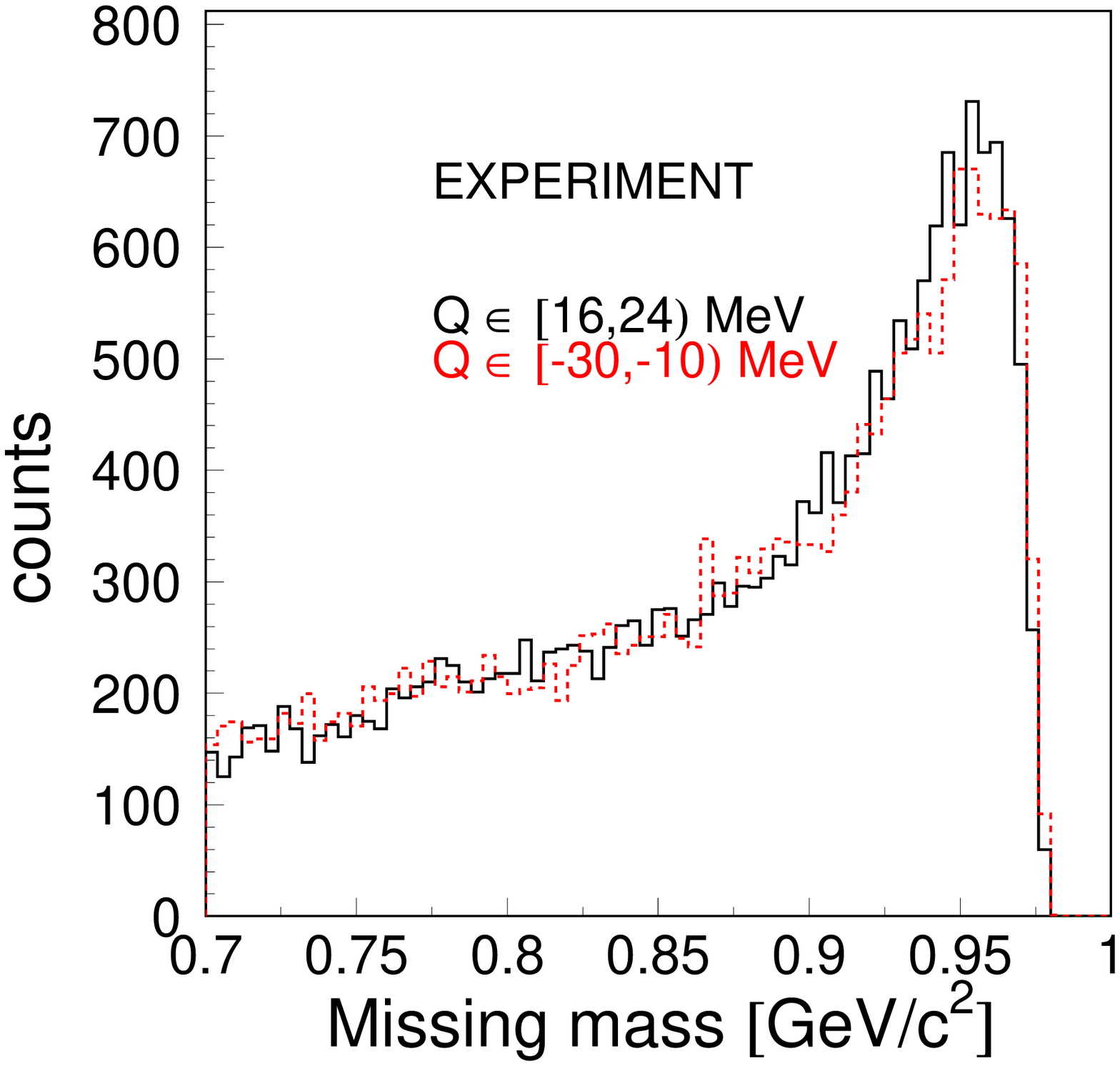}}\hfill
\parbox{.5\textwidth}{\caption{Experimental missing mass distributions of the $pn \to pnX$
         reaction (solid black line). The distributions were obtained
         for ranges of the excess energy $Q_{cm}$, from 0 to 8~MeV
         {\bf (upper panel left)}, from 8~MeV to 16~MeV {\bf (upper panel right)}
         and from 16~MeV to 24~MeV {\bf (lower panel)}. The
         corresponding background spectra (dashed red line) were
         constructed as described above.\label{mm_da}}}
\end{figure}
Afterwards, to each distribution for $Q_{cm} \ge 0$ the corresponding background
spectrum has been normalised for mass values smaller than 0.25~GeV/c$^2$
where no events from the $\eta^{\prime}$ are expected. In this region events
correspond to the one pion production, for which the total cross section remains nearly constant
in the shown range of the excess energy.
Figure~\ref{mm_da} shows the experimental mass distribution for signal (solid black
histogram) as determined for the excess energy ranges $[0,8)$~MeV, $[8,16)$~MeV, and $[16,24)$~MeV.
The dashed red line depicts the background spectrum, shifted to the kinematical
limit and normalised accordingly.
\begin{figure}[H]
\parbox{0.45\textwidth}\centering{\includegraphics[width=.45\textwidth]{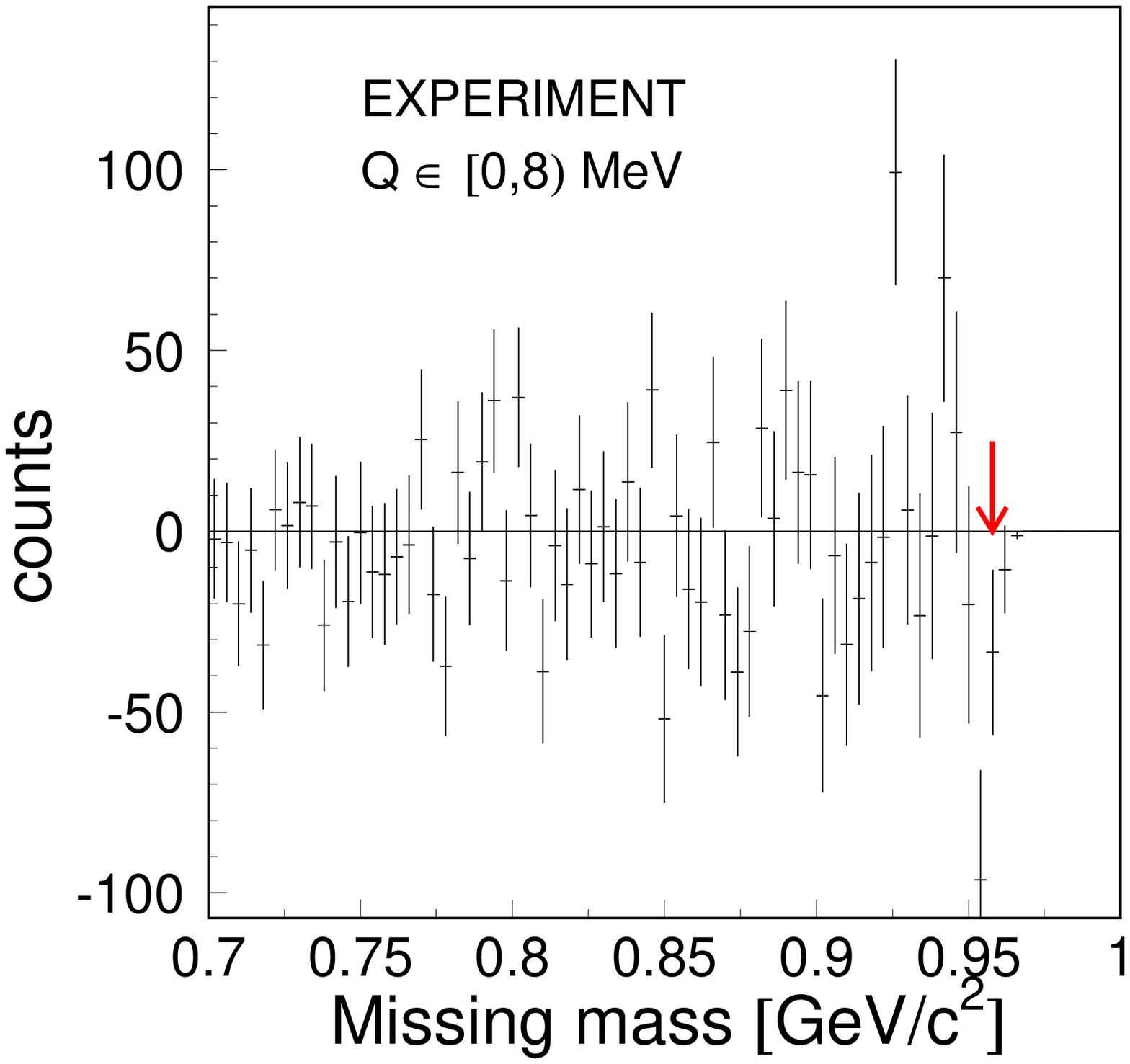}}
\parbox{0.45\textwidth}\centering{\includegraphics[width=.45\textwidth]{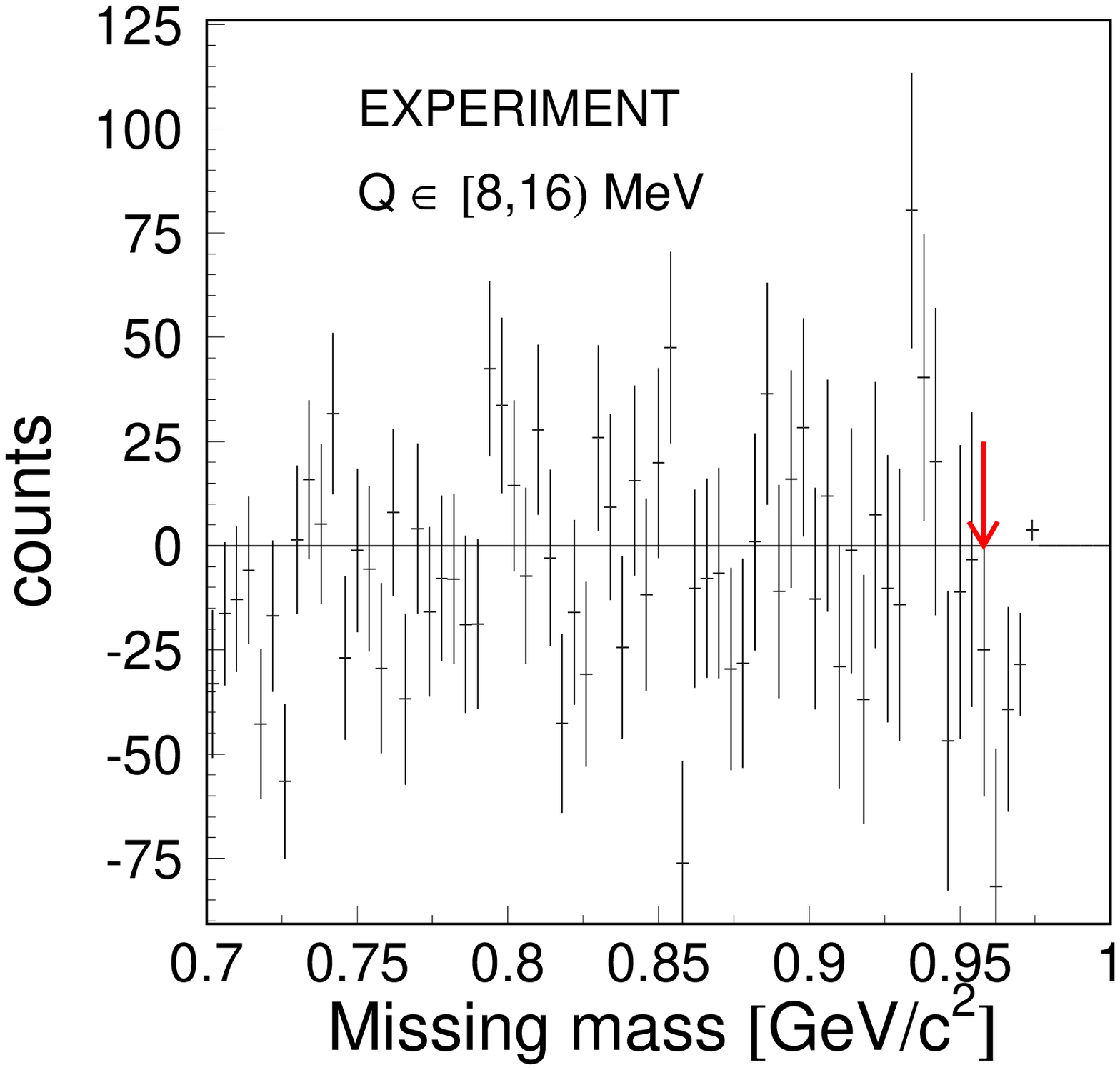}}
\parbox[c]{0.45\textwidth}{\centering\includegraphics[width=.45\textwidth]{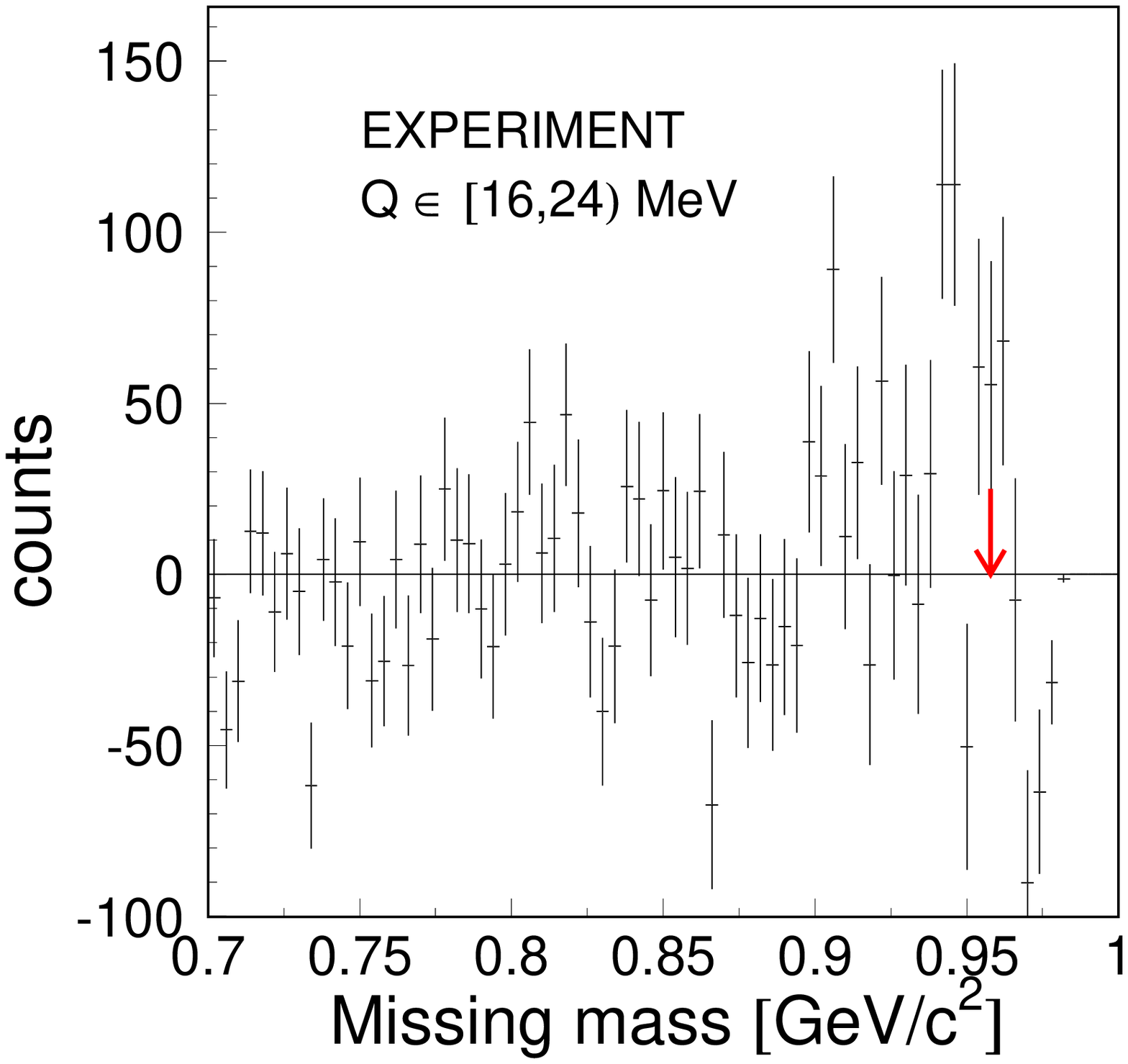}}\hfill
\parbox{.5\textwidth}{\caption{Experimental missing mass distributions of the $pn \to pn\eta^{\prime}$
         reaction for ranges of the excess energy $Q_{cm}$, from 0 to 8~MeV
         {\bf (upper panel left)}, from 8~MeV to 16~MeV {\bf (upper panel right)}
         and from 16~MeV to 24~MeV {\bf (lower panel)} determined after the
         background subtraction. Vertical bars indicate statistical
         errors. Arrows depict the nominal value of the $\eta^{\prime}$ mass. \label{mmbeztla} }}
\end{figure}

Figure~\ref{mmbeztla} shows distributions of the missing mass for the
$pn \to pn\eta^{\prime}$ reaction as determined after subtraction of the background. Due to the
low statistics and very low signal-to-background ratio the signal from
$\eta^{\prime}$ meson created in the proton-neutron collision is statistically
insignificant. Therefore, in this thesis we can only estimate the upper limit for the
$\eta^{\prime}$ meson production in the $pn \to pn\eta^{\prime}$ reaction.
\par
\begin{figure}[H]
\parbox{0.45\textwidth}\centering{\includegraphics[width=.45\textwidth]{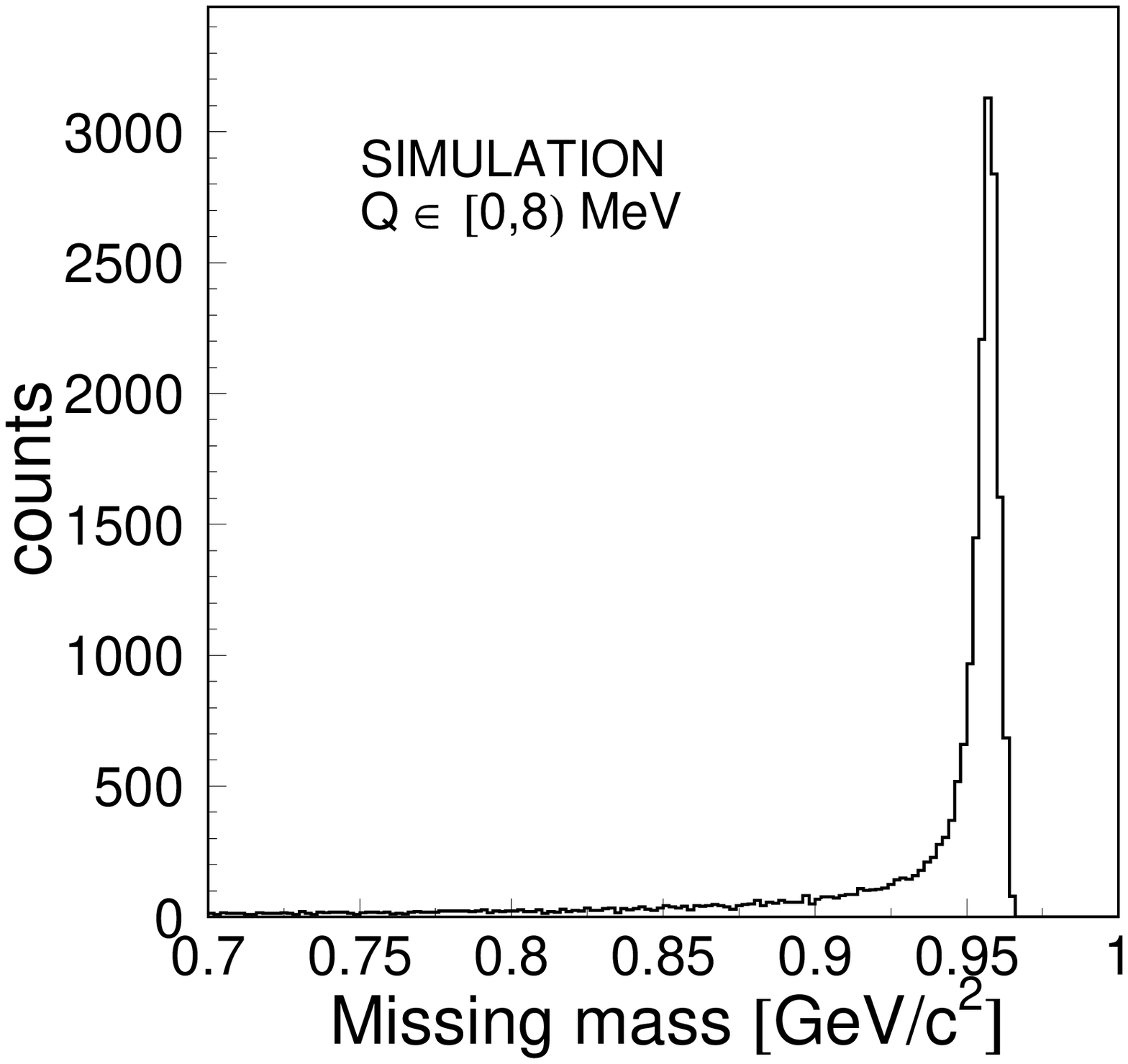}}
\parbox{0.45\textwidth}\centering{\includegraphics[width=.45\textwidth]{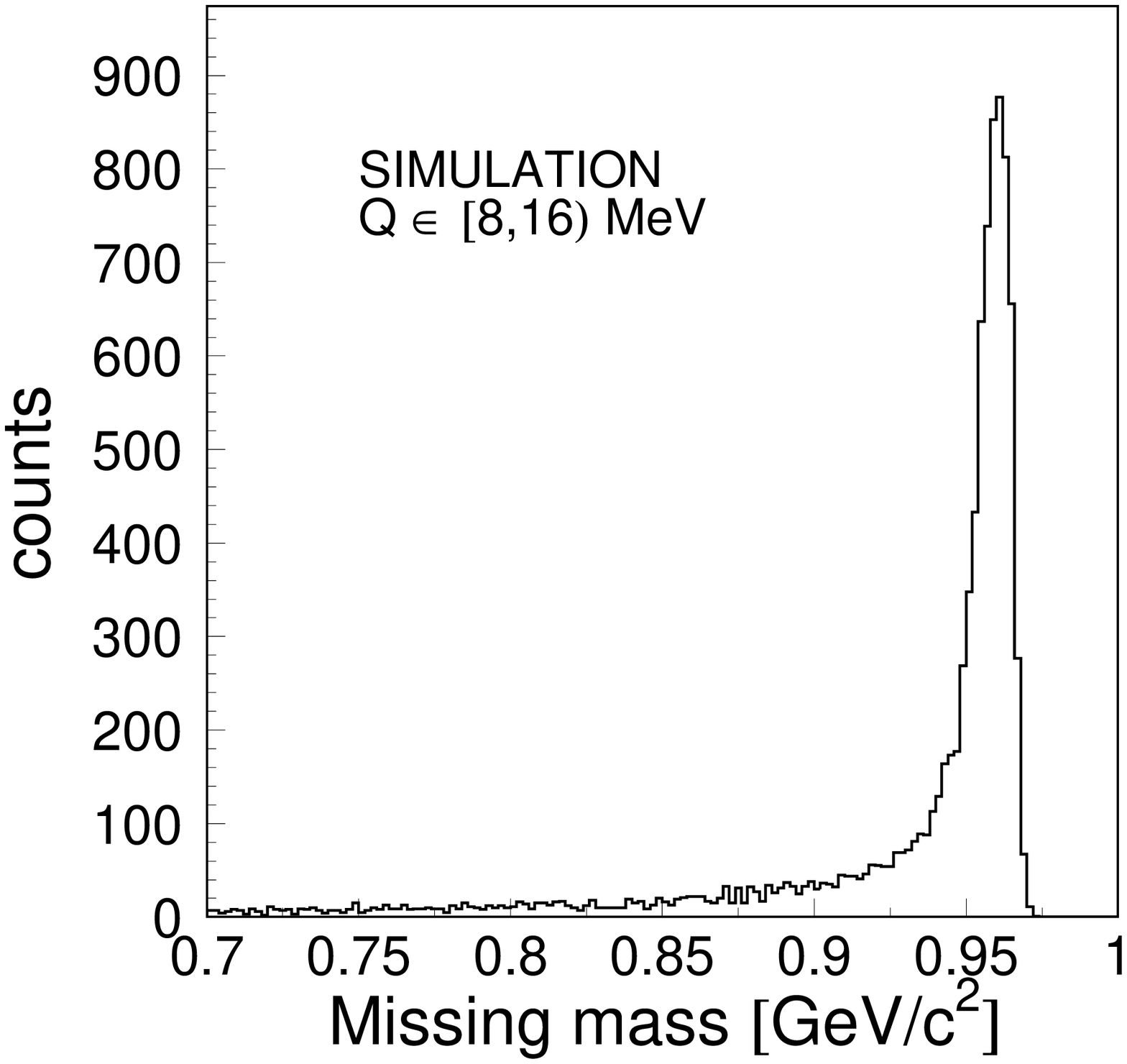}}
\parbox[c]{0.45\textwidth}{\centering\includegraphics[width=.45\textwidth]{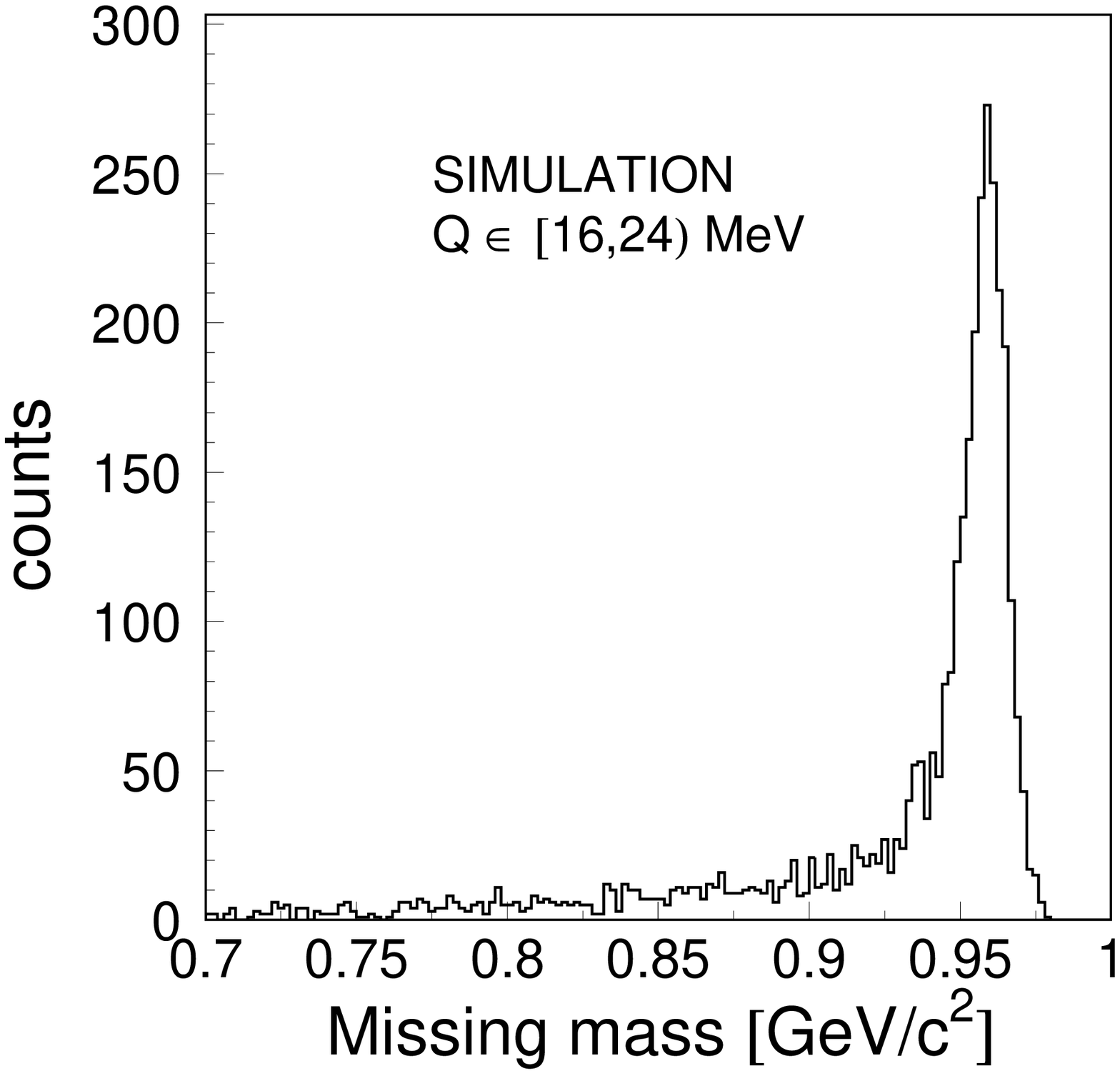}}\hfill
\parbox{.5\textwidth}{\caption{Simulated missing mass distribution of the $pn \to pn\eta^{\prime}$
         reaction for three ranges of the excess energy $Q_{cm}$, from 0 to 8~MeV
         {\bf (upper left)}, from 8~MeV to 16~MeV {\bf (upper right)} and from
         16~MeV to 24~MeV {\bf lower panel}.\label{mm_sym}}}
\end{figure}

Expected missing mass distributions from the Monte Carlo simulations are presented
in figure~\ref{mm_sym}. In the simulation, based on the GEANT-3 package we took
into account the momentum spread of the beam, beam and target dimensions, Fermi motion
of nucleons inside the deuteron target, the geometry of the COSY--11 detector setup,
multiple scattering of particles and other known physical effects were included in the GEANT-3
package~\cite{geant}. The simulated data
have been analysed in the same way as the experimental one, taking into consideration
position-, time- and energy-resolution of all detector components.

In order to estimate the systematic error of the number of the $pn \to pnX$ events,
the change of the missing mass spectra has been studied by varying different parameters describing
the experimental conditions (position of the detectors, time offsets, etc.)
in the analysis and in the simulation.
Change of the global time offset of the neutral particle detector by one standard
deviation of its time resolution $(\sigma_{t}^{N}~=~0.4$~ns) resulted in systematic
error of the missing mass distribution of 3\%. As was disscused above, the systematic
error due to the method used for the background subtraction is equal to 1\%.
The uncertainty due to the model dependence calculation of the Fermi momentum
distribution is equal to 2\%~\cite{czyzyk-dt} and
was estimated as the difference between results determined using the Paris~\cite{lacombe-plb101}
and the CD-Bonn~\cite{machleidt-prc63} potentials. A displacement of the spectator
detector (fixed based on the experimental data) by 1~mm (standard deviation of
its position accuracy) changes the number of events by 5\%.
 

\chapter{ Luminosity determination}
\label{lumi}
\markboth{\bf Luminosity determination}
         {\bf Luminosity determination}

\begin{figure}[H]
\centerline{\includegraphics[height=.4\textheight]{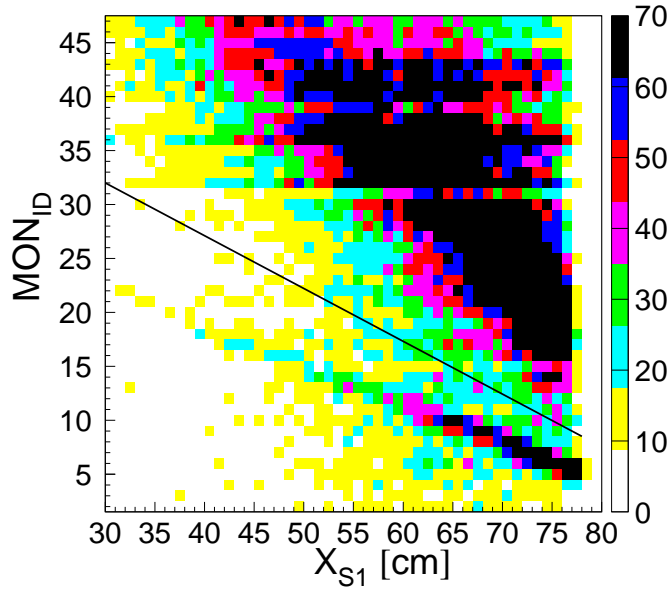}}
\caption{The hit position of the recoil particles measured
         in the monitor detector versus the position in the S1 detector
         of forward scattered particles. The pads number 15, 32 and 39 have lower
         counting rate due to a failure in their signal amplification.
         This imperfection was taken into account in the analysis.}
\label{s1id_monitor}
\end{figure}

For the calculation of the absolute total cross section
of the $pn \to pn\eta^{\prime}$ reaction the knowledge
of the integrated luminosity is mandatory. The determination
of the luminosity is based on the registration of the
quasi-free $pp \to pp$ reaction, which was measured simultaneously
with the $\eta^{\prime}$ meson production in proton-neutron collisions.
In this kind of scattering, the proton from the beam interacts with the proton
bound inside the deuteron target. The outgoing protons were measured in coincidence
with the S1 and S4 scintillator detectors (see figure~\ref{cosy11}). In the S4 scintillator
and the position sensitive silicon detector - referred to as the monitor detector -
the recoil proton is registered. The forward proton is bent in the magnetic field
of the dipole towards the drift chambers and S1 scintillator detector.
Figure~\ref{s1id_monitor} shows the measured correlation between the position
in the S1 scintillator and the position in the monitor detector.
The lower band arises from the free proton-deuteron scattering, whereas
the broad distribution results from the quasi-free proton-proton scattering events.

\begin{figure}[H]
\includegraphics[height=.29\textheight]{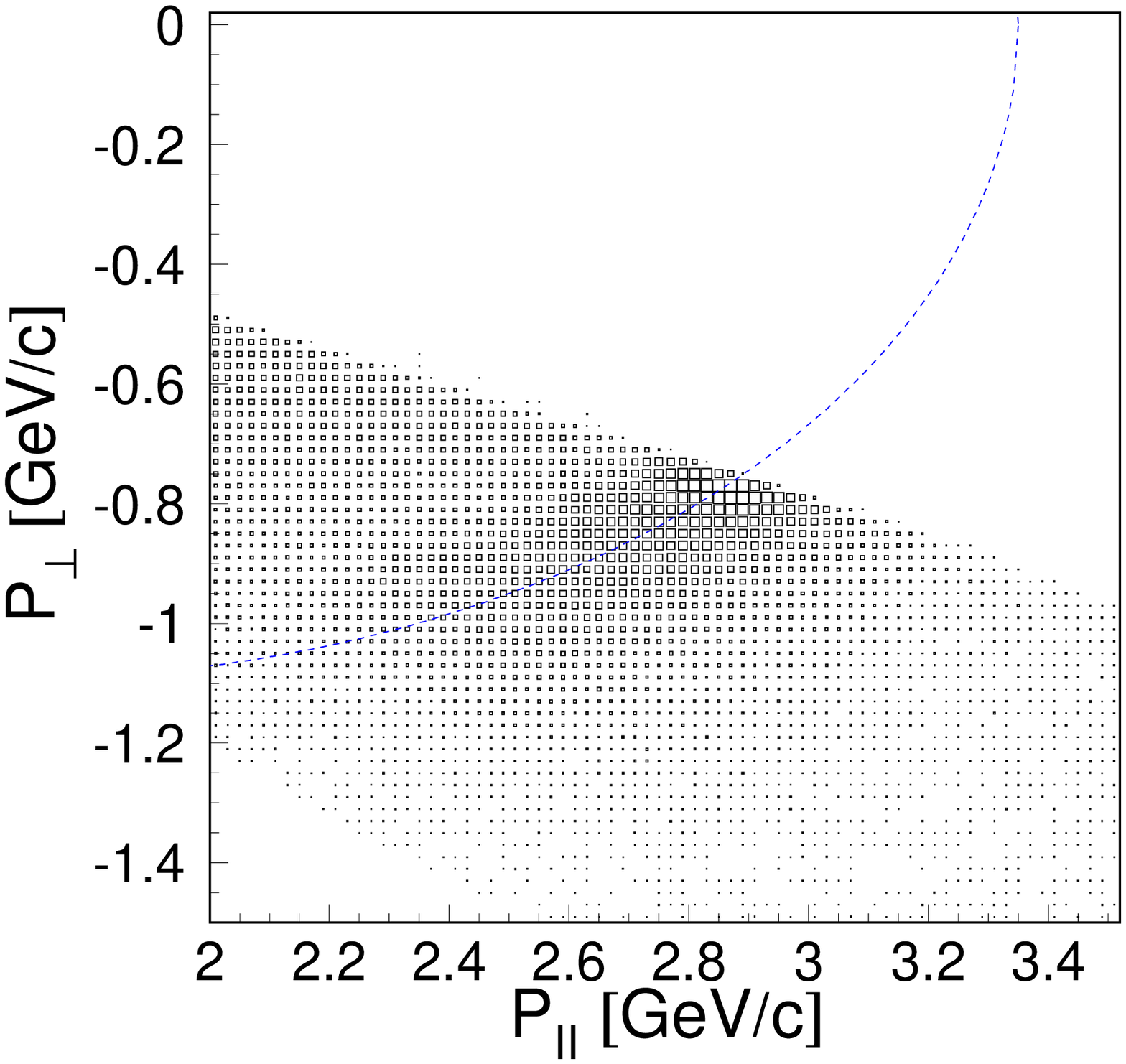}
\includegraphics[height=.29\textheight]{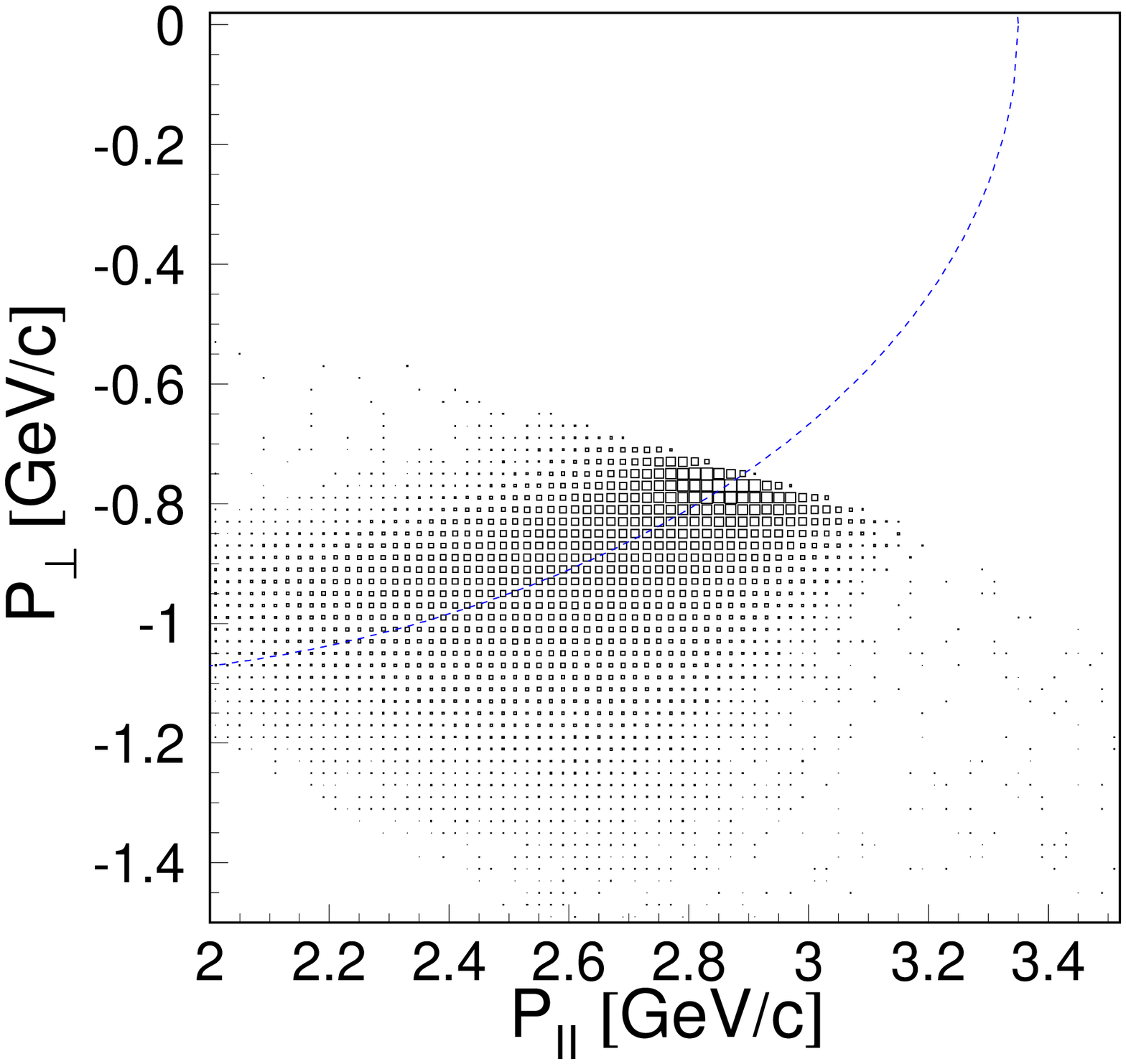}
\caption{Transversal versus parallel momentum component
          of the reconstructed forward proton momentum as obtained
          in the experiment {\bf (left)} and in the simulation {\bf (right)}}
\label{elipsa}
\end{figure}

In the off-line analysis we have separated the quasi-free proton-proton
from the proton-deuteron scattering applying a cut on the histogram as depicted
by black line in figure~\ref{s1id_monitor}. For the further analysis only events above the
black line were taken. The determination of the luminosity
by means of the quasi-free proton-proton reaction measurement takes advantage of the
availability of the precise cross sections determined by the EDDA group~\cite{edda01}.
The normalisation uncertainty of the EDDA differential cross sections is relatively
small (equal to circa 4\%~\cite{edda01,edda02}), whereas the systematic error arising
from the determination of the corresponding differential cross section for the
proton-deuteron scattering is in order of 10\%~\cite{adam-prc75}.\\
The momentum of the fast proton is determined by tracking back the
trajectory reconstructed from the signals in the drift chambers through the
magnetic field to the target point.
Figure~\ref{elipsa}~(left) shows the parallel versus the transversal component of
the reconstucted momentum of the forward scattered proton. Events corresponding
to the elastically scattered protons are seen near the kinematical ellipse,
which is marked as a dashed line. The right panel of figure~\ref{elipsa} presents
an analogous spectrum obtained from the Monte Carlo simulations taking into account
the acceptance of the detection system, the Fermi motion of the nucleons,
and the variation of differential cross section for the elastic scattering
as a function of the scattering angle and energy as it will be described later
in this section.

\begin{figure}[H]
\centerline{\includegraphics[height=.4\textheight]{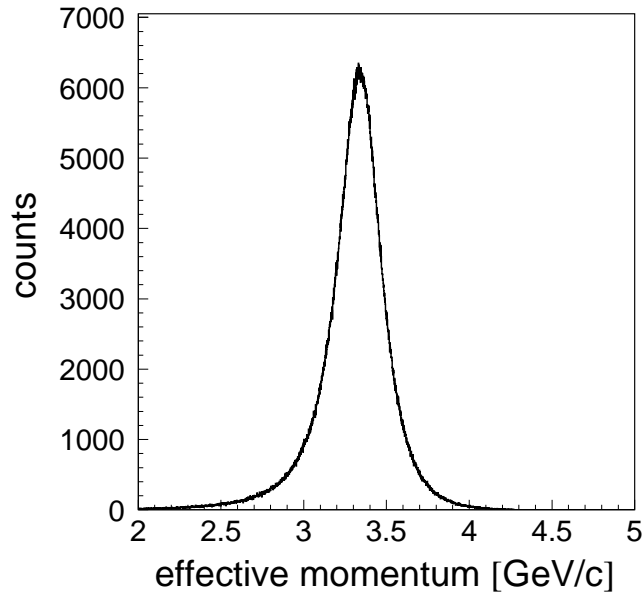}}
\caption{Distribution of the effective beam momentum for the quasi-free
         proton-proton scattering at the nominal proton beam
         momentum of $p_{beam}$~=~3.35~GeV/c as it is seen from the proton
         in the target.}
\label{beam_eff}
\end{figure}

In proton-proton experiments the elastic scattering $pp~\to~pp$
is taken as a reference reaction due to the wealth of
high-precision data available from the measurements of the EDDA collaboration~\cite{edda01}.
The additional advantage of selecting this channel is the relatively high number of
events, being clearly selectable
by a coincident detection of both ejectiles.
To determine the luminosity, the known differential cross
sections for elastically scattered protons are compared to the number of
scattered protons from the experimental data.\\
In case of
quasi-free elastic scattering we have to deal with the Fermi motion of
the nucleons inside the deuteron.
This motion implies that the value of the total energy in
the centre-of-mass system as well as the direction of the centre-of-mass
velocity varies from event to event.
To demonstrate the momentum spread caused by the Fermi motion let us
consider the equivalent effective beam momentum as it is seen from the
proton inside the deuteron.
Figure~\ref{beam_eff} shows the distribution of the effective beam momentum
for the quasi-free proton-proton scattering at the nominal value of the
beam momentum equal to 3.35~GeV/c. The momentum changes significantly
from 2.2~GeV/c up to 4.5~GeV/c, therefore this effect cannot be neglected.
At this point we have to emphasise that due to this effect,
in a single  subrange of the
scattering angle in the laboratory system there are events
originating from scattering at different values of the total energy $\sqrt{s}$,
as well as different scattering angles in the proton-proton centre-of-mass
system.\\
Therefore in order
to calculate the integral luminosity we have to perform simulations taking
into account these effects. Each simulated event has been associated with a weight
corresponding to the differential cross section which is a  function of the scattering
angle and the total energy in the proton-proton centre-of-mass system
$\sqrt{s}$~\cite{moskal-czyzyk-aip950}.
For this purpose we have used cross section values for the $pp \to pp$ reaction computed by means of the SAID
programme~\cite{said} because the accessible data base of the EDDA collaboration was insufficient.
As it is seen in figure~\ref{beam_eff} the effective beam momentum which is seen from the nucleon
inside the deuteron changes from 2.2 GeV/c up to 4.2 GeV/c whereas EDDA
measurements were performed in the beam momentum range from 0.712 GeV/c
to 3.387 GeV/c.

\begin{figure}[H]
\centerline{\includegraphics[height=.32\textheight]{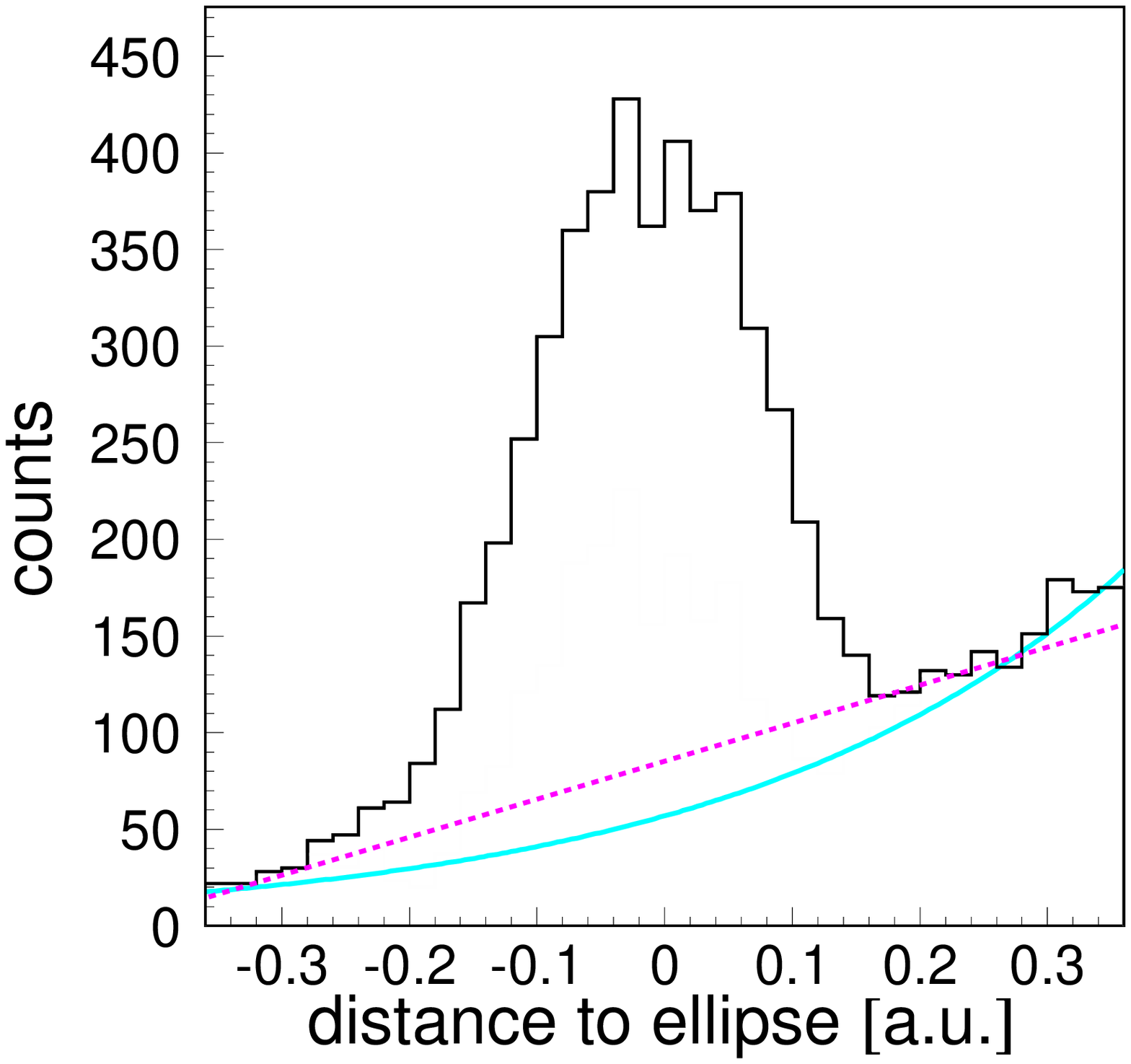}
\includegraphics[height=.32\textheight]{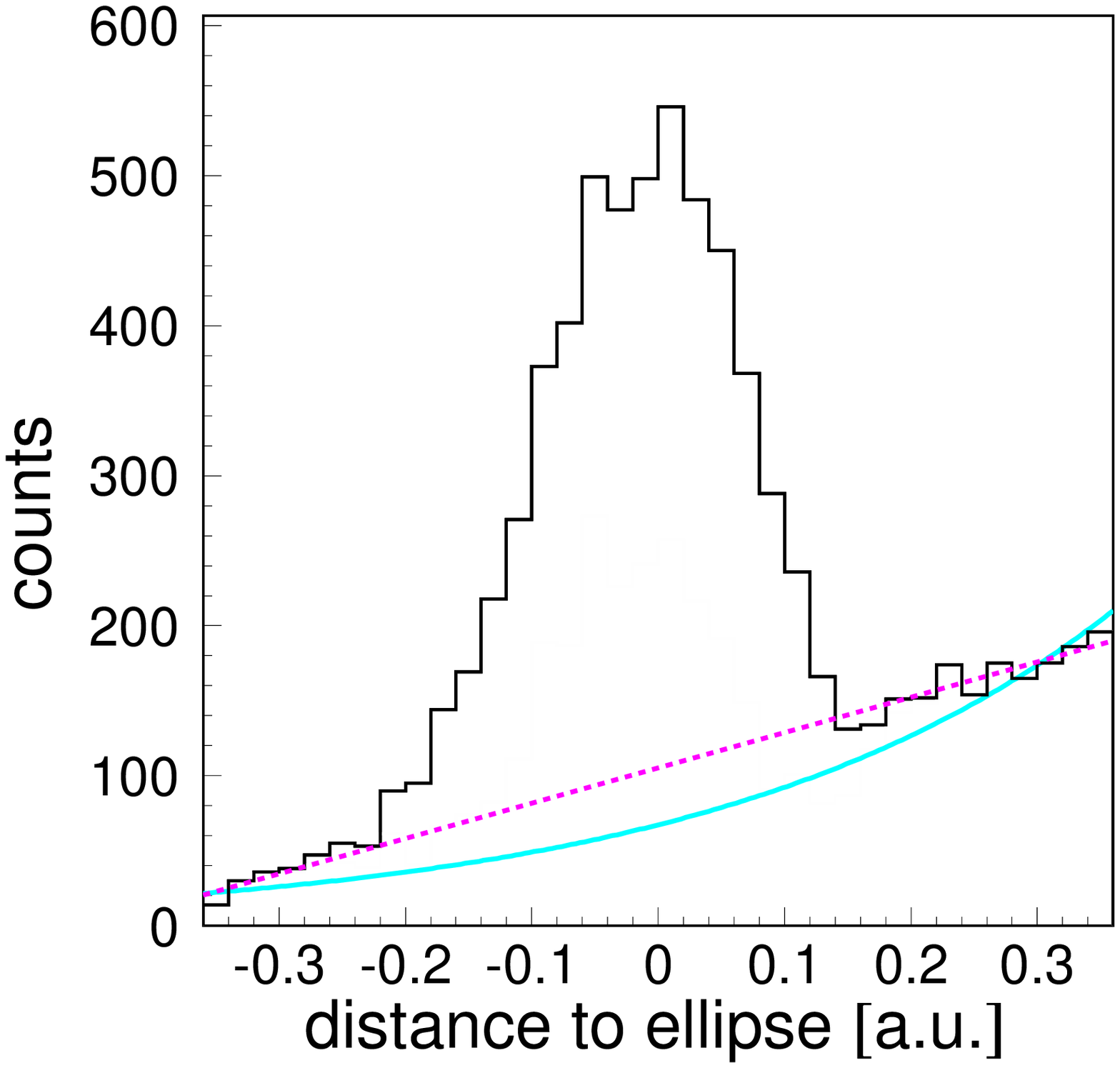}}
\caption{ Projection of the experimental event distribution
          from fig.~\ref{elipsa}~(left) along the expected kinematical
          ellipse for two subranges of the S1 detector. The dash lines depict the
          background as interpreated under the assumption that it is linear.
          The solid lines outline the background described
          by an exponential function.}
\label{dist_exp}
\end{figure}

In order to calculate the integrated luminosity, available range
of the x-coordinate along the S1 detector has been divided into four subranges.
Further on, in order to separate the background
originating from multi-particle reactions, the projection
of the distance of the points from the kinematical ellipse was extracted for each
subrange separately.
The results for two subranges are shown in figure~\ref{dist_exp}. Next, after the
background subtraction the real number of scattered events
$\Delta N_{exp}(\theta_{lab})$ into a given subrange of the S1 detector has been obtained.

\begin{figure}[H]
\includegraphics[height=.29\textheight]{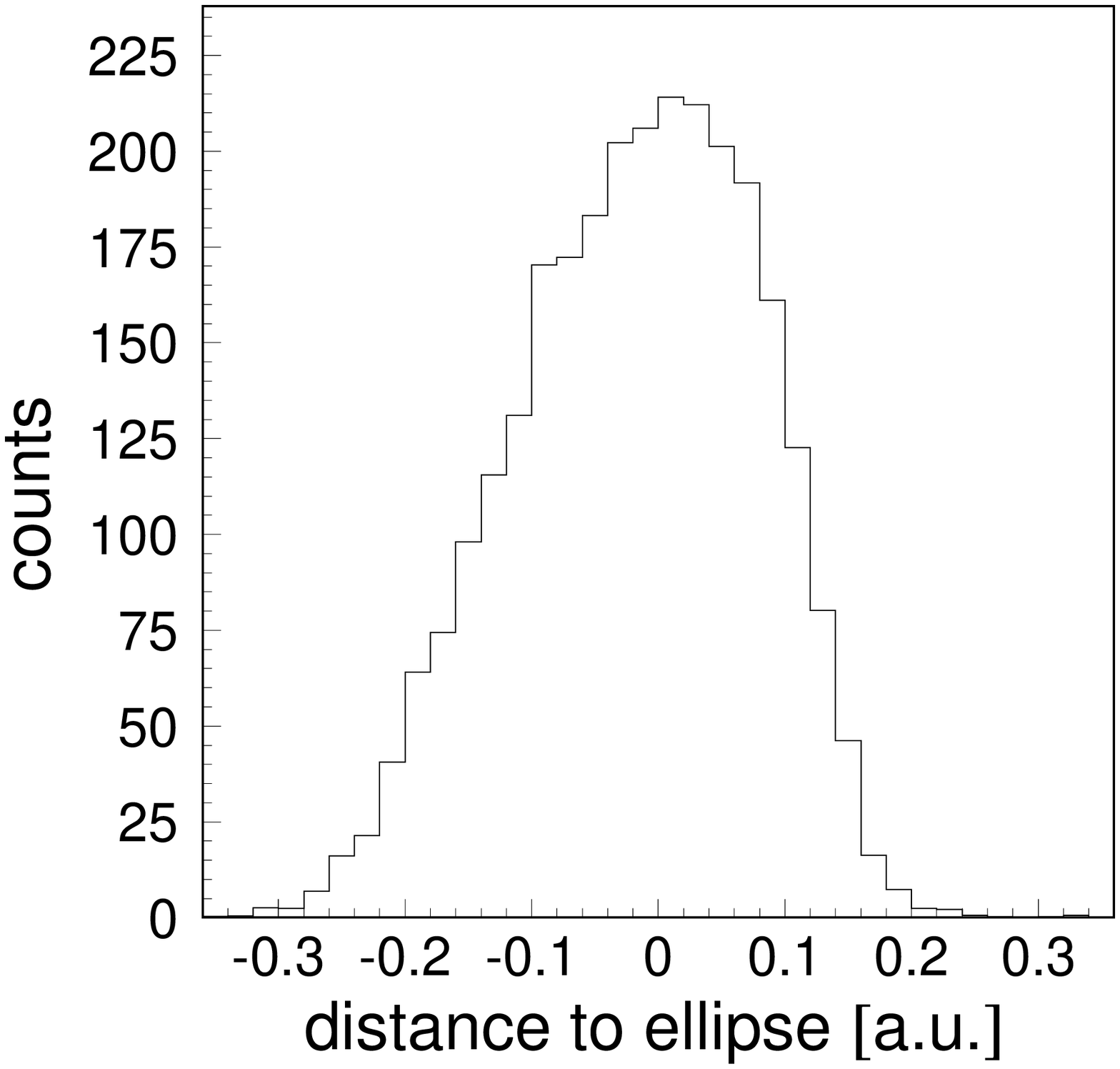}
\includegraphics[height=.29\textheight]{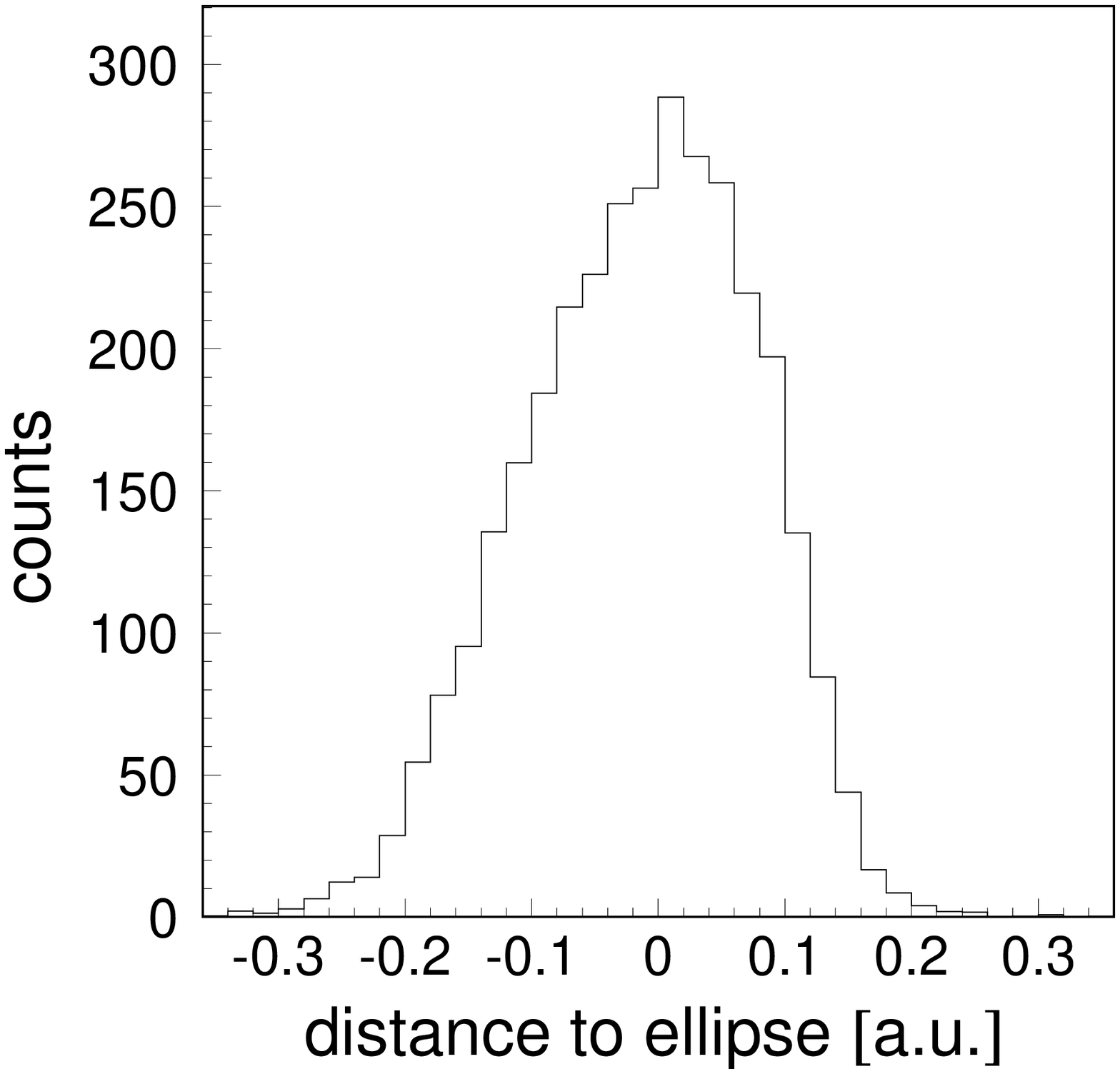}
\caption{Projection of the simulated event distribution
         from fig.~\ref{elipsa}~(right) onto the expected kinematical
         ellipse for two subranges of the S1 detector.}
\label{dist_sym}
\end{figure}
In order to determine the $\Delta N_{MC}(\theta_{lab})$ number,
we have simulated  $N_0 = 10^8$ quasi-free $pp \to pp$ events,
which have been analysed using the same procedure as in case of
the experimental data and which were weighted with the values of the differential
cross sections expressed in units of mb. The results for analogous subranges as for
the experimental distributions are shown in figure~\ref{dist_sym}.
\par
The integrated luminosity was calculated as follows~\cite{moskal-czyzyk-aip950}:
$$L = { N_{0} \over 2\pi} {{\Delta N_{exp}(\theta_{lab})} \over {\Delta N_{MC}(\theta_{lab})}}~=$$
\begin{equation}
{N_{0} \Delta N_{exp} \over 2 \pi \int_{\Delta \Omega (\theta_{lab}, \phi_{lab})}
{d\sigma \over d\Omega}(\theta^{\star}, \phi^{\star}, p_F, \theta_F, \phi_F)
f(p_F, \theta_F, \phi_F) dp_F dcos\theta_F d\phi_F d\phi^{\star} dcos\theta^{\star} }
\end{equation}
We have determined the integral luminosity for all subranges of the S1 detector
individually.
The weighted average integrated luminosity equals to
$L~=~(4.77~\pm~0.06)~\times~10^{36}cm^{-2}$.
The estimated uncertainty is statistical only.
\begin{figure}[H]
\centerline{\includegraphics[height=.4\textheight]{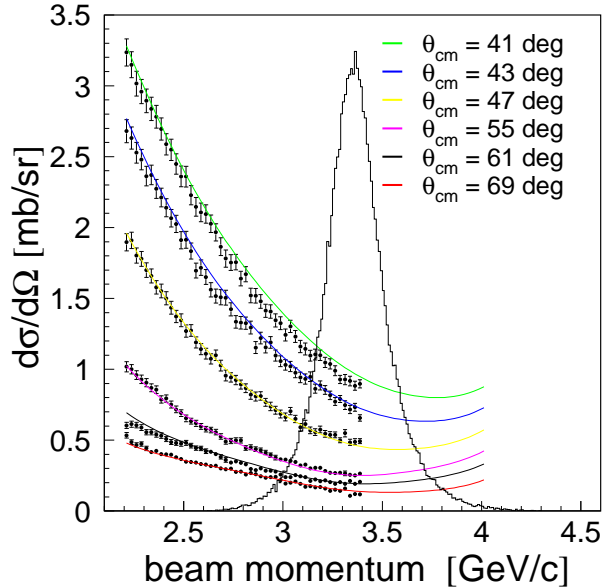}}
\caption{ Differential cross sections as a function of the beam momentum
          for a few values of the scattering angle $\theta_{cm}$ in the centre-of-mass
          system. Black points stand for EDDA collaboration
          data~\cite{edda01}, lines denote SAID calculations~\cite{said}.
          Distribution of the effective beam momentum for quasi-free $pp \to pp$
          reaction calculated at the beam momentum of 3.35 GeV/c is also superimposed
          in the figure.}
\label{said}
\end{figure}

In addition we have estimated the systematic uncertainty.
One of the possible sources of the systematic error may be attached to the
background subtraction. For the evaluation of this systematic uncertainty,
we have subtracted the background in two ways, namely in one case the background
was described by an exponential function, in the latter case we assumed a
linear background (compare the dash and solid lines in figure~\ref{dist_exp}).
Finally the systematic error due to the background
subtraction estimated as the difference between the results determined with the two methods
is not larger than $\pm 4\%.$ \\
The systematic error originating from the assumption of the potential model
of the nucleon bound inside the deuteron is equal to about 2\%~\cite{czyzyk-dt}.
Another source of the systematic error originates from the approximation
of the differential cross section by the calculations using the SAID procedure~\cite{said}.
The total cross section for the proton-proton scattering process has been measured
up to beam momentum $p_{beam}~\approx~3.4$~GeV/c, whereas the effective beam momentum
for the studied reaction ranges up to 4.2~GeV/c. Therefore for events with effective
$p_{beam}~\ge~$3.4~GeV/c the differential cross sections were interpolated using
the SAID programme.
Figure~\ref{said} shows a comparison of the existing
differencial cross section from the EDDA measurement and the SAID calculations.
In the same figure the distribution of the effective beam momentum is also shown.
In order to estimate this uncertainty, we have calculated
the luminosity assuming a linear dependence between the ${d\sigma \over d\Omega}$
and the momentum for events with effective beam momentum larger than 3.0~GeV/c.
Performed calculations show that  the result obtained under this assumption differs
from the result presented above by 5\%.

\begin{figure}[H]
\centerline{\includegraphics[height=.4\textheight]{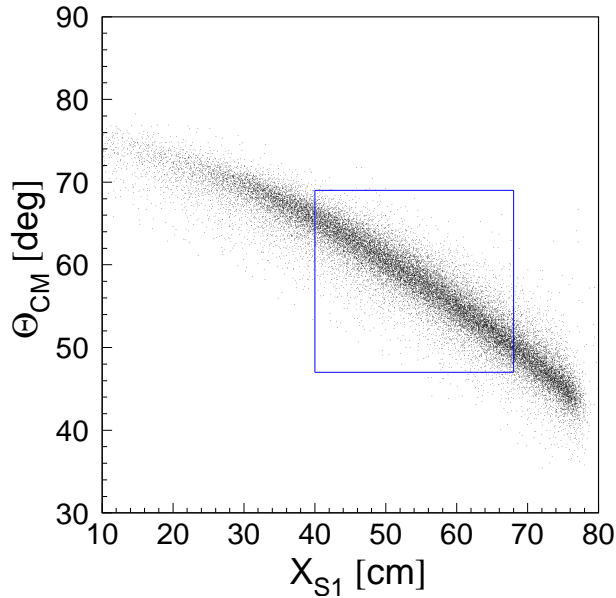}}
\caption{Relation between the centre-of-mass scattering angle and
         the position in the S1 detector for the quasi-free
         proton-proton scattering at $p_{beam} = 3.35$ GeV/c as
         has been obtained from Monte Carlo simulations of the quasi-free
         $pd \to ppn_{sp}$ reaction.}
\label{theta_s1id}
\end{figure}
To reduce this systematic error as much as possible,
events with such scattering angles in the centre-of-mass
system $\theta_{cm}$ have been selected for which the differential cross sections calculated
using the SAID programme are in agreement with the ${d\sigma \over d\Omega}(p)$
distributions measured by the EDDA collaboration (see figure~\ref{said}).
Figure~\ref{theta_s1id} shows the relation between the available range of the
$\theta_{cm}$ angles and the position in the S1 detector for the quasi-free
proton-proton scattering at $p_{beam} = 3.35$ GeV/c  as has been obtained in
the Monte Carlo simulation. The rectangular area denotes the applied cut on the
position in the S1 detector, which restricts the $\theta_{cm}$ angle to the
range from 48 to 69 degrees a good agrement between the
SAID calculations and EDDA data is given.\\
Summarizing, we have estimated the overall systematic error of the integrated
luminosity to be 7\%.


\chapter{ Excitation function}
\markboth{\bf Excitation function}
         {\bf Excitation function}

In the following section we estimate the upper limit of the total
cross section for the quasi-free $pn \to pn \eta^{\prime}$ reaction
as a function of the excess energy. Furthermore, an upper limit of the ratio
$R_{\eta^{\prime}}~=~{{\sigma(pn \to pn\eta^{\prime})} \over {\sigma(pp \to pp\eta^{\prime})}} $
will be determined and compared to the analogous ratio calculated for the $\eta$
meson~\cite{calen-prc58,moskal-prc79}.

\section{Upper limit of the total cross section}

\begin{figure}[H]
\centerline{\includegraphics[height=.5\textheight]{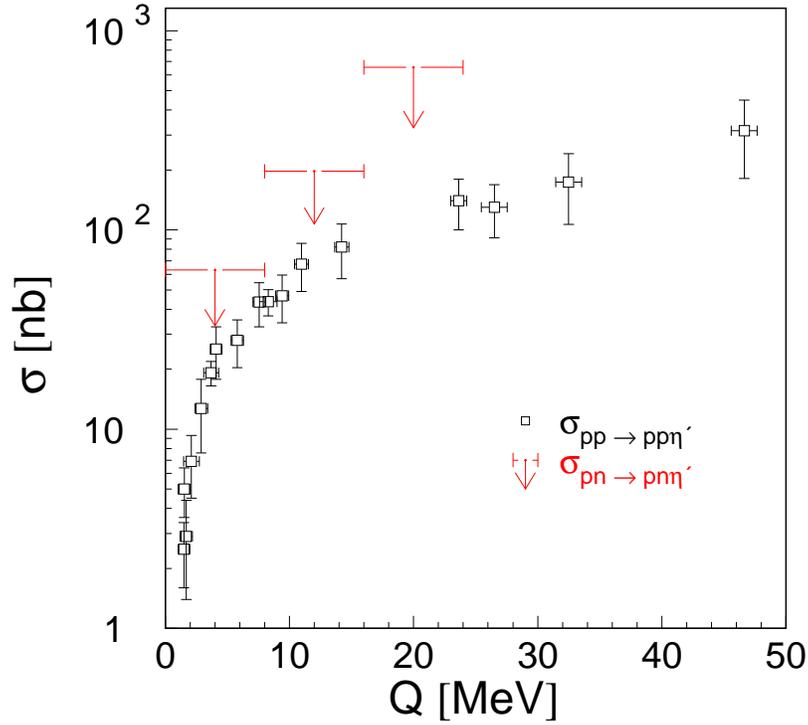}}
\caption{Total cross sections for the $pp \to pp\eta^{\prime}$
         reaction as a function of the excess energy (black open squares).
         Upper limit for the total cross section for the $pn \to pn\eta^{\prime}$
         reaction as a function of the excess energy (red dots).}
\label{cross_pn}
\end{figure}

The total cross section for the $pn \to pn \eta^{\prime}$
reaction was calculated using the formula:
\begin{equation}
\sigma^{TOT}(Q_{cm}) = {N^{\eta^{\prime}} \over {L_{int} \times E_{eff}}},
\end{equation}
where $N^{\eta^{\prime}}$ is the number of $\eta^{\prime}$ mesons for a given
excess energy $Q_{cm}$, $L_{int}$ is the luminosity integrated over the time of the measurement
and  $E_{eff}$ is the detection efficiency including the geometrical acceptance of the
detector system.\\
The luminosity was established from the number of the quasi-free proton-proton scattering
events~(see chapter~\ref{lumi}). The acceptance of the detector setup and efficiency was
determined based on Monte Carlo studies.
After subtracting the missing mass distributions for the negative values of $Q_{cm}$ from
the spectra for positive values of $Q_{cm}$ -- due to the very low signal-to-background ratio --
the signal from the $\eta^{\prime}$ meson was found to be statistically insignificant.
Therefore, we  estimate
the upper limit of the total cross section for the quasi-free
$pn \to pn \eta^{\prime}$ reaction. \\
The number of the $\eta^{\prime}$ mesons can be calculated as the difference between the number of
events $(N^{SIG})$ in the peak for the signal (solid line in figure 6.11) and the number of
events $(N^{BACK})$ of the background in the peak (dash line in figure 6.11):
\begin{equation}
N^{\eta^{\prime}} = N^{SIG} - N^{BACK};~~~~~~~~ \sigma N^{\eta^{\prime}} = \sqrt{N^{SIG}+N^{BACK}}
\end{equation}
The range for the integration has been chosen based on the simulation of the missing
mass distributions (see figure~6.13).
Assuming that in the experiment no $\eta^{\prime}$ mesons are observed $(N^{\eta^{\prime}}~=~0)$
the $\sigma N^{\eta^{\prime}}$ value was used to determine the upper limit of the total cross section
for the $pn \to pn\eta^{\prime}$ reaction at a 90\% confidence level. 
The result is shown
in figure~\ref{cross_pn} and in table~\ref{table_pn}.

\begin{table}[H]
\begin{center}
\begin{tabular}{|c|c|}
\hline
               & upper limit of \\
$Q_{cm}$ [MeV] & $\sigma (pn \to pn\eta^{\prime})$ \\
               & at 90\% CL [nb]\\
\hline
$[0,8)$        & 63  \\
$[8,16)$       & 197 \\
$[16,24)$      & 656 \\
\hline
\end{tabular}
\end{center}
\caption{Upper limit of the total cross section for the $pn \to pn\eta^{\prime}$ reaction
         as a function of the excess energy. The excess energy intervals correspond
         to the binning applied.}
\label{table_pn}
\end{table}
The horizontal  bars in figure~\ref{cross_pn} represents the intervals of the excess energy, for which
the upper limit of the total cross section was calculated.

The efficiency and the acceptance
of the COSY--11 detector system was estimated based on Monte Carlo simulations.
The error of the detection efficiency ($\sigma E_{eff}$) is determined to be 3\%.
The main contribution of this systematic
error comes from the uncertainty in the efficiency in the neutron registration, as was
discussed in section~\ref{efficiency_sim}.\\
The total systematic error of $\sigma^{TOT}(Q_{cm})$ was estimated as the quadratic sum
of independent systematic errors originating from luminosity calculation
($\sigma L_{int}~\approx~7\%$), determination of the efficiency ($\sigma E_{eff}~\approx~3\%$)
and estimation of the number of $\eta^{\prime}$ mesons ($\sigma N^{\eta^{\prime}}~\approx~6\%$)
and is equal to $\approx 10\%$.

\section{Ratio}

\begin{figure}[H]
\centerline{\includegraphics[height=.5\textheight]{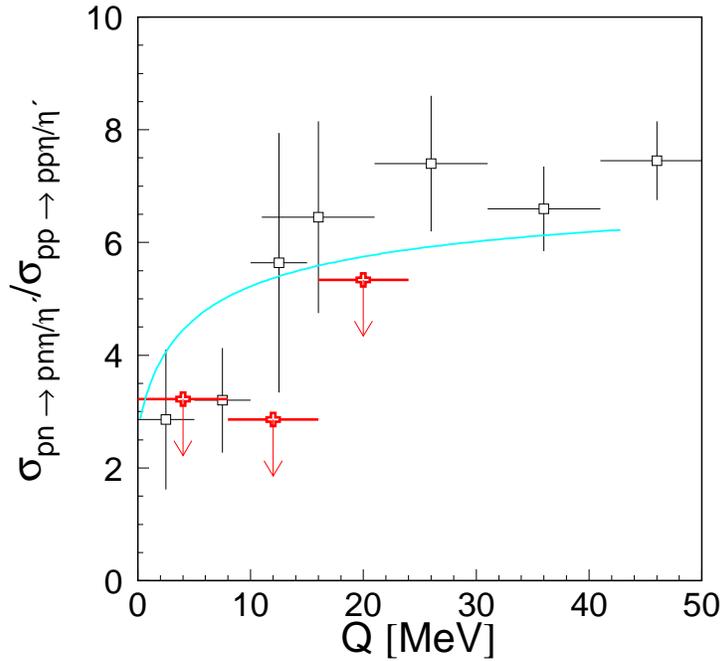}}
\caption{Upper limit of the ratio $(R_{\eta^{\prime}})$ of the total cross sections for the
         $pn \to pn\eta^{\prime}$ and $pp \to pp\eta^{\prime}$ reactions (red crosses)
         in comparison with the ratio $(R_{\eta})$ determined for the $\eta$ meson (open squares).
         The superimposed line indicates a result of the fit to the $R_{\eta}$ data
         taking into account the final
         state interaction of nucleons~\cite{moskal-prc79}.}
\label{ratio_etap}
\end{figure}

Figure~\ref{ratio_etap} presents the upper limit of the ratio
$R_{\eta^{\prime}}~=~{{\sigma(pn \to pn\eta^{\prime})} \over {\sigma(pp \to pp\eta^{\prime})}} $
of the total cross section for the $pn \to pn\eta^{\prime}$ and $pp \to pp\eta^{\prime}$ reaction
as a function of the excess energy (red crosses). Analogous ratios  but for the $\eta$ meson
are also shown as open squares.
The total cross section for the $pp \to pp\eta^{\prime}$ reaction was measured in former
experiments~\cite{moskal-plb474,khoukaz-epj,balestra-plb491,moskal-prl80,hibou-plb438}.
It reveals a strong
excess energy dependence, especially very close to threshold and hence this dependence
must be taken into account when comparing to the results for the $pn \to pn\eta^{\prime}$ reaction
which were established for 8~MeV excess energy intervals. Therefore, for a given
interval of excess energy, we have determined the mean value of the total cross section
for $pp \to pp\eta^{\prime}$ reaction using the parametrisation of F{\"a}ldt and
Wilkin fitted to the experimental data~\cite{wilkin-plb382,wilkin-prc56}~(for details
see appendix~\ref{appendixB}).

\begin{table}[H]
\begin{center}
\begin{tabular}{|c|c|c|c|}
\hline
               & upper limit of  &   & upper limit of \\
$Q_{cm}$ [MeV] & $\sigma(pn \to pn\eta^{\prime})$ & $\sigma(pp \to pp\eta^{\prime})$ [nb] & $R_{\eta^{\prime}}$ \\
     & at 90\% CL [nb]  &   & at 90\% CL\\
\hline
$[0,8)$   & 63  & 19.6 & 3.2\\
$[8,16)$  & 197 & 68.7 & 2.9\\
$[16,24)$ & 656 & 122.7 & 5.3\\
\hline
\end{tabular}
\end{center}
\caption{The columns of the table include from left to right: Excess energy range; Upper limit of
         the total cross section for the $pn \to pn\eta^{\prime}$ reaction;
         Mean value of the $pp \to pp\eta^{\prime}$ total cross section according to
         the parametrisation given in section~\ref{wilkin_eq};
         Upper limit of the ratio $R_{\eta^{\prime}}$ as a function of the excess energy.}
\label{table_ratio}
\end{table}

The determined $pp \to pp\eta^{\prime}$ total cross section values together with
the upper limit of the ratio $R_{\eta^{\prime}}$ are presented in table~\ref{table_ratio}.
In case of the $\eta$ meson, the ratio of the total cross sections for the reactions
$pn~\to~pn\eta$ and $pp~\to~pp\eta$ was determined to be $R_{\eta}$~$\approx~$~6.5 at
excess energies larger than $\approx 15$~MeV~\cite{calen-prc58}, what suggests the dominance
of the isovector meson exchange
in the production mechanism, whereas close to threshold the ratio falls down
to about $R_{\eta}$~$\approx~$~3~\cite{moskal-prc79}.
This decrease may be explained by the different energy
dependence of the proton-proton and proton-neutron final state
interactions~\cite{wilkin-plb382,wilkin-prc56}.
For the $\eta^{\prime}$ meson the upper limit of the ratio for the excess energy range from 0~MeV to 8~MeV
is nearly equal to values of the ratio obtained for the $\eta$ meson,
whereas for larger excess energies (from 8~MeV to 24~MeV) the ratio is notably lower.

   
\chapter{ Comparison with model predictions}
\markboth{\bf Comparison with model predictions}
         {\bf Comparison with model predictions}

The total cross section for the $pn \to pn\eta^{\prime}$ reaction close to the
threshold appeared to be too small to reveal a signal in the missing mass
spectrum with the statistics achieved in the discussed experiment. \\
The established upper limit does not allow to verify the scenario of the production
via the possible gluonic component in the $\eta^{\prime}$ wave function~\cite{bass-pst99,bass-app11}.
However, the results
also do not exclude this hypothesis and may be treated as a promissing finding in this context
since the established upper limits for the ratio R$_{\eta^{\prime}}$ are lower than the values
of the ratio for the $\eta$ meson.
\par
\begin{figure}[H]
\centerline{\includegraphics[height=.29\textheight]{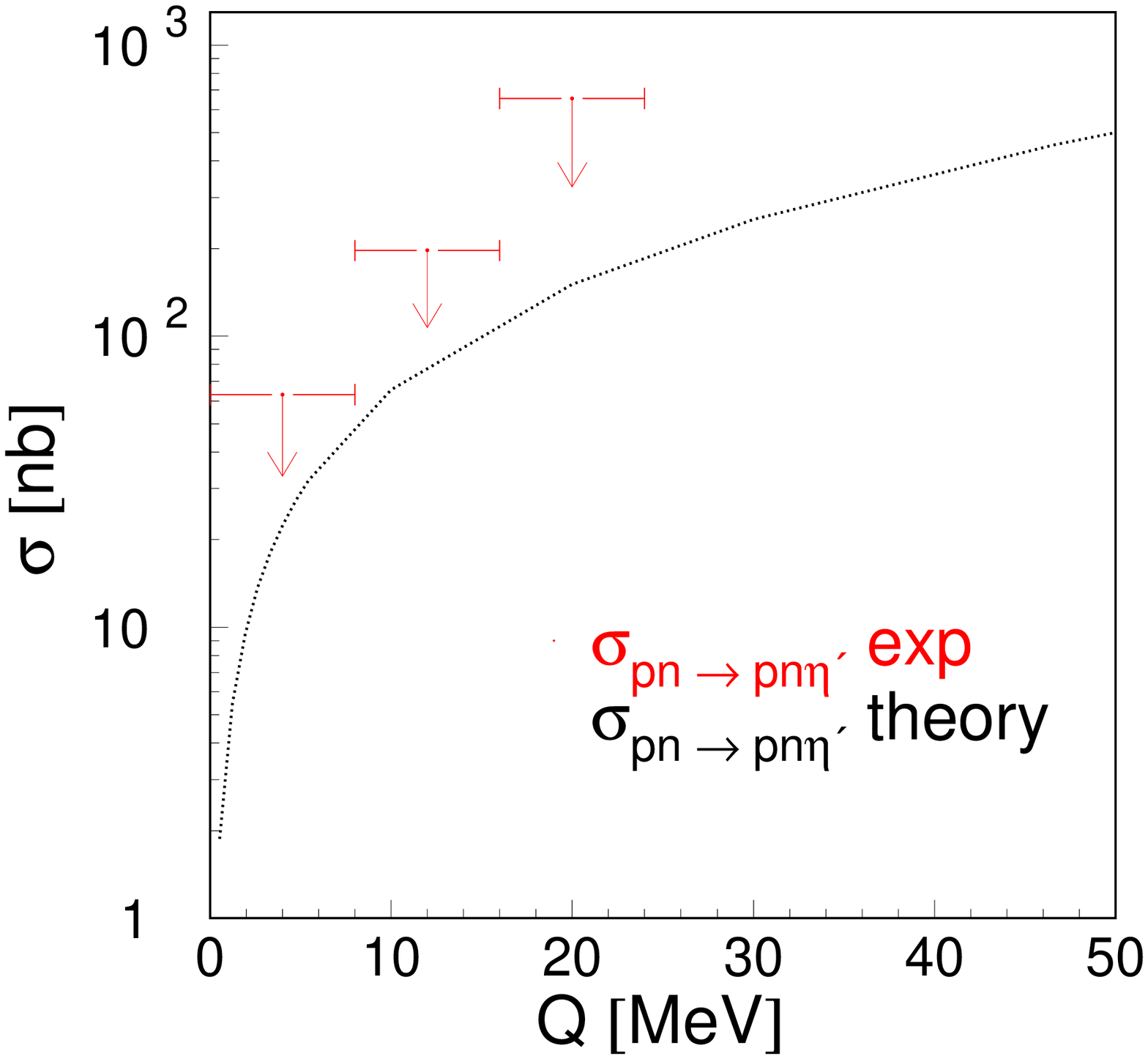}
           \includegraphics[height=.29\textheight]{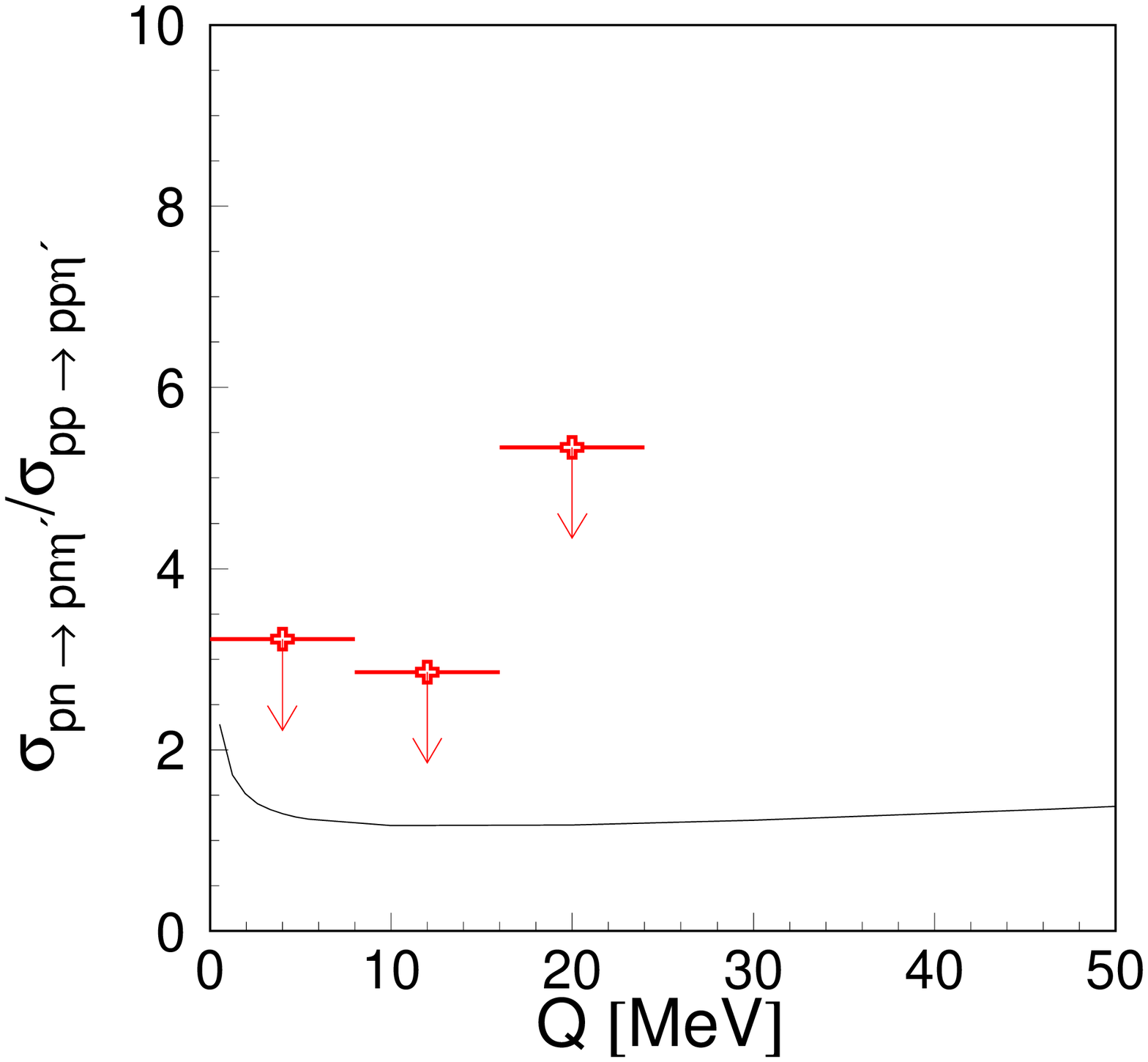}}
\caption{{\bf (Left:)} Upper limit of the total cross section for the
         $pn \to pn\eta^{\prime}$ reaction in comparison to calculations assuming
         that the $\eta^{\prime}$ production mechanism is associated with meson
         exchange~\cite{kampfer-ep}.
         {\bf (Right:)} Upper limit of the ratio $(R_{\eta^{\prime}})$ of the total
         cross sections for the $pn \to pn\eta^{\prime}$ and $pp \to pp\eta^{\prime}$
         reactions (red crosses). The dotted line represents results of calculations
         from~\cite{kampfer-ep}.}
\label{ratio_ep}
\end{figure}

Recently, X.~Cao and X.~G.~Lee~\cite{cao-prc78} performed an analysis of $pN \to pN\eta^{\prime}$ and
$ \pi N \to N\eta^{\prime}$ reactions within an effective Lagrangian approach, assuming
that the S$_{11}$(1535) resonance is dominant in the $\eta^{\prime}$ production.
Numerical results
show that the $\pi$ exchange is most important in the $pN \to pN\eta^{\prime}$ process
and predict a large ratio of $\sigma(pn \to pn\eta^{\prime})$ to $\sigma(pp \to pp\eta^{\prime})$
equal to R$_{\eta^{\prime}}$~=~6.5~in the excess energy range from 0~MeV to 160~MeV.
Thus the authors assumed that the production of the $\eta^{\prime}$ meson proceeds in the same
way as the production of the $\eta$ meson, and as a consequence predicted a R$_{\eta^{\prime}}$ value
equal to the experimentally established R$_{\eta}$ value.\\
Yet, the experimentally determined upper limit of R$_{\eta^{\prime}}$ being the result
of this thesis is considerably lower than the R$_{\eta^{\prime}}$~=~6.5 predicted in reference~\cite{cao-prc78}.
Therefore, on this stage
of knowledge we can exclude the hypothesis that the dominant production mechanism of the $\eta^{\prime}$ meson
in nucleon-nucleon collisions is associated with the excitation and de-excitation of the S$_{11}$(1535)
resonance.\\
Another model has been elaborated by L.~P.~Kaptari and B.~K{\"a}mpfer~\cite{kampfer-ep},
considering the production of the $\eta^{\prime}$ meson
in the $pp \to pp\eta^{\prime}$ and $pn \to pn\eta^{\prime}$ reactions close to threshold
by analysing the data within a covariant effective meson-nucleon theory by including meson
and nucleon currents with resonances S$_{11}$(1650), P$_{11}$(1710) and P$_{13}$(1720).
In the framework of this model
a reasonable agreement with data on $pp \to pp\eta^{\prime}$ has been achieved with a
contribution of the meson conversion currents. The total contribution of nucleonic and
resonance currents was found to be small. \\
The predictions for the $pn \to pn\eta^{\prime}$ reaction, based on this model~\cite{kampfer-ep} are seen as a
dotted line in figure~\ref{ratio_ep}, and
are confronted with the upper limits of the total cross sections for the
$pn \to pn\eta^{\prime}$ reaction (left) and the upper limits of the ratio R$_{\eta^{\prime}}$ (right).
The experimentally determined values
of the total cross sections for the $pn \to pn\eta^{\prime}$ reaction, as well as the results achieved
for the ratio R$_{\eta^{\prime}}$ do not exclude the calculations of ref.~\cite{kampfer-ep}.
\par
For the $pp \to pp\eta^{\prime}$ reaction, calculations of K.~Nakayama and
collaborators~\cite{nakayama-prc69} yields the S$_{11}$(1650) resonance as the dominant contribution
to the production current, whereas the P$_{11}$(1880) resonance, mesonic and nucleonic currents
were found to be much smaller. The combined analysis of the $pp \to pp\eta^{\prime}$ reaction and
photoproduction reactions was crucial for these findings. The details of the excitation mechanism
of the S$_{11}$ resonance, however, are not constrained by the currently existing data.
As the authors pointed out~\cite{nakayama-prc69}, to learn more about the relevant excitation
mechanism, observables other than the cross sections, such as the invariant mass distribution~\cite{pklaja-phd}
are necessary.


\chapter{Summary and perspectives}
\markboth{\bf Summary and perspectives}
         {\bf Summary and perspectives}

The main goal of this dissertation is the determination of the excitation function
of the total cross section for the quasi-free $pn \to pn\eta^{\prime}$ reaction
near the kinematical threshold. The motivation of the presented experiment was the
comparison of the $pp \to pp\eta^{\prime}$ and
$pn \to pn\eta^{\prime}$ total cross sections in order to learn about the production mechanism
of the $\eta^{\prime}$ meson in the channels of isospin $I~=~1$ and $I~=~0$ and to investigate
aspects of the gluonium component of the $\eta^{\prime}$ meson.\\
The experiment, which is described in this thesis has been performed at the cooler synchrotron
COSY in the Research Centre J{\"u}lich by means of the COSY--11 detector system.
For the purpose of this experiment the standard COSY--11 detector setup was extended by
the neutral particle and spectator detectors. The quasi-free
$pn \to pn\eta^{\prime}$ reaction has been induced by a proton beam with a momentum of 3.35~GeV/c in
a deuteron target. All outgoing nucleons have been registered by the COSY--11 detectors, whereas
for the $\eta^{\prime}$ meson identification the missing mass technique was applied.\\

The upper limit of the total cross section for the $pn \to pn\eta^{\prime}$ reaction
in the excess energy range between 0 and 24~MeV determined for the first time ever
is presented in this thesis. Having the total cross section for the $pp \to pp\eta^{\prime}$ reaction
in the same energy range, measured in previous experiments, the upper limit of the ratio
R$_{\eta^{\prime}}$~=~$\sigma(pn \to pn\eta^{\prime})$~/~$\sigma(pp \to pp\eta^{\prime})$ has been estimated.
The achieved result differ from the analogous ratio R$_{\eta}$ obtained for the $\eta$ meson. This allows to
conclude that the production mechanism of these mesons in nucleon-nucleon collisions are different
and the meson $\eta^{\prime}$ is not
dominantly created via the excitation of the S$_{11}$(1535) resonance as it is the case for the $\eta$ meson.\\
It is also shown, that the theoretical predictions assuming the production
mechanism to be associated with the fusion of virtual mesons or with glue excited in the interaction
region of the colliding nucleons are not excluded by the experimental data. However, it is not possible
to distinguish between these mechanisms based on the existing data. This conclusion calls for
further theoretical as well as experimental investigations. \\
For quantitative tests of these mechanisms an order of magnitude larger statistics and
a larger energy range would be required. This can be reached with the WASA-at-COSY facility.
\par
In parallel to the investigation of the $pn\to pn\eta^{\prime}$ reaction,
presented in this thesis, we have extended the experimental
studies to a pure isospin zero state of the interacting nucleons by the measurement
of the quasi-free $pn\to d\eta^{\prime}$ reaction. This experiment was conducted using the
proton beam with a momentum of 3.365~GeV/c and a deuteron target.
Assuming that the ratio of the cross sections for the $pn\to d\eta^{\prime}$
and $pp\to pp\eta^{\prime}$ reactions will be in the same order of magnitude as the ratio already
established~\cite{calen-prc58} for the $pn\to d\eta$ and $pp\to pp\eta$ reactions it is expected
to identify about 1000 $pn\to d\eta^{\prime}$ events in the data sample~\cite{rejdych-aip950}.\\
Together with the results of the  $pp\to pp\eta^{\prime}$ and $pn\to pn\eta^{\prime}$
reactions we would then complete the study of the
$\eta^{\prime}$ meson production cross section in nucleon-nucleon collisions.
This will further reduce significantly the ambiguities of theoretical models and will lead to
a better understanding
of the production mechanism of the $\eta^{\prime}$ meson in nucleon-nucleon collisions.
The result will also be of importance for rate estimates for studying  the $\eta^{\prime}$ meson
decays with the WASA-at-COSY facility~\cite{adam-ep,zielinski-ar}.\\


\renewcommand{\theequation}{A.\arabic{equation}}
\renewcommand{\thefigure}{A.\arabic{figure}}
\renewcommand{\thetable}{A.\arabic{table}}
\setcounter{equation}{0}  
\setcounter{figure}{0}    
\setcounter{table}{0}     

\appendix
\chapter{Precision of the neutron momentum determination}
\markboth{\bf Appendix }{\bf Appendix }
\label{appendixA}

The experimental precision of the missing mass determination of the $pn~\to~pnX$
reaction strongly relies on the accurate measurement of the momentum of neutrons.
The neutron detector is designed to deliver the information about the time at which
the registered neutron or gamma quantum induced a hadronic or electromagnetic reaction.
The time of the reaction combined with this information allows to calculate the
time--of--flight (TOF$^N$) of the neutron (or gamma) on the 7.36 m distance between the target
and the neutron detector. Therefore the accuracy of the time--of--flight depends on $a)$
the neutron detector time resolution and $b)$ the time resolution of the S1 counter.
The neutral particle detector time resolution was determined to be 0.4~ns~\cite{rozek-ar2003}
which corresponds to the time of flight in the scintillator material for a gamma quantum
on a distance of a size of a single module. The neutron
path (l) is defined as the distance between the position of the centre of the target-beam
overlap and the centre of the module which provided a signal as the first one. The accuracy
of this measurement amounts approximately to half the size of the module
which is $4.5$~cm.
In case of neutrons the absolute value of the momentum (p)  can be expressed as:
\begin{equation}
p = m \cdot {l \over {TOF^N}} \cdot  {1 \over {\sqrt{1 - ({l \over {TOF^N}})^2/c^2}}},
\label{ped}
\end{equation}
where $m$ denotes the mass of the neutron, $l$ stands for the distance between the target
and the neutral particle  detector and $TOF^N$ is the time--of--flight of the particle.
Monte Carlo studies of the quasi-free $pp\to pp$ reaction via the
$dp~\to~ppn_{spec}$ process have been performed in order
to establish the momentum resolution of the neutron detector.
A spectator neutron has been denoted as $n_{spec}$.
This reaction was chosen since it could be identified in the data taken during
previous COSY-11 measurements conducted with the deuteron beam~\cite{cezary-fzj},
and hence the simulation results could
have been corroborated by the experimental data~\cite{przerwa-dt}.
The detection of both outgoing protons from the $dp~\to~ppn_{sp}$ reaction
proceeds via the well established method~\cite{brauksiepe-nim} and the spectator neutron
is measured by the neutral particle detector~\cite{przerwa-dt}.
\par
\begin{figure}[H]
\centerline{\includegraphics[height=.32\textheight]{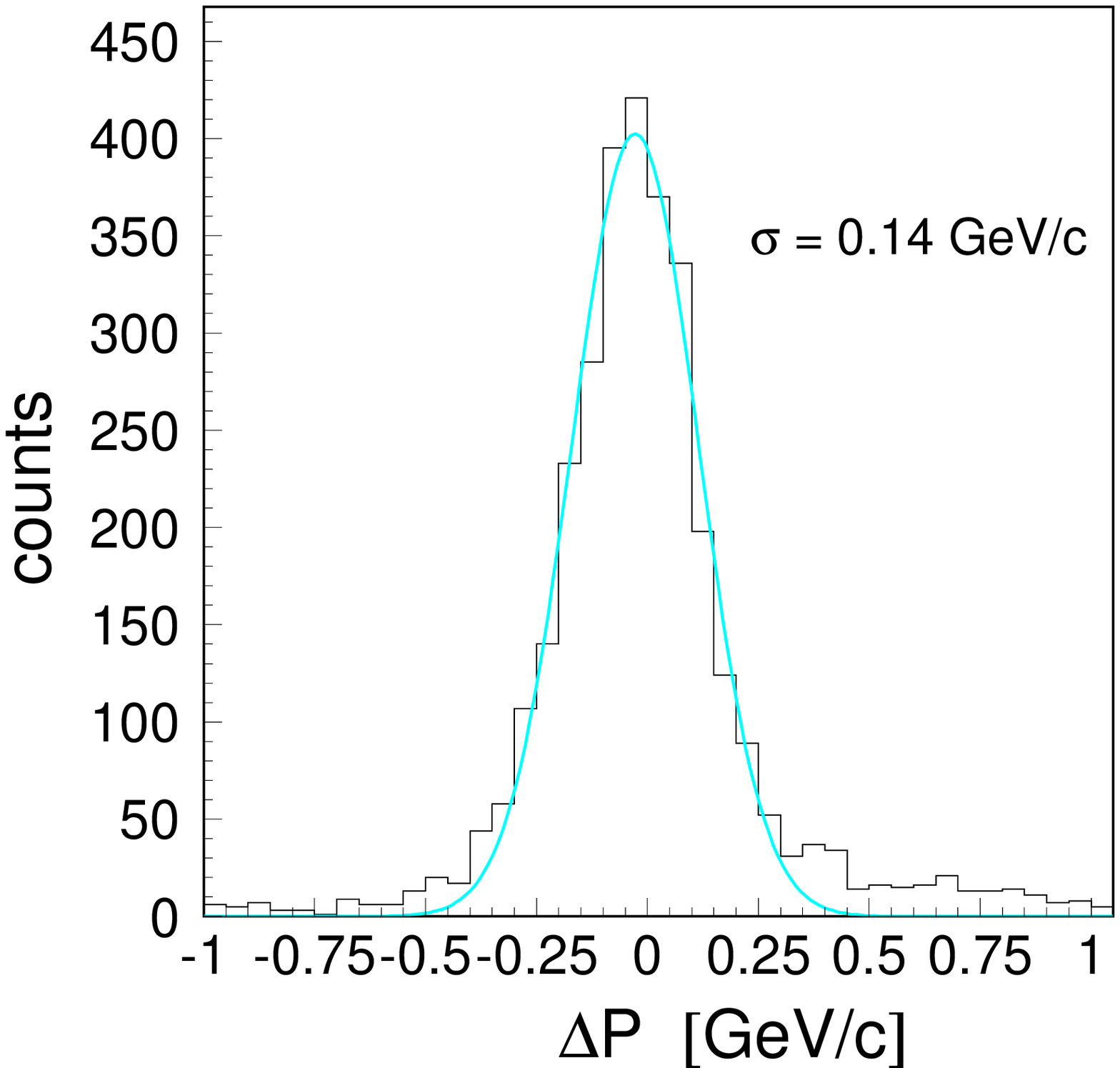}
\includegraphics[height=.29\textheight]{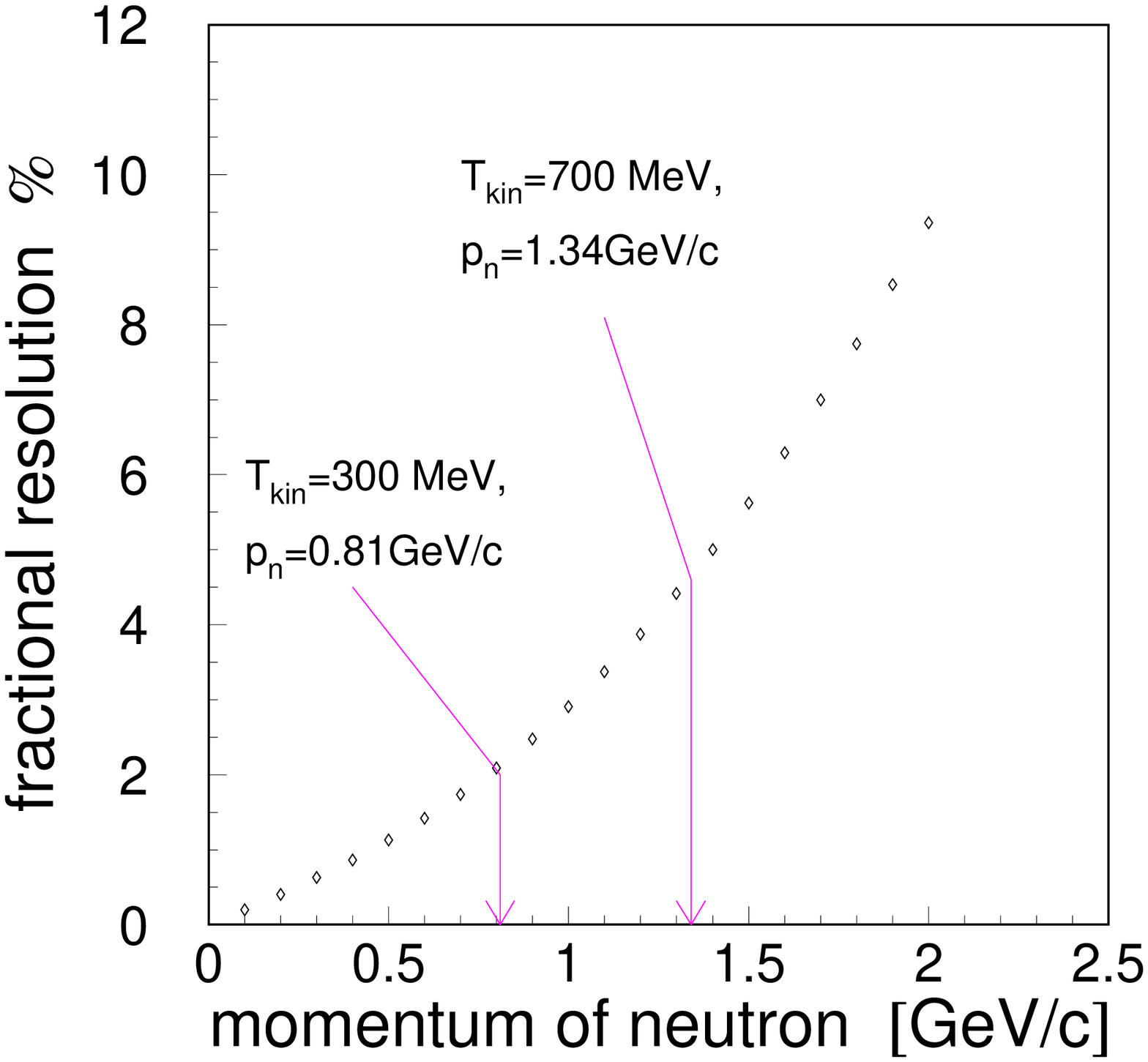}}
\caption{{\bf Left:} Difference between generated ($P_{gen}$) and reconstructed ($P_{rec}$)
         neutron momenta for the $dp~\to~ppn_{spec}$ reaction. {\bf Right:}
         Fractional momentum resolution of the neutral particle detector as a function
         of neutron momentum. Arrows depict the kinetic energy and momentum range
         of neutrons originating from $pd \to p_{sp}pn\eta^{\prime}$ reaction at the
         beam momentum of 3.35~GeV/c.}
\label{dp_p}
\end{figure}

Figure~\ref{dp_p}~(left)  presents the difference ($\Delta P = P_{gen} - P_{rec}$)
between the generated neutron momentum ($P_{gen}$) and the reconstructed neutron momentum
from signals simulated in the detectors ($P_{rec}$). The value of ($P_{rec}$) was calculated
taking into account the time resolution of the neutral particle detector
($\sigma$~=~0.4~ns)~\cite{rozek-ar2003}
as well as the time resolution of the S1 counter ($\sigma$~=~0.25~ns)~\cite{moskal-phd}.
The time at which the proton hits the S1 detector is needed to determine
the time of the reaction in the target place.
The distribution of $\Delta P$  --~for neutrons possessing a momentum of 1.6~GeV/c~--
was fitted by a Gaussian function resulting in a momentum
resolution of $\sigma(P)$~=~0.14~GeV/c~\cite{przerwa-dt}.
The accuracy of the $TOF^{N}$ mesurement is approximately independent of the momentum of neutron,
however, the relative resolution for determining the neutron momentum
alters significantly as its momentum changes.
This fractional momentum resolution --~which is
shown in figure~\ref{dp_p}~(right) as a function of the neutron momentum~--
is given by the equation:
\begin{equation}
{{\sigma p} \over p} = {{{{dp} \over {dt}} \cdot \sigma_t} \over p},
\end{equation}
where $\sigma_t  = \sqrt {\sigma_n^2 + \sigma_{S1}^2}$ accounts
for the time resolution of both the neutral particle and the S1 detectors.
The result presented in figure~\ref{dp_p} was obtained assuming that
$\sigma_n$~=~0.4~ns, $\sigma_{S1}$~=~0.25~ns,
and $l$~=~7.54~m. The value of $l$ corresponds to the distance between the target
and the centre of the neutron detector.

This fractional resolution depends on the neutron momentum and e.g.
for the neutrons produced at threshold for the $pn~\to~pn\eta$ reaction
($p$~=~0.76~GeV/c) it amounts to 2$\%$
and at threshold of the $pn~\to~pn\eta^{\prime}$ reaction
($p$~=~1.06~GeV/c) it is equal to 3.2$\%$


\renewcommand{\theequation}{B.\arabic{equation}}
\renewcommand{\thefigure}{B.\arabic{figure}}
\renewcommand{\thetable}{B.\arabic{table}}
\setcounter{equation}{0}  
\setcounter{figure}{0}    
\setcounter{table}{0}     

\chapter{Parameterisation of the $pp \to pp\eta^{\prime}$ total cross section }
\markboth{\bf Appendix }{\bf Appendix }
\label{appendixB}

\begin{figure}[H]
\centerline{\includegraphics[height=.4\textheight]{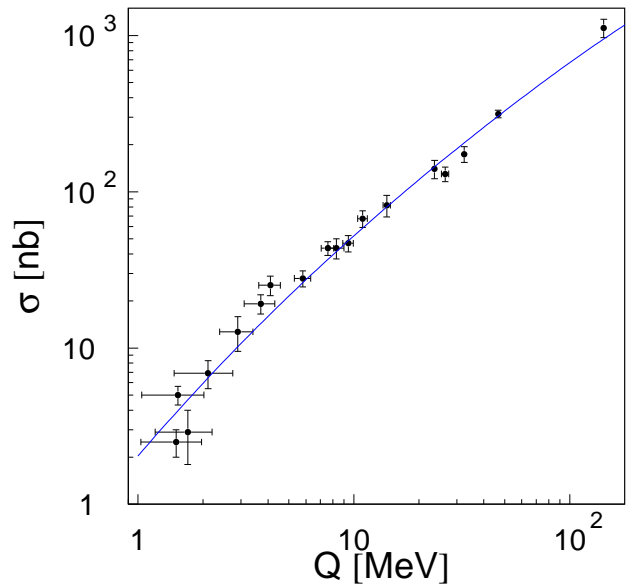}}
\caption{Total cross sections for the $pp \to pp\eta^{\prime}$
         reaction as a function of the excess energy derived from the COSY-11,
        DISTO and SPESIII measurements~\cite{moskal-plb474,khoukaz-epj,balestra-plb491,
        moskal-prl80,hibou-plb438}. The solid line depicts a parameterisation of the total
        cross section using formula~\ref{wilkin_eq}}
\label{pp_wilkin}
\end{figure}

In general the total cross section is defined as an integral of the
probabilities to populate a given phase space  interval over the whole kinematically
available range normalised to the flux factor $F$ according to formula:

\begin{equation}
\sigma~(Q)~=~{1 \over F} \int dV_{ps}|M|^2,
\end{equation}
where $|M|^2$ denotes the production amplitude and $V_{ps}$ is the phase space volume.
Figure~\ref{pp_wilkin} shows the total cross section for the $pp \to pp\eta^{\prime}$
reaction as a function of the excess energy. It is clearly seen that near thereshold
the $\eta^{\prime}$ production strongly depends on the energy.

The solid line depicts a parameterisation of the total cross section utilizing
a factorisation of the production matrix element $|M|$ into the short range
primary amplitude and the initial and final state interaction. The proton-proton FSI
has been taken into account according to the F{\"a}ldt and Wilkin
model~\cite{wilkin-plb382,wilkin-prc56} using the following formula:
\begin{equation}
\sigma~(Q)~=~ C_1~{V_{ps} \over F}~\Big(~1+\sqrt{1+\frac{Q}{\epsilon}}~\Big)^{-2}~=~
C_2~{Q^2 \over {\sqrt{\lambda(s,m_{1}^2,m_{2}^2)}}}~\Big(~1+\sqrt{1+\frac{Q}{\epsilon}}~\Big)^{-2}
\label{wilkin_eq}
\end{equation}
where $C_1$, $C_2$ and $\epsilon$ have to be determined from the data,
and the triangle fuction $\lambda(x,y,z)$~\cite{kajantie} is defined as:
\begin{equation}
\lambda(x,y,z)~=~x^2~+~y^2~+~z^2~-~2xy~-~2xz~-~2yz.
\end{equation}
The function~\ref{wilkin_eq} was fitted to the data and as a result
values of the free parameters $C_2$ and $\epsilon$ have been obtained~\cite{moskal-ijmpa22}:
\begin{center}
$ \epsilon~=~(0.62~\pm~0.13)$~MeV \\
$C_2~=~(84~\pm~14)$~mb
\end{center}
Applying the parameterisation from equation~\ref{wilkin_eq} with known values of
$C_2$ and $\epsilon$ one can compute the total cross section for the $\eta^{\prime}$ meson
production at any value of the excess energy close to threshold.


\newpage
\clearpage
\pagestyle{plain}

\begin{flushleft}

At the end of this thesis, I would like to express my gratitude to a large number\\
of people without whom this work wouldn't have been possible.\\

\vspace{0.5cm}

First of all, I want to express my profound gratitude to Prof. Pawe{\pl} Moskal for
the great help and invaluable advices toward bringing this dissertation to fruition,
for the enourmous patience and for encouraging me to succeed in achieving high goals.
It was a great pleasure but above all a honour to work with You, Pawe{\l}.
You are an example to follow as a scientist and educator.

\vspace{0.5cm}

I would like to express my deeply gratitude to Prof. Walter Oelert
for giving me the opportunity to work within the COSY-11 group. I am also indebt
for comments, valuable advice and encouragement.\\

\vspace{0.5cm}

Less direct but not less important has been the inspiration of Dr. Steven D. Bass.\\

\vspace{0.5cm}

I have also benefited from the great help and advice of Dr. Rafa{\l} Czy{\.z}ykiewicz.\\

\vspace{0.5cm}

I am also very grateful to Prof.~Pawe{\l}~Moskal, Prof.~Walter~Oelert, Prof.~Jerzy~Smyrski,
Dr.~Dieter~Grzonka and Dr.~Steven~D.~Bass for careful reading and correcting this manuscript.\\
\vspace{0.5cm}

I want to express my appreciation to my Friends and  Colleagues from the COSY-11 group
with whom I have the great fortune to interact and work.\\

\vspace{0.5cm}

Dzi{\pe}kuj{\pe} mojej Mamie za wsparcie podczas tych kilku lat studi{\'o}w i nieustann{\pa} trosk{\pe}.
Dzi{\pe}kuj{\pe} bratu Micha{\l}owi, {\.z}e na mnie poczeka{\l}.

\vspace{0.5cm}

Za mi{\l}o{\'s}{\'c}, anielsk{\pa} cierpliwo{\'s}{\'c} i "doping" dzi{\pe}kuj{\pe}
mojemu kochanemu m{\pe}{\.z}owi.
\end{flushleft}

   \cleardoublepage

   \def\bibname{References}
\newpage
\clearpage
\pagestyle{plain}
\pagestyle{myheadings}

\end{document}